\pdfoutput=1 %arXiv admin added this line
\newcommand{\eqnref}[1]{Eq.\ (\ref{#1})}

\newcommand{\eqnarefs}[2]{Eqs.\ (\ref{#1}) and (\ref{#2})}

\newcommand{\secref}[1]{Sect.\ \ref{#1}}
\newcommand{\secsref}[2]{Sects.\ \ref{#1} and \ref{#2}}

\newcommand{\figref}[1]{Fig.\ \ref{#1}}

\newcommand{\figsref}[2]{Figs.\ \ref{#1} and \ref{#2}}
\newcommand{\figdref}[2]{Figs.\ \ref{#1} -- \ref{#2}}
\newcommand{\subfigref}[2]{Fig.\ \ref{#1}(#2)} 

\newcommand{\figsize}[0]{2.6in}
\newcommand{\figsizelarge}[0]{0.7\linewidth}
\newcommand{\subfigwidth}[0]{1.6in}
\newcommand{\Ham}{\mathcal{H}}
\newcommand{\etal}{\emph{et al}.}

\newcommand{\ie}{\emph{i.e.}}
\newcommand{\myfig}[5]{
\begin{figure}[{#5}]
\includegraphics[keepaspectratio,width=#4,angle=0]{#1}%
\caption{#2}\label{#3}%
\end{figure}}
\newcommand{\myfigw}[5]{
\begin{figure*}[{#5}]
\includegraphics[keepaspectratio,width=#4,angle=0]{#1}%
\caption{#2} \label{#3}%
\end{figure*}}
\newcommand{\ab}{_{\alpha\beta}}

\pdfoutput=1
\documentclass[epj]{svjour}
\usepackage{graphicx,dcolumn,bm,verbatim,hyperref,subfigure,amsmath,amssymb,color}
\begin{document}

%\bibliographystyle{apsrev} % Choose Phys. Rev. style for  iography
%\bibliographystyle{nature} % Choose Phys. Rev. style for bibliography

%\preprint{APS/123-QED}
\title{Detecting hidden spatial and spatio-temporal structures in glasses and 
complex physical systems by multiresolution network clustering}
\titlerunning{Detecting hidden spatial and spatio-temporal structures in glasses}
%\titlerunning{Multiresolution network clustering}

%\title{Characterizing large-scale structures in glasses and complex amorphous 
%materials\\
%by multiresolution network clustering}

\author{P. Ronhovde \inst{1} \and S. Chakrabarty \inst{1} \and D. Hu \inst{1} \and M. Sahu \inst{1} 
\and K. K. Sahu \inst{1,2} \and K. F. Kelton \inst{1} \and N. A. Mauro \inst{1} \and Z. Nussinov \inst{1,3} 
\thanks{Corresponding author: zohar@wuphys.wustl.edu}}
\institute{Department of Physics, Washington University in St. Louis, Campus Box 1105, 1 Brookings Drive, St. Louis, MO 63130, USA \and Metal Physics and Technology, ETH, 8093 Zurich, Switzerland \and Kavli Institute for Theoretical Physics, Santa Barbara, CA93106, USA}
% \affiliation{Department of Physics, Washington University in St. Louis, Campus Box 1105, 1 Brookings Drive, St. Louis, MO 63130, USA}
% \affiliation{Kavli Institute for Theoretical Physics, Santa Barbara, CA93106, USA }
% \email[Corresponding author:]{zohar@wuphys.wustl.edu} 
%S. Chakrabarty, M. Sahu, K. K. Sahu, K. F. Kelton,  N. Mauro, and Z. Nussinov $^{*} $}
%\address{Department of Physics, Washington University in St. Louis, Campus Box 1105, 1 Brookings Drive, St. Louis, MO 63130, USA}

%\author{M. Widom}          %\email{}
%\affiliation{Department of Physics, 5000 Forbes Avenue, 
%Carnegie Mellon University, Pittsburgh, Pennsylvania 15213, USA}%
%\author{Peter Ronhovde}  %\email{ronhovde@hbar.wustl.edu}
%\author{Saurish Chakrabarty} %\email{}
%\author{Mousumi Sahu}        %\email{}
%\author{Kisor Sahu}          %\email{}
%\author{Ken F. Kelton}          %\email{}
%\author{Michael Woodard}          %\email{}
%\author{Zohar Nussinov}  %\email{zohar@wuphys.wustl.edu}

\date{\today}

\abstract
{We elaborate on a general method that we recently introduced for characterizing the ``natural'' structures 
in complex physical systems via a {\em multi-scale} network based approach for
the data mining of such structures. The approach is based on ``community detection'' wherein
interacting particles are partitioned into ``an ideal gas'' of optimally decoupled groups of particles. 
Specifically, we construct a set of network representations (``replicas'') 
of the physical system based on interatomic potentials and apply a multiscale 
clustering (``multiresolution community detection'') analysis using 
information-based correlations among the replicas.
Replicas may be (i) different representations of an identical static system or (ii) embody dynamics by when considering replicas to be time separated
snapshots of the system (with a tunable time separation) or (iii) encode general correlations when different replicas
correspond to different representations of the entire history of the system as it evolves in space-time.  
Inputs for our method are the inter-particle potentials or experimentally measured two (or higher order) particle
correlations. 
We apply our method to computer simulations of a binary Kob-Andersen 
Lennard-Jones system in a mixture ratio of A$_{80}$B$_{20}$,
%which has been shown to exhibit glassy behavior.
%We also apply the method to a computer simulation of 
a ternary model system with components ``A'', ``B'', and ``C'' 
in ratios of A$_{88}$B$_7$C$_5$ (as in Al$_{88}$Y$_{7}$Fe$_{5}$), and to atomic coordinates
in a Zr$_{80}$Pt$_{20}$ system as gleaned by reverse Monte Carlo 
analysis of experimentally determined structure factors. 
We identify the dominant structures (disjoint or overlapping) and general length scales 
by analyzing extrema of the information theory measures. 
We speculate on possible links between (i) physical transitions or crossovers
and (ii) changes in structures found by this method as well as phase transitions associated with 
the computational complexity of the community detection
problem. We briefly also consider continuum approaches and 
discuss the shear penetration depth in elastic media;
this length scale increases as the system becomes increasingly rigid. 
% # 89.75.Hc is more tangential 
%PACS, the Physics and Astronomy Classification Scheme.
}
%BEGIN BODY TEXT
\maketitle

\section{Introduction} \label{sec:glassesintroduction}

This article constitutes a longer companion work to an earlier summary \cite{brief} in which the basic notions
to be detailed here for complex physical systems were succinctly outlined. 
We begin by briefly reviewing a special class of complex physical systems that is of great fundamental and technological
importance -- that of amorphous materials. From a practical standpoint, these materials often have industrial processing 
and preparation advantages \cite{ref:physicsofAS,ref:BMGintroMRS} relative to
crystalline systems enabling, e.g., greater solubility of pharmaceuticals \cite{ref:pharmasolubility} 
%\cite{ref:pharmasolubility,ref:yuADDR},
and many other advantages \cite{ref:physicsofAS,ref:caseforBMG}. Below, we list
several specific complex amorphous systems. 
(i) Metallic glasses can be stronger than their respective crystalline structure and exhibit interesting 
electrical, chemical, and magnetic properties \cite{ref:BMGintroMRS}.
(ii) Phosphate glasses are of great use in biomedical applications and chalcogenide glasses are of vital
importance in optical recording media such as Blu Ray technology \cite{wut}.  (iii) Far
more recent and exotic challenges involve incommensurate complex electronic structures
found in systems such as the high temperature superconductors \cite{jan,rmp_steve}. 
The understanding of the character of such non-trivial structures is a
problem of considerable interest in disparate
arenas.

In perfect crystals, the natural system scales are evident by the regular 
ordering of the lattice. 
The fundamental unit cells of a crystal typically involve several atoms 
that are replicated in a simple pattern to span the entire system. 
There are no intermediate scale structures within the system 
from the atomic scale of the lattice up to the complete single crystal.  
Identifying the basic periodic unit cells is vital to the understanding 
of all crystalline solids. This simplicity of structure enables an understanding of crystalline solids in great detail. 
Early on, the existence of specific unit cell structures were postulated
to exist in crystals based on the sharp facets and other macroscopic properties
of large crystals.

There are more complex systems in which new structures appear 
on additional intermediate scales between the atomic-scale and the 
macro-scale of the system. 
In recent years, scientific exploration has endeavored to understand 
a vast array of such complex materials that do not have a simple 
theoretical starting point. 

As alluded to above, some of the best known complex materials are glasses. 
Liquids that are rapidly cooled (``supercooled'') below their melting 
temperature avoid crystallization and instead %, at sufficiently low temperatures, 
become quenched into an amorphous state. 
On supercooling, liquids may veer towards {\em local} low energy structures
(that cannot, on their own, be globally replicated to fill space without the inclusion of other structures)
before being quenched into an amorphous state.
Local low energy structures such as those formed by icosahedral packing are indeed observed in metallic glasses 
\cite{ref:schenkISRO,ref:keltonfirst}.
Due to the lack of a simple crystalline reference, the structure 
of glasses is extremely hard to quantify beyond local 
scales. 

Many theories of glasses rely on the hypothesis of natural structures 
in the glass \cite{ref:nelsonGF,ref:sadocmosseri,ref:tarjusAPT,ref:nussinovAPT}.
Actually finding such structures in a general way has been more elusive. 
{\em How then may we detect and best characterize the most notable structures 
in amorphous systems?}
In the current work, we introduce a general framework to answer this 
question with specific applications to model glass formers.

The outline of this article is as follows. Initially, we present some background information
concerning the pursuit of characterizing structures in glasses and the basic features of
our method  in \secref{sec:background}. Details of the primary simulated systems are presented 
in \secref{sec:simulation}.
The community detection and multi-scale community detections methods are explained in \secref{sec:MRAandCD}.
Our multiresolution method is applied to physical models 
in \secref{sec:application}. We summarize our findings in \secref{sec:amorphousconclusion}.
Details regarding the applied information measures are given in Appendix \ref{sec:vi} 
and those concerning our overlapping dynamics procedure in Appendix \ref{sec:edgediscussion};  we used 
this overlapping community detection to augment 
the community partitions determined by the methods in \secref{sec:MRAandCD}.
Appendix \ref{sec:Sofqprepeaks} highlights the simple (yet often overlooked) fact that prepeaks in the 
structure factor do not
constitute a necessary condition for medium range order. 
In Appendix \ref{sec:overlapmethod}, we discuss how we handle ``overlapping nodes''-
nodes that are common to one or more ``communities'' within the community detection
methods that we employ. The remaining 
appendices elaborate on additional test cases and various facets
of our method. We conclude by discussing structures in space-time and their
general properties. We discuss the detection of multi-scale structure in 
space-time in Appendix \ref{sec:actionmodel}. In Appendix \ref{sec:action}, we consider continuum approaches and 
discuss the divergence of a general length scale-- the ``shear penetration depth'' --
as a supercooled liquid becomes quenched into a rigid glass.

\section{Background} \label{sec:background}
\subsection{Lightning summary of numerous current approaches}

Existing work in the pursuit of understanding the glass transition is vast, 
spanning decades and affecting many fields of science and engineering.
Glass formers exhibit several unique common features \cite{ref:lubchenkowolynes}. 
Glasses demonstrate short range order (SRO) and medium range order 
(MRO) structures, but no easily discernible static long range order exists.
A striking property of glass formers (especially of the so-called
``fragile'' glass formers) is that their relaxation times (as measured, by, e.g., their viscosity) can increase by many orders 
of magnitude over a relatively narrow temperature range. This dramatic slowing down is not associated with the usual 
distinguishing measures of conventional thermodynamic phase transitions. 
These systems have rich energy landscapes with an exceptionally high number of metastable states
\cite{ref:angelaniPEL,ref:parisiglasstrans,ref:sastry,ref:debenedetti,ref:lubchenkoaging,ref:johnsonMRS,ref:doyestructures}. 

Given the broad appearance of glass-related states, 
different frameworks have been explored to work towards a ``universal'' 
characterization of the glass transition. Many theoretical approaches, e.g., \cite{ref:lubchenkowolynes,ref:ktw,ref:tm,ref:mode_coupling,ref:davidr,ref:dyn_con,ref:nussinovAPT,ref:tarjusAPT} have been developed over the years. The notable theory of random first order transitions (RFOT) 
investigates mosaics of local configurations
\cite{ref:lubchenkowolynes,ref:ktw}.  RFOT is
related \cite{ref:nussinovAPT} to theories of locally preferred structures \cite{ref:nussinovAPT,ref:tarjusAPT,ref:dk,ref:nelsonGF,ref:sadocmosseri}.  
Other theories seek a similar measures of structure. Amongst many others, these approaches include spin glass type analysis \cite{ref:tm},
theories of topological defects and 
kinetic constraints
\cite{ref:nussinovAPT,ref:tarjusAPT,ref:ritortsollich,ref:cvetkovicNZ,ref:aharonov},
and numerous ingenious approaches summarized in excellent reviews, e.g., 
\cite{ref:debenedetti,ref:rev_bert,ref:chandler}.
Formally, as demonstrated in \cite{ref:montanariCL}, a growing static length scale is associated with
the diverging relaxation times in supercooled liquids. 
Some works indeed found indications of increasing correlation lengths (static and those
describing dynamic inhomogeneities) as the temperature was lowered
\cite{ref:tanakacritical,ref:mosayebiCLSGT,ref:berthierCL,ref:karmakarsastry}.
Static correlation lengths were amongst other approaches, notably, examined in terms of  (i)``point-to-set'' 
correlations \cite{ref:biroliverrocchio,ref:birolibouchaud} as well as (ii) pattern
repetition \cite{ref:kurchanlevineCL}.  We will briefly discuss these measures later on.

In metallic glasses, early work to ascertain local structural used 
a dense random packing model \cite{ref:bernalrandom}. 
It was later established that such structures are better 
represented by an efficient cluster packing (ECP) model \cite{ref:miracleECP,ref:miraclestructure,ref:miraclestructuralMRS}.
SRO features were thought to pivot on the existence of local 
icosahedral structures centered around solute atoms.
Various idealized SRO configurations were presented 
in \cite{ref:miracleclusters}.
Schenk \etal{} experimentally verified icosahedral short range 
order (ISRO) in undercooled liquids \cite{ref:schenkISRO}. 
Kelton \etal{} were the first to experimentally establish a connection 
between ISRO and the nucleation barrier \cite{ref:keltonfirst}. 
Later work further established the importance of ISRO in glasses 
\cite{ref:luoISROglass,ref:ganeshISRO,ref:shenicosahedral}.

Many structural characterizations are oriented toward static viewpoint 
of the system, but some dynamical features have also been examined.
Analysis of ``free volume'' (unoccupied space between atoms) fluctuations
\cite{ref:miraclestructuralMRS} have been used. %(related to ECP)
Shear stress calculations investigate dynamical processes in glass forming materials \cite{ref:johnsonMRS,ref:delGadoIS}.
Dynamic heterogeneities involving cooperative motion of structures
in a glass have also been studied 
%\cite{ref:kobDH,ref:weeksDH,ref:edigerdynamic,ref:widmercooperHF,ref:stevensonSW}.
\cite{ref:kobDH,ref:weeksDH,ref:widmercooperHF,ref:stevensonSW,ref:glassesoverview}.

Viable characterizations of SRO and MRO structures were advanced for low \cite{ref:sheng} 
and high \cite{ref:shenicosahedral} solute concentrations,
binary \cite{ref:sheng,ref:shenicosahedral}, 
and multicomponent systems \cite{ref:miraclestructure}.

Some methods of characterizing local structures include 
Voronoi tesselation \cite{ref:finneyRSP,ref:aharonov,ref:sheng}, 
Honeycutt-Andersen indices \cite{ref:honeycuttHA}, 
and bond orientation ordering parameters \cite{ref:steinhardtBOO}.
These local measures center on an atom or a given link and, by definition, 
are restricted from detecting more complex longer range general structures.  
%A long-standing challenge addressed in this proposal is the direct 
%detection of general structures in amorphous physical systems

Experimental means to directly measure MRO structures are given 
in \cite{ref:HufnagelSROMRO,ref:treacyMRO}.
Some potential MRO clusters were examined 
\cite{ref:doyestructures,ref:sheng,ref:shenicosahedral}.
Some approaches to understand MRO use pattern matching to idealized 
MRO structures often constructed as agglomerations of perfectly 
ordered SRO features.

% new text here ----------------------------------------------------------
A very useful experimentally driven approach for ascertaining MRO has been 
to look for ``prepeaks'' in X-ray and neutron scattering data [that is, peaks in the 
structure factor $S(q)$ that appear for wavenumbers $q$, corresponding 
to the inverse interatomic distances, which are lower in magnitude than 
that corresponding to the dominant $S(q)$ peak].
We remark here that while this approach may capture general MRO structures, 
it is possible to have such structures without structure factor prepeaks 
(see Appendix \ref{sec:Sofqprepeaks}).
% new text here ----------------------------------------------------------

\subsection{Preliminaries concerning our method}

Our unbiased structure characterization method extends \emph{multiresolution} 
ideas \cite{ref:rzmultires} that have generally been reserved for network science applications 
(analyzing graphs in myriad social and biological networks) to complex materials.
Any complex physical system may be expressed as a network composed 
of nodes that code basic units of interest (e.g., atoms, electrons, etc.).
Weighted links may capture the strength of the interactions between the 
different nodes or experimentally determined correlations (e.g., covariance 
or partial correlation contributions to the structure factor). 
After casting the system as a network, we then search
for ``communities'' of nodes (e.g., clusters of atoms) that are more tightly 
linked to each other (or --  in the case of the use
of correlation functions as link weights --  are more
strongly correlated with one another) than to nodes in other clusters \cite{ref:rzlocal}.

Our multiresolution method employs the notion of ``community detection'', see , 
e.g., \cite{GN,fortunato,fortunato1,newman_girvan,blondel,newman_fast,RB,newman-vector,dynamicshighd,osc,RosB,book_comm},
to quantitatively identify the ``best'' scale (or \emph{scales})
for a complex physical system.
%for the network.
Our approach \emph{does not} rely on intuition or a knowledge of expected 
``important'' features.
Rather, it quantitatively estimates the best scale(s) through 
information-theory-based correlations, such as the variation of 
information (VI) \cite{ref:vi} or normalized mutual information (NMI), 
among different solutions. [In Appendix \ref{sec:vi}, we review these information theory measures.]
In essence, different copies of the community detection problem are 
given to different solvers (``replicas'').
If many of these solvers strongly agree regarding certain features 
of the solution, then these aspects are more likely to be correct.
Similarly, a large discrepancy may indicate large fluctuations on 
a particular scale (or scales).
Extrema in NMI or VI among the results of these independent solvers
then indicate the best scales for the network.

Multiple extrema can therefore indicate the existence of \emph{multiple relevant length/time scales}.
Although in most physical instances there is only a single dominant correlation length,
there are many other cases in which more than one length/time scale is present \cite{explain_scales}.

One distinction between our work and some other established studies 
of local structures in glasses is that our analysis is not looking 
strictly at the positional structure.
Rather, it evaluates structures in terms of the potential energies 
(i.e., the internal binding energies of the clusters, 
see also \cite{ref:doyestructures}).
Our method can encapsulate weights that represent general statistical 
(pair of higher order) stress (or other) correlation functions,
relative atomic displacements, etc.

% new text -------------------------
Our approach provides a perspective different from the ``point-to-set''
\cite{ref:mezardRTGT,ref:franzKGM,ref:montanariCL}
%[M. M´ezard, A. Montanari, J. Stat. Phys. 124, 1311 (2006);
% S. Franz, A. Montanari, J. Phys. A: Math. Theor. 40, F251 (2007);
%A. Montanari, G. Semerjian, J. Stat. Phys. 125, 22 (2006)]
and other methods \cite{ref:kurchanlevineCL}.
%[J. Kurchan, D. Levine, Correlation length for amorphous systems, arXiv:0904.4850].
The point-to-set method examines the overlap between configurations
in a given volume (a ``cavity'') in an equilibrated system and compares 
those to configurations in the same cavity of the equilibrated system 
in which the boundary of the cavity was held fixed.  
Physically, it probes how probable it is to have a particular configuration 
within a disk or ball of a particular diameter given the boundary conditions.
If many small clusters exist inside a sphere of fixed radius, then, a change 
in the boundary conditions %of outside the sphere 
will not significantly alter the bulk cluster distribution within the sphere. 
Conversely, if the sphere radius is smaller than the natural correlation 
length, then the number of configurations compatible with the boundary will 
be small and the overlap will be large.
A different but perhaps related approach, that of Ref.\ \cite{ref:kurchanlevineCL},
%[J. Kurchan, D. Levine, Correlation length for amorphous systems, arXiv:0904.4850] 
examines the distribution of structures inside a given volume to identify 
the correlation length.
The method examines whether the distribution of configurations inside the 
volume occurs with a random frequency (when the linear scale of the volume 
is \emph{larger} than the correlation length) or not (when the linear scale 
of the volume is \emph{smaller} than the correlation length).

Our method does not search for overlap at different scales for a
multitude of configurations nor does it examine their frequency.
Rather, the pertinent structures are revealed by the information theory 
extrema between different copies of the entire system.
We do not  tabulate possible configurations and their
occurrence probabilities or examine the system in restricted volumes.

Furthermore, the basic structures that we find may be used as the 
natural units in a renormalization group type analysis or coarse grained theories where clusters are 
replaced by single nodes and an effective energy can be written that entails 
interactions between the different clusters alone.
Finding the basic units is not trivial in amorphous systems such as glasses 
and numerous disordered systems. 
In these systems, there is generally no symmetry or any other obvious natural key that may determine 
how to optimally partition the system on different scales.

\section{Simulations of model glasses} \label{sec:simulation}

We examine a model glass former derived from a three-component AlYFe 
metallic glass \cite{ref:sahuAlYFe} which we designate as ``A'', ``B'', 
and ``C'' in mixture ratios of $88\%$, $7\%$, and $5\%$, respectively.
The presence of the different components B and C assists in the formation 
of a glassy state \cite{ref:egamiatomistic} since few pure compounds 
manifest a glassy state except under extreme preparation conditions.

\subsection{Ternary model glass former} \label{sec:ternarysimulation}

\myfig{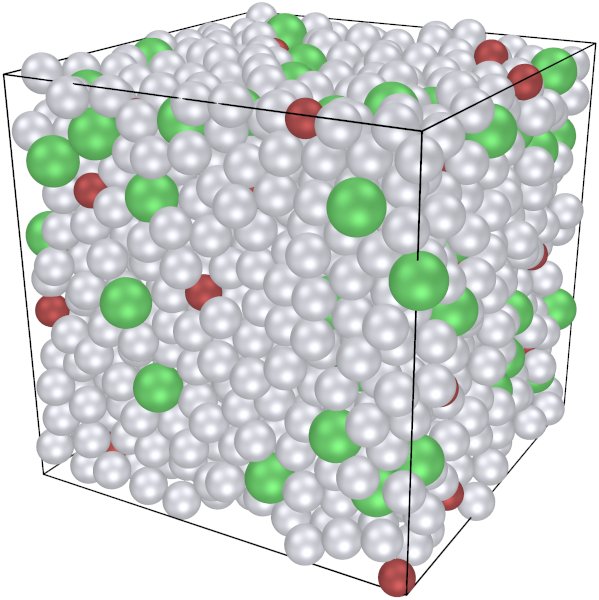}{A depiction of our simulated model glass 
former with three components ``A'', ``B'', and ``C'' with mixture ratios 
of $88\%$, $7\%$, and $5\%$, respectively.
The $N=1600$ atoms are simulated via IMD \cite{ref:imd} in cube 
of approximately $31$ \AA{} in size with periodic boundary conditions.
%The identities of the atoms are Fe (red), Al (silver), Y (green) 
The identities of the atoms are C (red), A (silver), B (green) 
in order of increasing diameters.}
{fig:fullAlYFesystem}{0.85\linewidth}{t!}

\begin{table}[t]
\begin{tabular}{| c | c | c | c | c | c | c | c |}
\hline
& $a_0$ & $a_1$ & $a_2$ & $a_3$ & $a_4$ & $a_5$  \\
\hline
AA & * & * & * & * & * & *  \\
%AA & 2.11* & 9.49* & -32.3* & 3.66* & -10.6* & 6.20* \\
%AA & 2.11  & 9.49  & -32.3  & 3.66  & -10.6  & 6.20  \\
AB & 1.92 & 17.4 & 6.09 & 3.05 & -4.68 & 3.48 \\
AC & 2.38 & 8.96 &-14.9 & 3.11 & -3.88 & 4.38 \\
%BB & 2.01* & 4.95* & 5.01* & 2.74* & -2.26* & 3.00* \\
BB & * & * & * & * & * & *  \\
BC & 1.88 & 8.00 &-3.42 & 2.53 & -1.25 & 3.00 \\
%CC & 2.75* & 15.3* &-6400* & 2.38* & -4.69* & 8.71* \\
CC & * & * & * & * & * & *  \\
\hline
\end{tabular}
\caption{Fit parameters for \eqnref{eq:VrAlYFe} obtained from fitting 
configuration forces and energies to ab-initio data \cite{ref:mihalkovicEOPP,ref:vaspwww}.
The units of the parameters are such that given $r$ in $\AA$, $\phi(r)$ is in $eV$.
(That is, the parameters $a_1$, $a_4$ and $a_5$ are dimensionless, $a_0$ is in $\AA$, $a_2$ is in $eV\AA^{a_5}$ and
$a_3$ is in $\AA^{-1}$.)
The same-species (*) data is replaced by a suggested potential derived 
from generalized pseudo-potential theory \cite{ref:moriartyGPT} 
(see also Appendix \ref{sec:overlapmethod}).}
\label{tab:ppfit}
\end{table} 

\myfig{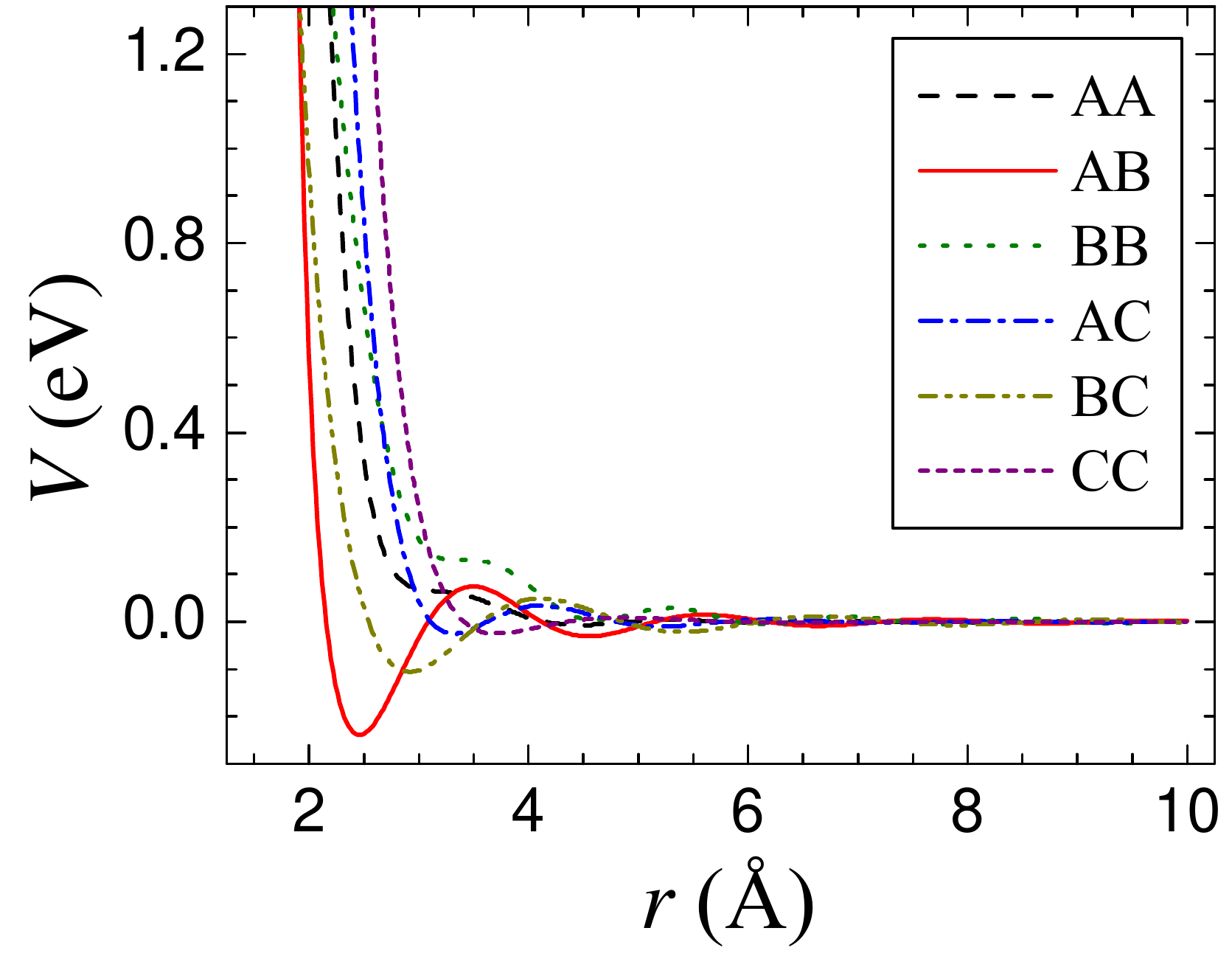}{The pair potentials for our 
three-component model glass former (see \figref{fig:fullAlYFesystem}).
We indicate the atomic types by ``A'', ``B'', and ``C'' which are included 
with mixture ratios of $88\%$, $7\%$, and $5\%$, respectively.
The units are given for a specific candidate atomic realization (AlYFe)
discussed in the text.
The same-species data uses a suggested potential derived from generalized 
pseudo-potential theory \cite{ref:moriartyGPT} (see also Appendix \ref{sec:overlapmethod}).}
{fig:ppfit}{0.85\linewidth}{b}

One system that we examine is derived from a three-component AlYFe
metallic glass.
As depicted in \figref{fig:fullAlYFesystem}, it is a model glass former 
with components designated as ``A'', ``B'', and ``C'' in mixture ratios 
of $88\%$, $7\%$, and $5\%$, respectively.
% The presence of the different components B and C assists in the formation of a glassy state \cite{ref:egamiatomistic}.

We use classical molecular dynamics (MD) \cite{ref:imd} to simulate 
the system dynamics.
For this, we need accurate effective pair potentials that portray the 
pairwise interactions between the atoms in the system.
Our model potential energy function is \cite{ref:mihalkovicEOPP}
\begin{equation} \label{eq:VrAlYFe}
 \phi(r) = \left(\frac{a_0}{r}\right)^{a_1} 
      + \frac{a_2}{r^{a_5} }
        \cos\left( a_3 r + a_4 \right),
\end{equation}
where $r$ is the distance between the centers of two atoms.
This potential form incorporates a realistic weak long range interaction.
Table \ref{tab:ppfit} summarizes the parameter values $a_i$ 
which depend on the specific types for a pair of interacting 
atoms, and \figref{fig:ppfit} shows the respective potential plots.

The interaction parameters $\{a_i\}_{i=1}^{5}$ %in Eq (\ref{eq:VrAlYFe}) 
were determined \cite{ref:mihalkovicEOPP} by fitting configuration 
forces and energies to ab-initio data \cite{ref:vaspwww}. 
The same-species model interactions are finally replaced by that 
suggested by generalized pseudo-potential theory (GPT) \cite{ref:moriartyGPT}. 
As illustrated in \figref{fig:fullAlYFesystem}, we simulate $N=1600$ atoms 
in a cubic system approximately $31$ \AA{} in size using periodic 
boundary conditions.
This width is approximately twice the size of any suspected MRO structures.

The system is initialized at a temperature 
of $T=1500$ K and allowed to equilibrate for a long time using a constant 
number of atoms (N), a constant volume (V), and a constant energy (E). That is, we
work within the NVE ensemble.
After allowing for system equilibration, we save $s$ high temperature 
configurations separated by a fixed period of simulation time. 
Prior to cooling, the length scales in the system are changed by $1\%$ 
to account for the increase in density as a result of cooling since we 
choose to cool the system in an NVT ensemble to control the temperature.
The system is then rapidly quenched to a temperature of $T=300$ K, and it is allowed 
to equilibrate (in a mostly frozen state) in an NVE 
ensemble.
We again save $s$ separate low temperature configurations separated 
by a long period of simulation time.

\subsection{Lennard-Jones glass} \label{sec:LJsimulation}

We additionally test the ubiquitous Lennard-Jones potential using the 
Kob-Andersen (KA) $80$:$20$ binary liquid \cite{ref:kobandersenOne} 
which lies in the glass-forming mixture region \cite{ref:valdesMixing}.
The potential is
\begin{equation} \label{eq:LJVr}
 \phi\ab(r) = 4\epsilon\ab\left[ \left(\frac{\sigma\ab}{r}\right)^{6} 
         - \left(\frac{\sigma\ab}{r}\right)^{12} \right]
\end{equation}
where $\alpha$ or $\beta$ designate one of two atomic types A and B.
Specifically, in accord with KA we set the dimensionless units
$\epsilon_{AA} = 1.0$, $\epsilon_{AB} = 0.50$, $\epsilon_{BB} = 1.5$, 
$\sigma_{AA} = 1.0$, $\sigma_{AB} = 0.88$, and $\sigma_{BB} = 0.80$.

As in the ternary glassy system above, we use MD \cite{ref:imd} 
to simulate a LJ system of $N=2000$ atoms.
The system is initialized at a temperature of $T=5$ (using energy units 
where the Boltzmann constant $k_B=1$) and allowed to evolve for a long 
time.
%Following an intermediate equilibration period at $T=1000$ K,
We save $s$ high temperature configurations separated by $1000$ time 
steps.
The time step size is $\Delta t = 0.0069$ in LJ time units.
Then, the system is rapidly quenched to a temperature of $T=0.01$ which is well below the glass 
transition temperature of the KA-LJ system.
The system is evolved in this mostly frozen state, and we save 
$s$ low temperature configurations separated by $1000$ steps of simulation 
time.

\section{Multiresolution clustering on amorphous materials} \label{sec:MRAandCD}

Our idea is to apply, for the first time, multiresolution network analysis 
methods to ascertain {\em all} pertinent structures in complex amorphous 
materials.  
A key subcomponent of this analysis is the community detection method 
itself.  We first explain these ideas in network analysis and their physical 
analogs.

\subsection{Physical Motivation}

In an {\em ideal decomposition} of a large system into decoupled subsystems (communities), there is no interaction between 
different communities, and the system is effectively that of an ideal gas
of disjoint communities.
Stated differently, in the simplest setting in which the Hamiltonian would 
be block diagonal, the evolution of nodes (e.g., atoms) in each community 
would be decoupled from all other nodes in other communities.
We next consider a fundamental Newtonian many particle setting.
If the total force on a cluster is zero (i.e., if the cluster is strongly decoupled from all others) $\vec{F}_{cluster} =0$ then, 
the particles in that cluster will drift (on average) with a uniform 
velocity (the center of mass velocity of that cluster), 
%$\vec{v}_{cluster}$ is optimally constant as a function of time, 
i.e., $\vec{v} \approx \vec{const}$. \cite{commentv}

In such instances, we may treat each community as a different particle 
in an ideal gas of non interacting such particles.
The general problem is to find (the time dependent) permutation that 
renders the pair interaction strengths and/or correlations between particles into the
best possible block diagonal form (on the time scale 
chosen). Community detection emulates this for graphs. 
In the atomic realization that we discuss in this article, 
this emulates a partition of the system into
optimally decoupled clusters such that the system
may be viewed as an {\em ``ideal gas'' of decoupled
communities}.

For slow cooling of a liquid which enables crystallization, a first order 
or critical transition appears in the community detection problem.
A similar transition is materialized in many slowly cooled liquids as they crystallize to form
nicely ordered solids. 
By contrast, in an extremely rapid cooling of a liquid, the interactions between the 
particles are similar to those in spin-glass systems (i.e., the particles are not nicely organized and consequently the 
inter-particle interactions harbor a large degree of randomness). It is notable that a spin-glass transition 
appears in the community detection problem for random graphs \cite{ref:huCDPT}.

\subsection{Community detection} \label{sec:hamiltonian}

In a graph such as that depicted in \figref{fig:networkexample}, 
nodes correspond to abstracted fundamental elements of the system, 
and edges represent defined relationships between the nodes.  
Community detection describes the problem of finding strongly connected 
groups of nodes.  
Nodes in different clusters are more 
weakly connected than the nodes belonging to the same cluster.
As alluded to earlier, we apply direct physical analogies where a node corresponds 
to a single atom.  
Edges, and their corresponding weights, are directly defined 
by the associated pair-wise interaction energy (or, in the absence
of known interactions, may
be set by measured inter-particle correlations).
Specifically, we use the interatomic potential energies 
in \eqnarefs{eq:VrAlYFe}{eq:LJVr}.
This potential model of the network edges is physically appealing 
in that finding the best partition for the network is akin 
to minimizing the cluster binding energies of the physical system.

\myfig{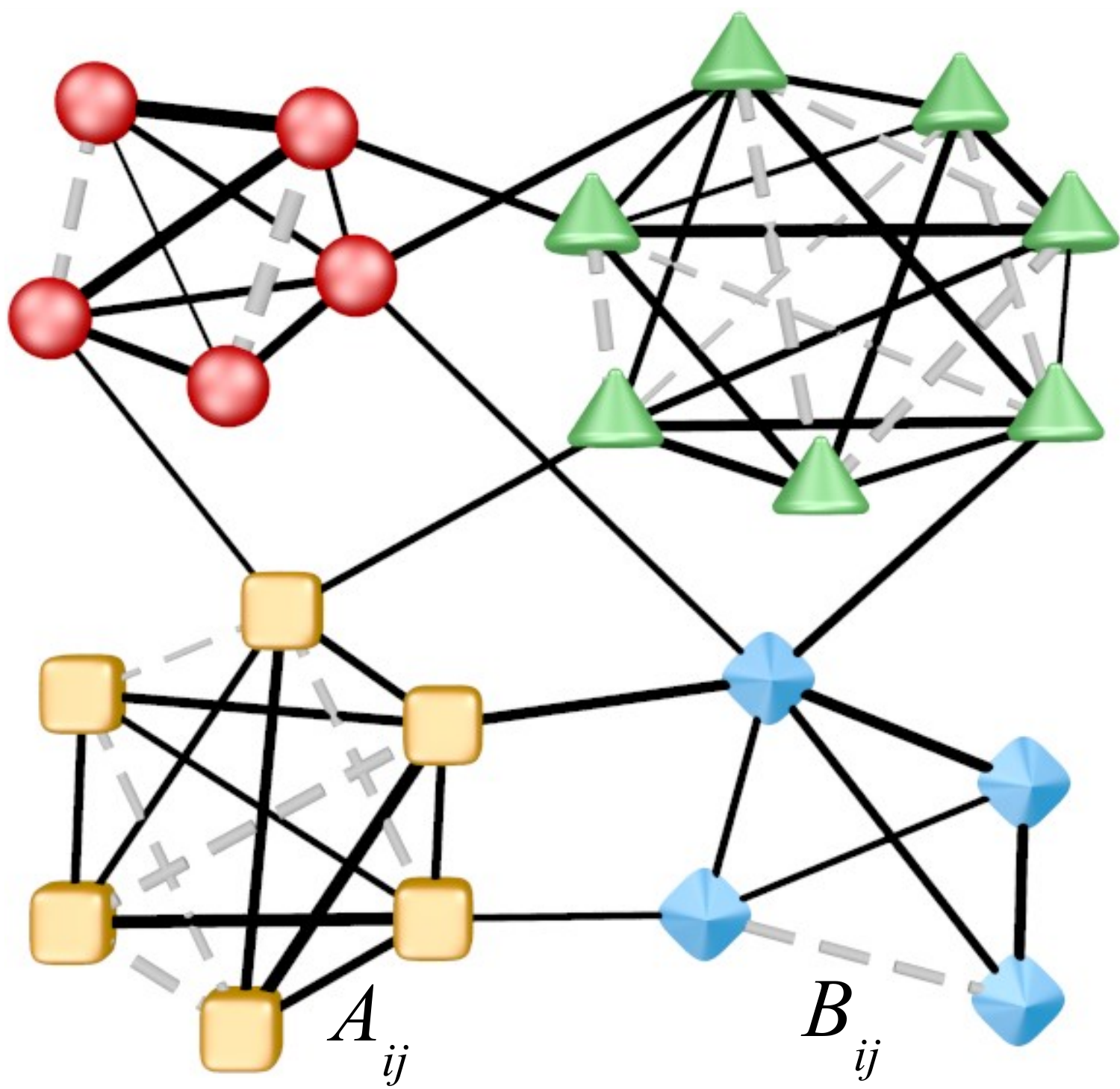}{An arbitrary weighted network with $4$ 
natural communities (strongly connected) depicted as distinct node shapes.
The goal in community detection is to identify any such strongly related 
clusters of nodes based on their defined edge relationships.
Solid lines depict weighted links corresponding to complimentary or attractive
relationships where $A_{ij}>0$ and $B_{ij}=0$ in \eqnref{eq:ourPottsmodel}.
Gray dashed lines depict missing, adversarial, or repulsive relationships
where $A_{ij}=0$ and $B_{ij}>0$ in \eqnref{eq:ourPottsmodel}.
In both cases, the relative link weight is indicated by the 
respective line thicknesses.
For presentation purposes, missing intercommunity edges are not depicted.
In this paper, we directly relate the edges (attractive and missing/repulsive) 
to the interaction energy between pairs of atoms which implies that the 
natural groups would correspond to bound clusters of atoms.}
{fig:networkexample}{0.85\linewidth}{t!}

\begin{figure*}[t]
\begin{center}
\subfigure[]{\includegraphics[width=\figsize]{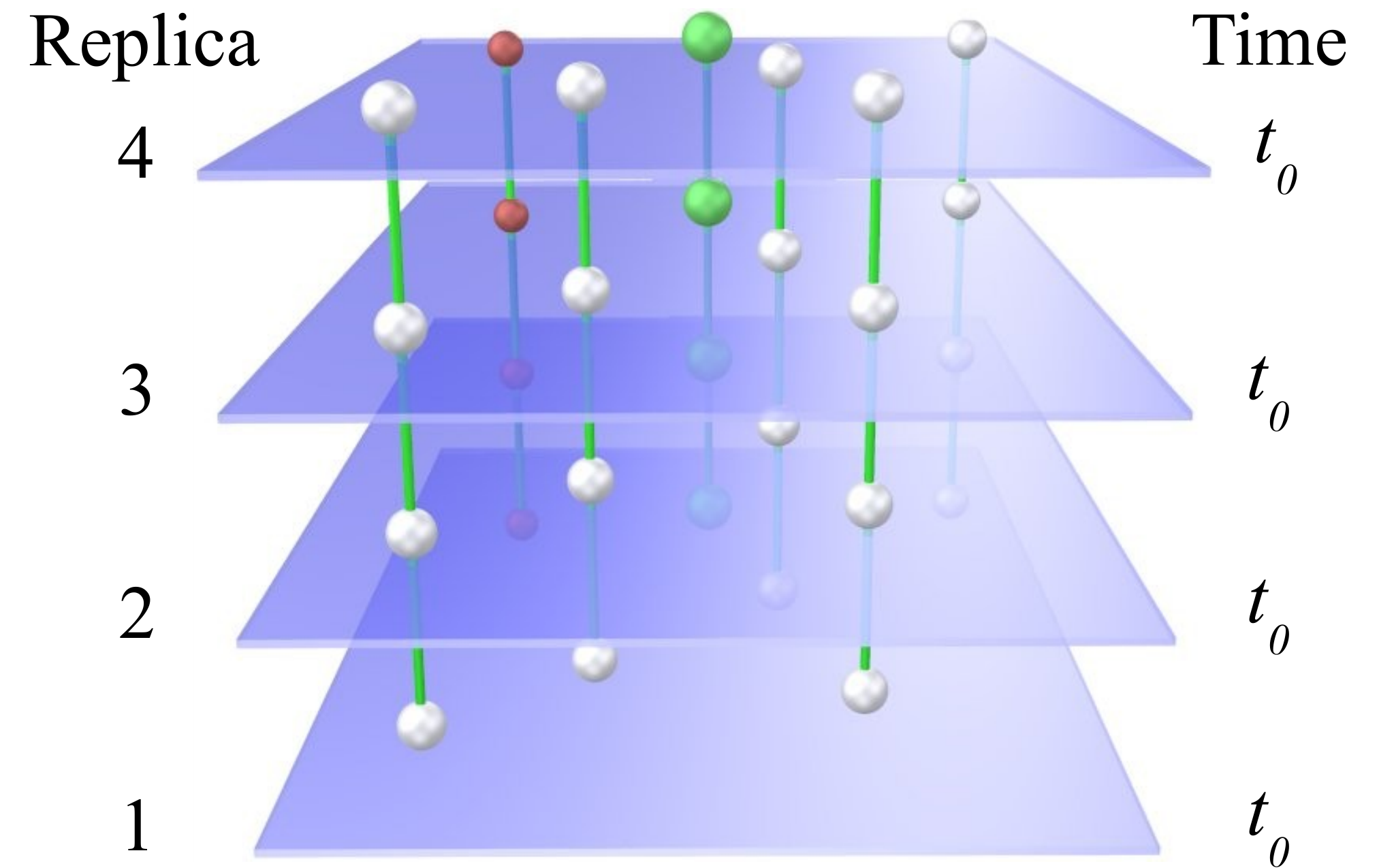}}
\subfigure[]{\includegraphics[width=\figsize]{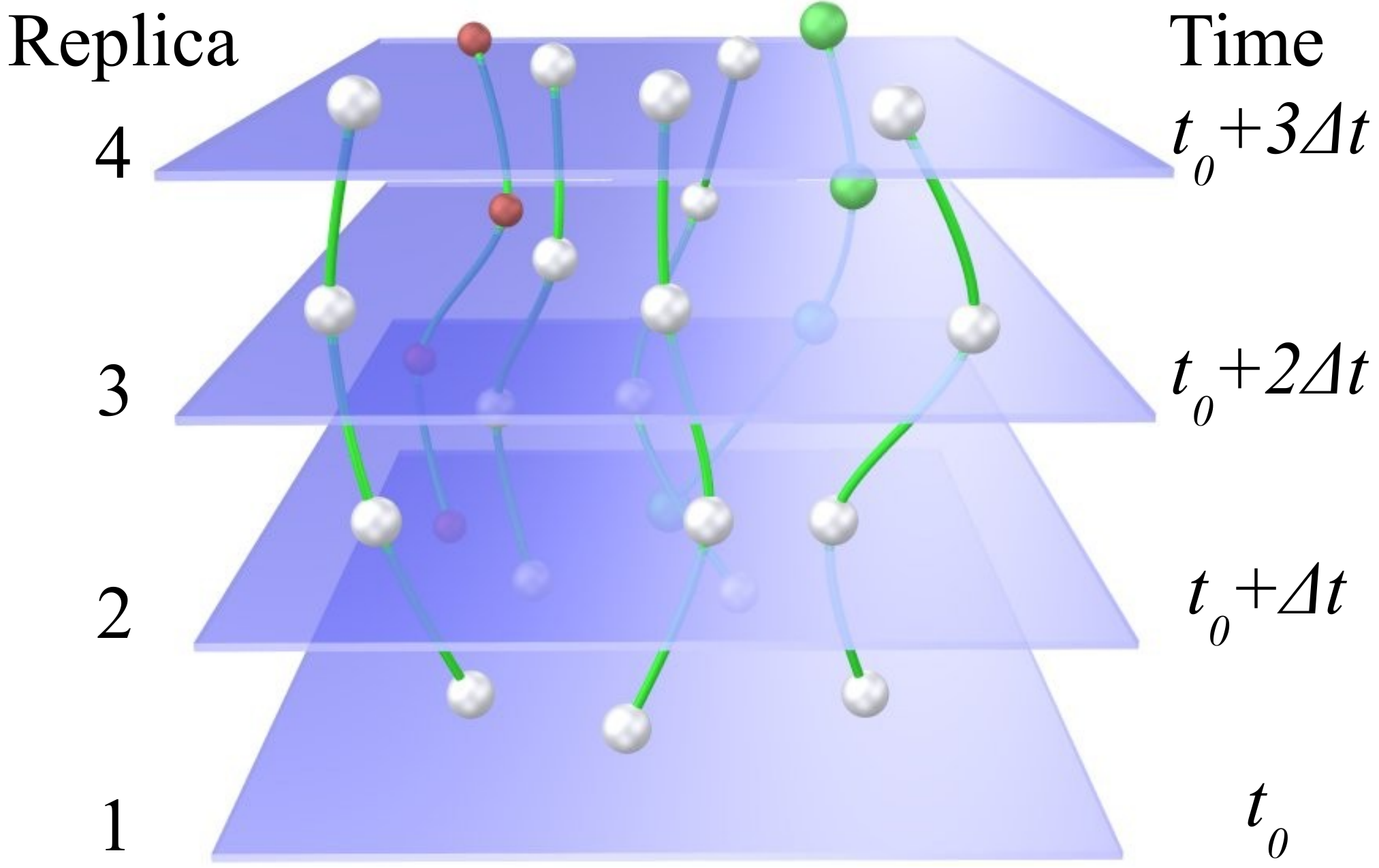}}
\caption{Panel (a) depicts our replica construction for the physical system 
at a ``static'' time $t_0$ (no time separation between replicas, 
see Appendix \ref{sec:StaticAlYFemodel}).
For presentation purposes, only a few nodes (atoms) are illustrated.  
Panel (b) depicts a similar set of replicas separated by a time $\Delta t$ 
between successive replicas.
In both panels (a) and (b), we generate a model network for each replica using 
the potential energy between the atoms as the respective edge weights for the 
network. Independently, within each replica, we then subsequently minimize  \eqnref{eq:ourPottsmodel} 
at a given value of $\gamma$ using the algorithm briefly described 
in \secref{sec:hamiltonian}. 
Afterwards, we use the information measures in Appendix \ref{sec:vi} to evaluate 
how strongly the set of replicas agree on the best partition.}
\label{fig:MRAreplicaspic}
\end{center}
\end{figure*}

Our Potts model Hamiltonian for community detection is \cite{ref:rzlocal}
\begin{equation}  \label{eq:ourPottsmodel}
	\Ham(\{ \sigma \} ) = -\frac{1}{2} \sum_{i\neq j}
	  { \big(  A_{ij} - \gamma B_{ij}  \big) 
	  \delta (\sigma_i,\sigma_j) }.
\end{equation}
Globally minimizing this Hamiltonian corresponds to identifying 
strongly connected clusters of nodes.
The elements of the matrices $A_{ij}$ and $B_{ij}$ are the edge weights 
and are defined as follows: 
an ``attractive'' weight has $A_{ij}>1$ if nodes $i$ and $j$ are connected, 
$A_{ij}=0$ if the nodes are not connected,
a ``repulsive'' weight has $B_{ij}\ge 0$ if the nodes are not connected,
and $B_{ij}=0$ if nodes $i$ and $j$ are connected.
In this paper, we define model the weights $A_{ij}$ and $B_{ij}$ 
by pair-wise potential energies (see \secref{sec:application}).
In principle, we can generalize the Hamiltonian to include $n$-body 
correlations or interactions.

We split the ``attractive'' (ferromagnetic) and ``repulsive'' 
(anti-ferromagnetic) contributions into two separate weighted matrices 
in order to insert the model weight $\gamma$ that adjusts the energy 
trade-off between the two types of interactions.
The parameter $\gamma$ allows us to vary the target scale of the community 
solution.
The spin states $\sigma_i$ designate the community
membership of each node $i$ with a range $1\le\sigma_i\le q$
where $q$ is the number of communities.
This number $q$ may be variable \cite{ref:rzmultires,ref:rzlocal}(such as in the multi-resolution 
that we perform in the current work) in order to find the optimal solutions or held fixed.  
Node $i$ is a member of community $k$ if $\sigma_i = k$.
In this Hamiltonian, each spin $\sigma_i$ interacts 
\emph{only} with other spins in its \emph{own} community.

Briefly, using \eqnref{eq:ourPottsmodel} our community detection 
algorithm rapidly moves nodes between communities based on the current 
lowest energy assignment until no more moves are possible (see also
Appendix \ref{sec:edgediscussion}).
We then attempt $t$ independent trials and select the lowest energy 
trial as the best division. \cite{ref:rzmultires,ref:rzlocal}
[The number of trials serves as our optimization parameter in this
greedy algorithm.
It effectively allows the algorithm to explore more possible
configurations before selecting the best configuration for a given
replica. The number of required trials in order to achieve a prescribed 
accuracy monitors the computational complexity (correlating with 
the number of local minima in which individual trials may get stuck). 
Somewhat better optimization could be obtained with a heat bath
solution algorithm \cite{ref:huCDPT} at a cost of substantially
increased computational effort.]
This community detection algorithm partitions the network into
communities by assigning a unique cluster membership for each node.
Further details are provided in the appendices. 

Local features in metallic glasses generally exhibit interconnecting 
short range structures \cite{ref:sheng}.
In our community detection problem, this feature corresponds to allowing 
``overlapping'' node memberships where atoms can be members of more than 
one local cluster.
We incorporate this effect by assigning a node as a secondary member 
of every community for which it has a negative binding energy in terms 
our Potts model in \eqnref{eq:ourPottsmodel} (see Appendix \ref{sec:edgediscussion}).

We can express the general partition function for a community partition 
with inter-community interactions which may correspond 
to the surface terms of clusters in Random First Order Transition theory (RFOT).
Our parameter $\gamma$ effectively plays the role of scaling the relation between 
surface and bulk terms in RFOT.
A high value of $\gamma$ corresponds to large surface effects while a small
$\gamma$ corresponds to dominant bulk effects.

Several strengths of our method are listed below.  
The analysis is independent of the type of structures that are 
being analyzed (structured, amorphous solid, and possibly even 
liquid systems).
It is robust to noise in the model network, and it yields very 
accurate results with a simple and fast greedy algorithm \cite{ref:rzlocal}. 
Because edge assignments are based on relative node positions
(through the interaction potential), our method should be robust 
with respect to translational or rotational motion of solid structures 
in the system (such as crystal nucleation).

\subsection{Multiresolution community detection} \label{sec:MRA}

\myfigw{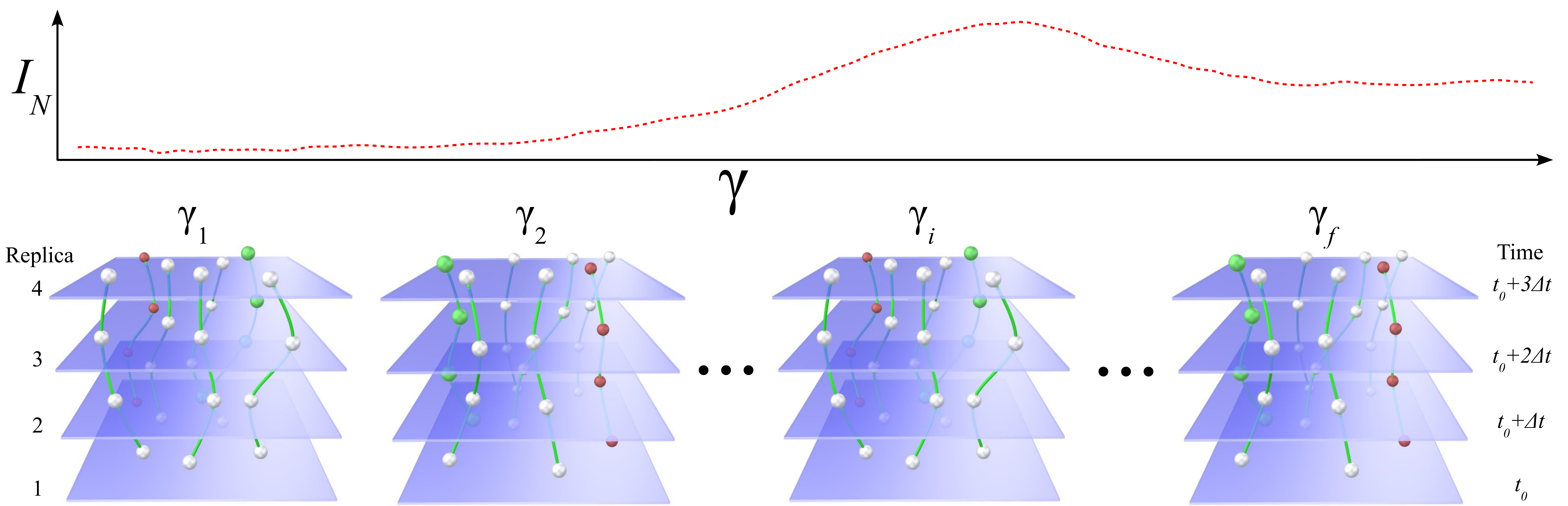}{A depiction of our multiresolution algorithm 
using the replicas of \subfigref{fig:MRAreplicaspic}{b} for a range of resolution 
parameters $\gamma$ in \eqnref{eq:ourPottsmodel}.
We solve the set of replicas at each $\gamma_i$ after which we utilize the information 
measures in Appendix \ref{sec:vi}, such as $I_N$ in the schematic, to measure the 
level of agreement among the replicas for each tested resolution.
The NMI $I_N$ or VI $V$ extrema (or plateaus in some instances) indicate preferred 
(or more ``stable'') resolutions.}
{fig:MRAschematic}{0.95\linewidth}{t}

%----------------------------------------------------------------
% begin MRA algorithm example figures for metallic glass *****
%\myfig{SNAlYFeT300Vavg.eps}{
\myfig{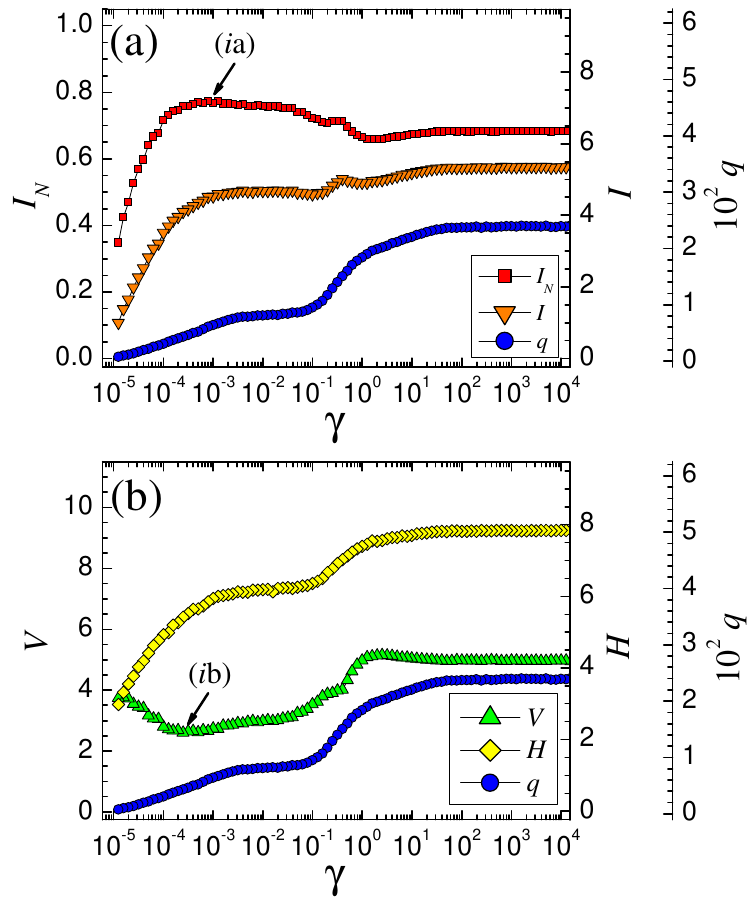}{Panels (a) and (b) show the plots of information 
measures $I_N$, $V$, $H$, and $I$ and the number of clusters $q$ 
(right-offset axes) versus the Potts model weight $\gamma$ 
in \eqnref{eq:ourPottsmodel}.
The ternary model system contains $1600$ atoms in a mixture of $88\%$ type A, 
$7\%$ of type B, and $5\%$ of type C with a simulation temperature of $T=300$ K 
which is well \emph{below} the glass transition for this system.
This system shows a strongly correlated set of replica partitions
as evidenced by the information extrema at ($i$) in both panels.
A set of sample clusters for the best resolution at $\gamma\simeq 0.001$ 
is depicted in \figref{fig:AlYFebestclusters}.}
{fig:AlYFeTLow}{0.84\linewidth}{t}

%\myfig{SNAlYFeT1500Vavg.eps}{
\myfig{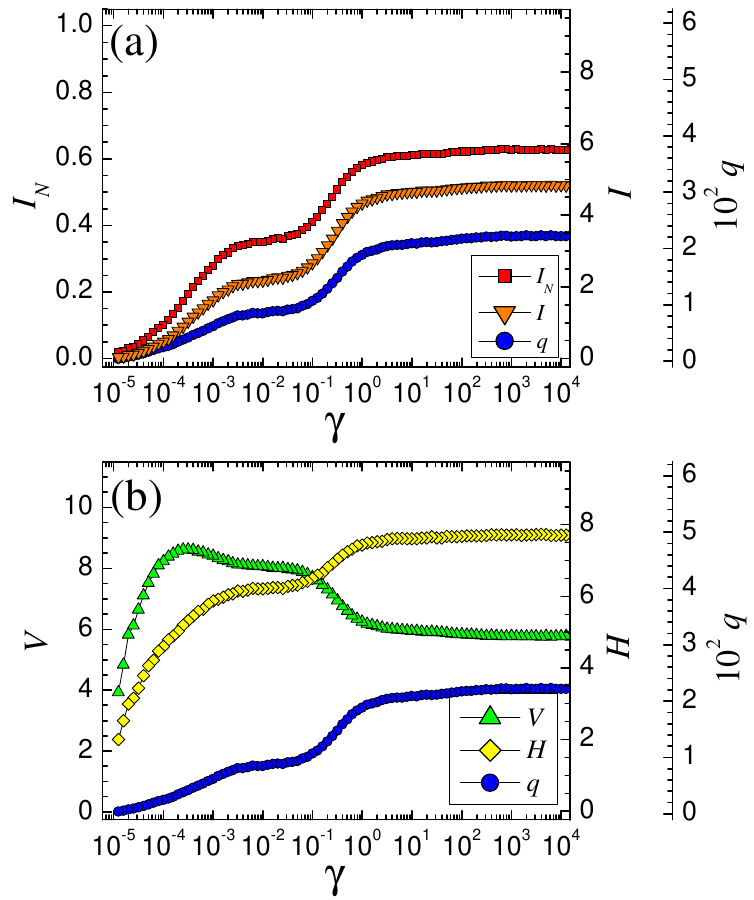}{Panels (a) and (b) show the plots of information 
measures $I_N$, $V$, $H$, and $I$ and the number of clusters $q$ 
(right-offset axes) versus the Potts model weight $\gamma$ 
in \eqnref{eq:ourPottsmodel}.
The ternary model system contains $1600$ atoms in a mixture of $88\%$ type A, 
$7\%$ of type B, and $5\%$ of type C with a simulation temperature of $T=1500$ K 
which is well \emph{above} the glass transition for this system.
At this temperature, there is no resolution where the replicas are 
strongly correlated.
See \figref{fig:AlYFeTLow} for the corresponding low temperature case
where the replicas are much more highly correlated at $\gamma\simeq 0.001$.}
{fig:AlYFeTHigh}{0.84\linewidth}{t}
% end MRA algorithm example figures for metallic glass *******

\myfig{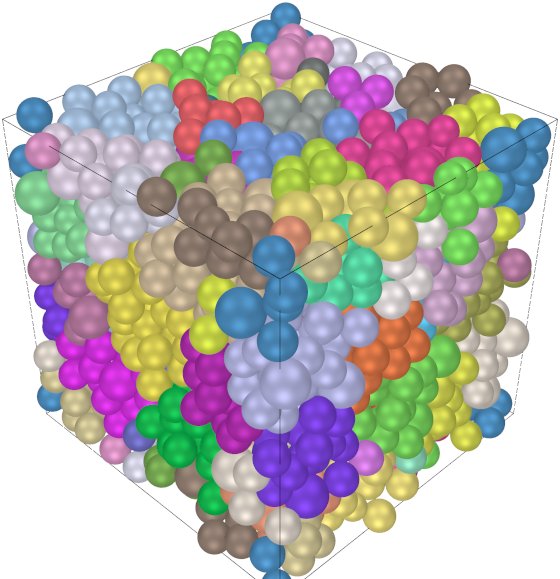}{A depiction of the full 
\emph{partitioned} system where unique cluster memberships are depicted 
as distinct colors (best viewed in color).
The atomic identities are B, A, C in order of increasing diameters.
Overlapping nodes (multiple memberships per node) are added to these
communities to determine the best interlocking system clusters.}
{fig:AlYFebestpartition}{0.85\linewidth}{t}

\begin{figure}[t]
\begin{center}
\subfigure[]{\includegraphics[width=\subfigwidth]{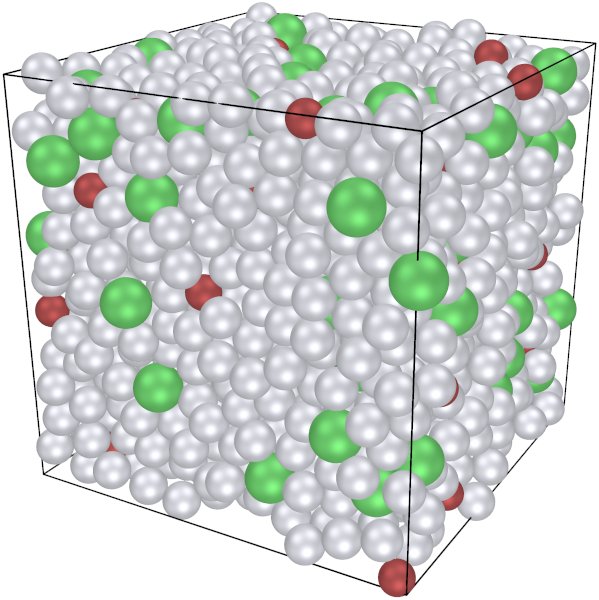}}
\subfigure[]{\includegraphics[width=\subfigwidth]{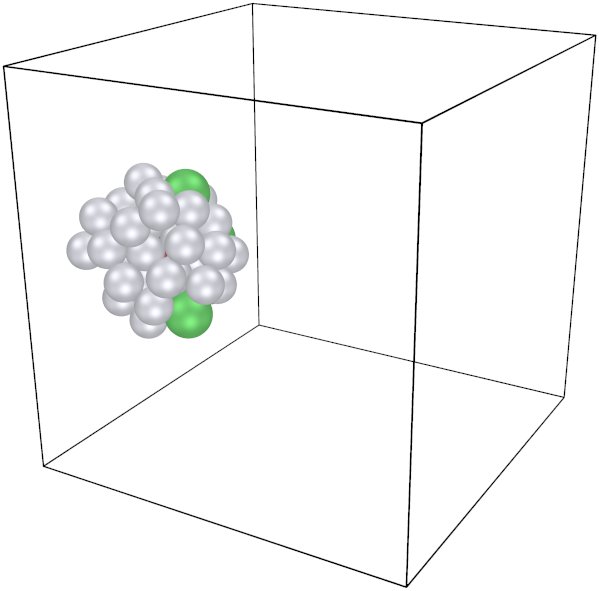}}
\subfigure[]{\includegraphics[width=\subfigwidth]{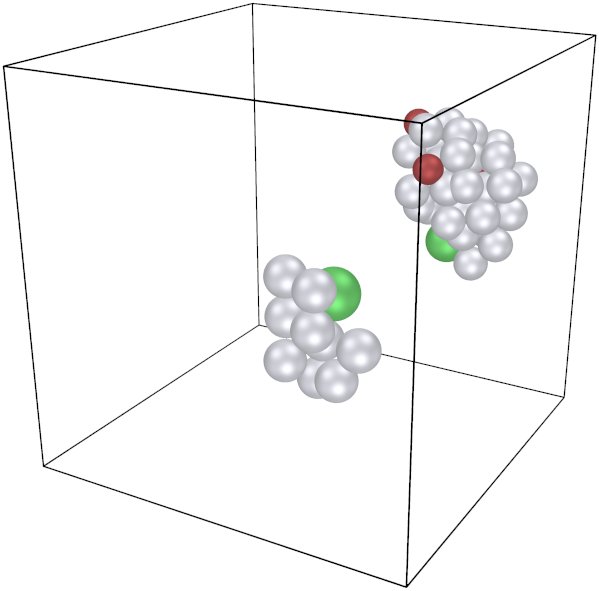}}
\subfigure[]{\includegraphics[width=\subfigwidth]{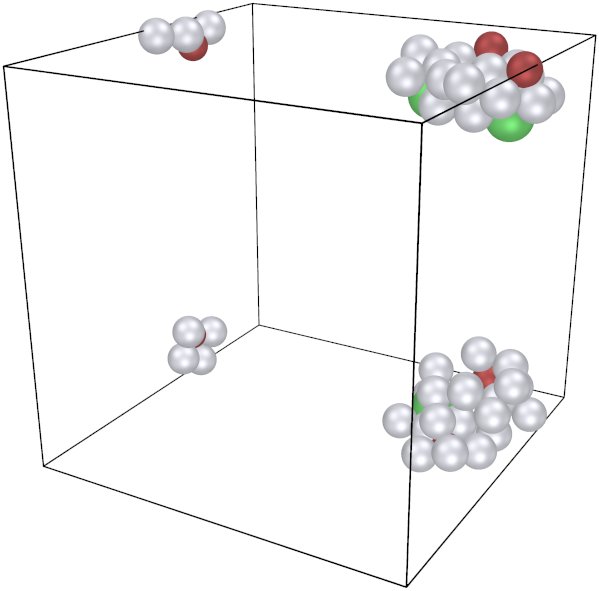}}
\caption{Panel (a) is the full system cube, and panels (b) -- (d) show three
sample clusters (one distinct cluster each using periodic boundary conditions) 
within the simulation box .  
Note that the algorithm can identify structures beyond immediate short range 
neighbors.}\label{fig:systemclusters}
\end{center}
\end{figure}

%\myfig{AlYFeClusters.eps}{
\myfig{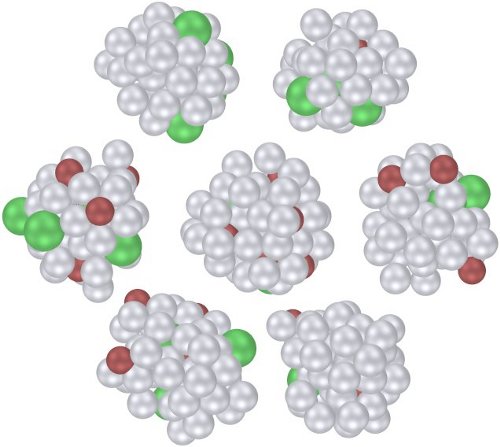}{A depiction of some of the best clusters 
of the low temperature ($T=300K$)  ternary system at the peak replica correlation at feature ($i$) in \figref{fig:AlYFeTLow}.
These clusters include overlapping node membership assignments
where each node is required to have an overall negative binding 
energy to the other nodes in the cluster.
The atomic identities are C (red), A (silver), B (green) in order 
of increasing diameters.}
{fig:AlYFebestclusters}{0.85\linewidth}{hb}
% end MRA algorithm example figures for metallic glass *******
% ------------------------------------------------------------

Multiresolution methods %in community detection 
\cite{ref:multires,ref:kumpulamultires,ref:lanc,ref:rzmultires,ref:muchaSCI}
extend the ideas of community detection to identify the ``best'' division(s) 
over a range of network scales (``resolutions''). 
We test $s$ independent realizations of the system (``replicas'') over all relevant network 
scales by specifying different values of $\gamma$ in \eqnref{eq:ourPottsmodel}.
Networks with a weakly defined structure will result in more diverse solutions 
among the different replicas.
Conversely, strongly defined structures will result in replica solutions 
that agree more strongly.
When replicas represent time-separated 
configurations, a strong agreement among the replicas corresponds to 
consistent physical structures over time. The time separation between
different snapshot can be tuned to find the corresponding pertinent structures
for general time scales. The particular case of vanishing time
separation corresponds to different representations of the same
static system. See Fig. \ref{fig:MRAreplicaspic}.  In the particular case of vanishing time separation, 
we permute, for identical nodes (atoms), the numbers labeling the nodes within the network;
this leads to replicas corresponding to different initial starting points for the algorithm described in
\secref{sec:hamiltonian}. 
[Another possibility now shown in Fig.  \ref{fig:MRAreplicaspic} is that of replicas
being different copies of the same system as it evolves in space-time (i.e., a replica not
corresponding to a given snapshot of the system at a fixed time but rather
monitoring the entire system as it evolves in space-time). This latter possibility
will be alluded to in Appendix \ref{sec:actionmodel}.]

Within this framework, we identify the best resolution(s) by analyzing how 
well the independent replica solutions agree with each other in terms 
of information content \cite{ref:rzmultires}. See Fig. \ref{fig:MRAschematic}
for a schematic involving time separated replicas. 
Extremal correlated resolution(s) identify the best 
division(s) of the network.
We apply the variation of information (VI) $V$ metric and the normalized 
mutual information (NMI) $I_N$ measure to evaluate the level of similarity 
among all pairs of the $s$ replicas (see Appendix \ref{sec:vi}).
Extrema correspond to locally stable solutions
that remain locally unchanged for variations in the system
scale. In the case of very stable system resolutions, 
local extrema can be replaced by plateaus in these 
information theory measures that indicate no change in the system
solution over an extended range of resolution scales $\gamma$
(as seen in the networks examined in \cite{ref:rzmultires} and, in the appendix of
the current work, in some crisp
lattice systems analyzed in Appendix \ref{sec:lattices}). 
We can further extract additional qualitative information about the 
``stability'' of network partitions across a range of resolutions 
by examining the average number of clusters $q$ 
\cite{ref:multires,ref:kumpulamultires,ref:fenndynamic},
mutual information $I$ \cite{ref:multires}, 
or the Shannon entropy $H$ \cite{ref:rzmultires,ref:fenndynamic}.

\subsubsection{Phase diagram of the community detection problem}
\label{phdiag}
Apart from trying to find the best possible division of communities in a particular system, we have also looked at the
ease at which such a solution can be found. This further indicates when, physically,  the system might be analyzed 
in terms of nearly decoupled communities and when it cannot. It also enables us, by varying the parameters and temperature
to map out when a solution is present or not,
to determine whether the solutions found can be trusted and are physically relevant.

When solving the system of \eqnref{eq:ourPottsmodel} 
at non-zero temperature for a given network (i.e., for an atomic configuration 
that is held fixed), entropic effects can, on their own, lead to a transition 
as the temperature is increased \cite{ref:huCDPT}. This transition
corresponds to a spin glass type transition for random
systems capturing a transition from easy to
hard computational problems.  These transitions may also be standard equilibrium transitions for sufficiently regular systems --  critical for several regular graphs (e.g., 
the square lattice viewed as a graph analyzed via a Potts model for $q \le 4$ communities) or first order
($q>4$ in the example above); for weak disorder, the latter transitions may be rounded off and further display
signatures of Griffiths type behavior \cite{griffiths}. We briefly elaborate on this idea below. As illustrated in the examples
analyzed in this work, spatially increasing low temperature structures
in a supercooled liquid as well as correlation lengths
in systems such as the Ising spin systems can be 
ascertained via our method. In a general decomposition
of an interacting system into optimally decoupled groups of
particles (``communities''), the partition function 
is approximated as 
\begin{eqnarray}
Z \simeq \sum_{\{\Lambda\}} g(\Lambda)  \prod_{c=1}^{q_{\Lambda}} z_{c}.
\label{decompose}
\end{eqnarray}
Here, $z_{c}$ is the partition function as computed with the Hamiltonian of
the entire system for the particles in community $c$, $g(\Lambda)$ is the frequency of obtaining a particular 
{\em set} of communities $\Lambda$ in a decomposition of the entire system
into optimally disjoint communities, and $q_{\Lambda}$ is the total number of communities
in the partition $\Lambda$. Thus, physical transitions or crossovers
may, in this approximation, be related to transitions (or divergent scales) or crossovers
in the communities found in $\Lambda$ themselves and/or phase transitions associated with 
the computational complexity of the community detection
problem (as further manifest via the distribution of partitions $g(\Lambda)$) \cite{ref:huCDPT}.
A decomposition of the type of Eq. (\ref{decompose}) is
exact for a (standard uniform) $q$ state Potts model ($H = - \frac{1}{2} \sum_{i \neq j} A_{ij} \delta_{\sigma_{i}, \sigma_{j}}$ where $A_{ij} = 0,1$) wherein the interaction 
energy between spins in different clusters 
(with each cluster/community given
by uniform value of $\sigma_{i}$) is identically zero ($\delta_{\sigma_{i},\sigma_{j}}=0$).

In Fig. \ref{fig:complexity}, we plot a specific measure \cite{ref:rzmultires} 
of the computational complexity as a function of the density of energetically attractive
inter-community links $p_{out}$ and temperature $T$
for a particular random graph. 
Within the “solvable” flat region (starting at low $p_{out}$ and $T$) the system can be effectively decomposed into 
decoupled elements, i.e., the partition function satisfies Eq. (\ref{decompose}) with a well-defined set of partitions 
$\{\Lambda\}$. This region is separated by “ridges” of high complexity in which the community detection problem 
becomes exceedingly hard from the “unsolvable” region in which no sensible community detection occurs (weak thermal 
fluctuations aid the system in avoiding metastable states while large thermal fluctuations are detrimental). While 
the specific phase diagram boundaries were found with the Hamiltonian of our method (discussed next), the phase diagram 
changes little when other methods are used. In several simple cases the phase boundaries coincide with those of the known 
cases (e.g., the phase transition of the Ising model on the square lattice when investigated as a network with links 
representing the strength of the spin exchange).

\myfig{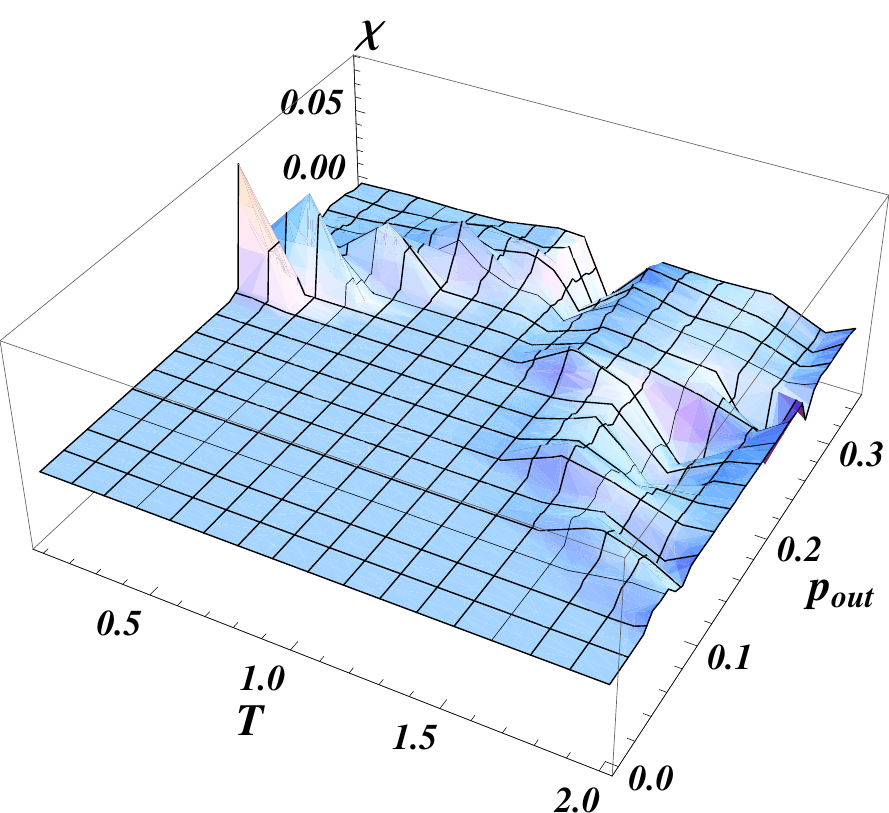}{
A measure of the computational complexity $\chi$ of the community detection problem as a function of  the temperature T and the density of 
inter-community links $p_{out}$.
}
{fig:complexity}{0.85\linewidth}{b}

\section{Multiresolution application to model glass formers} 
\label{sec:application}

We assign edges between the nodes (atoms) with the respective 
weights based on the empirical pair-potentials given 
by \eqnarefs{eq:VrAlYFe}{eq:LJVr}.
Specifically, we calculate the potential energy $\phi_{ij}$ between 
each pair of nodes $i$ and $j$ in the system and then shift each value 
by a constant $\phi_0$ to obtain $\phi'_{ij}=\phi_{ij}+\phi_0$ 
(assuming that $\phi_{ij}\to 0$ as the distance between particles $i$ and $j$ tends to infinity ($r\to\infty$)).
The shift $\phi_0>0$ is necessary for the community detection algorithm 
to properly partition the network of atoms since it provides an objective 
definition of which interatomic spacings are preferable for a well-defined 
cluster and which are preferred to be excluded from a cluster.

In our particular application here, we calculate the average potential 
energy of the system and set $\phi_0=-\phi_\mathrm{avg}$. %, and subtract 
For use in \eqnref{eq:ourPottsmodel}, we define an edge with a weight 
$A_{ij} = -\phi'_{ij}$ between nodes $i$ and $j$ if $\phi'_{ij}<0$,
and we weight any missing links (or ``repulsive edges'') 
by $B_{ij} = \phi'_{ij}$ if $\phi'_{ij}\ge 0$.
We then solve both model systems over a large range of $\gamma$ using
$s=12$ replicas and $t=10$ optimization trials per replica.

While $\phi_\mathrm{0}=-\phi_\mathrm{avg}$ is an intuitive shift 
that accomplishes the goal of an objective cluster definition here, 
it is not an appropriate shift for some problems.
For example, using $\phi_\mathrm{0}=-\phi_\mathrm{avg}$ turns out 
to be problematic in some cases for lattice models.
In a general setting, we examine a continuum of potential shifts $\phi_0$ 
and monitor extrema in the information theory measures as a function 
of both $\gamma$ in \eqnref{eq:ourPottsmodel} and $\phi_0$.

In addition to the systems tested below, we applied the algorithm 
to various test cases including square, triangular, and cubic lattice 
structures. 
The algorithm is able to correctly identify the natural leading
order scales (plaquettes and composites of plaquettes as ``cascades''
in the information theory correlations).
Further testing involved two-dimensional defects (dislocations, 
interstitials, etc.) and domain walls in a lattice.
Defects in triangular lattices occurred most frequently near cluster 
boundaries.

We also tested static configurations for the ternary model glass 
system where each replica is a model of the same configuration. 
There we detected structures in both low and high temperatures
where the high temperature ``structures'' are more fragile (that is,
harder to solve in the clustering problem).
This corresponds to identifying relevant transient features in a dense 
liquid.

\subsection{Ternary model glass results} \label{sec:modelresults}

In \figsref{fig:AlYFeTLow}{fig:AlYFeTHigh}, panels (a) and (b) 
show the information theory based correlations (averaged over all
replica pairs as in \cite{ref:rzmultires}) over a 
range of network resolutions.
The lower temperature system at $T=300$ K in \subfigref{fig:AlYFeTLow}{a} 
shows a peak NMI at ($i$a) with a corresponding VI minimum at ($i$b).
\figref{fig:AlYFebestpartition} depicts an example of the full system 
\emph{partition}. 
\figref{fig:systemclusters} shows some sample clusters within the simulation 
bounding box at resolution parameter value of $\gamma_{best}\simeq 0.001$ where we include overlapping node 
memberships (the replicas correlations are calculated on partitions), 
and \figref{fig:AlYFebestclusters} depicts additional samples of the best clusters.
%\figref{fig:AlYFebestclusters} depicts a sample of the best clusters at 
%$\gamma_{best}\simeq 0.001$ where we include overlapping node memberships
%(the replicas correlations are calculated on partitions).
The corresponding high temperature ($T=1500$ K) solutions
have a much 
lower NMI at $\gamma_{best}\simeq 0.001$ indicating significantly 
worse agreement among replicas.
That is, one would expect that the high temperature system $T=1500$ K 
is in a liquid state, so any observed features are not dynamically 
stable across all replicas (snapshots of the system over time).
At $T=300$ K, the best structures have consistent cluster sizes that 
are exclusively MRO.

The plateau regions for $\gamma>10$ are similar to the LJ plot 
in \figref{fig:LJTLow}, but in this system the NMI plateau is lower.
In the high temperature case in \figref{fig:AlYFeTHigh}, there are 
additional ``almost-plateaus'' for the range 
$0.001\lesssim\gamma\lesssim 0.1$.  
These plateaus represent a region of structural transition, but we are 
not concerned with them because the replica correlations are very low.

\subsection{Binary Lennard-Jones glass results} \label{sec:LJresults}

%-------------------------------------------------------------
% begin MRA algorithm example figures for LJ glass ***********
%\myfig{SNLJTLowVavg.eps}{
\myfig{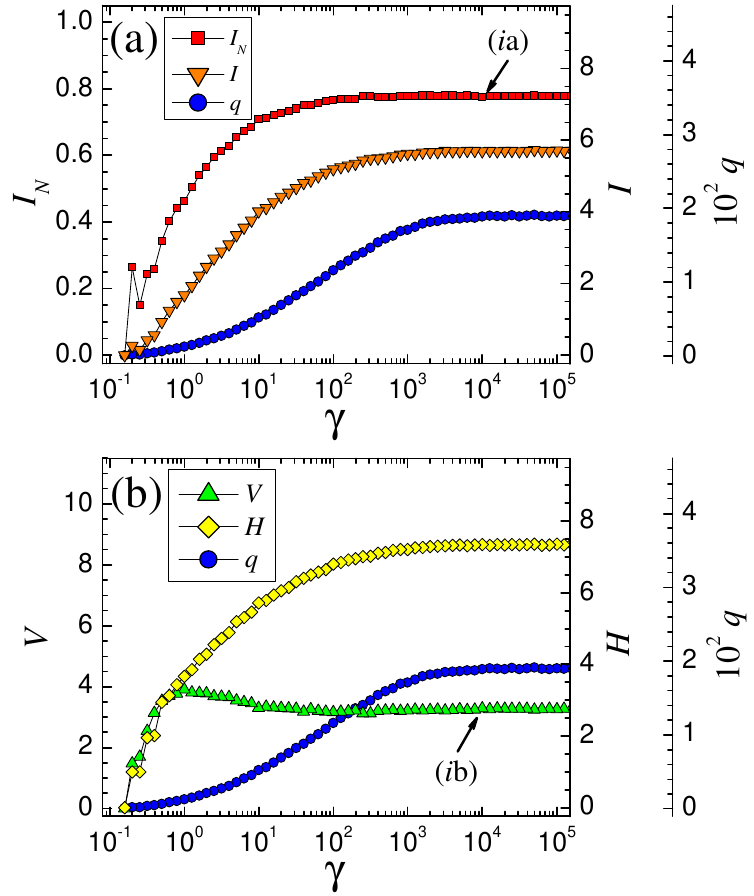}{Panels (a) and (b) show the plots of information measures 
$I_N$, $V$, $H$, and $I$ and the number of clusters $q$ 
(right-offset axes) versus the Potts model weight $\gamma$ 
in \eqnref{eq:ourPottsmodel}.
The LJ system contains $2000$ atoms in a mixture of $80\%$ type A and
$20\%$ type B (Kob-Andersen binary LJ system \cite{ref:kobandersenOne}) 
with a simulation temperature of $T=0.01$ (energy units) which is well 
\emph{below} the glass transition of $T_c\simeq 0.5$ for this system.
This system shows a somewhat strongly correlated set of replica 
partitions as evidenced by the information extrema at ($i$a,b)
in panels (a) and (b).
A set of sample clusters for the best resolution at $\gamma=10^4$ 
is depicted in \figref{fig:LJbestclusters}.}
{fig:LJTLow}{0.84\linewidth}{b}

%\myfig{LJClusters.eps}{
\myfig{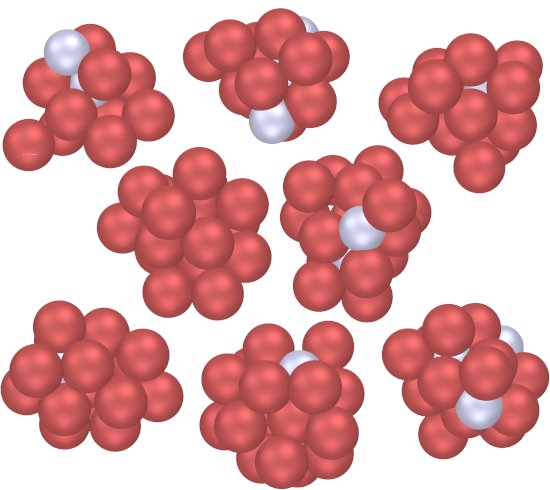}{Several of the best clusters for the peak 
replica correlation at feature ($i$) in \figref{fig:LJTLow}.
These clusters include overlapping node membership assignments
where each node is required to have a overall negative binding 
energy to the other nodes in the cluster.
The atomic identities are B (silver) and A (red) in order 
of increasing diameters.}
{fig:LJbestclusters}{0.85\linewidth}{t}
% end MRA algorithm example figures for LJ glass *************
% ------------------------------------------------------------

%\myfig{SNLJTHighVavg.eps}{
\myfig{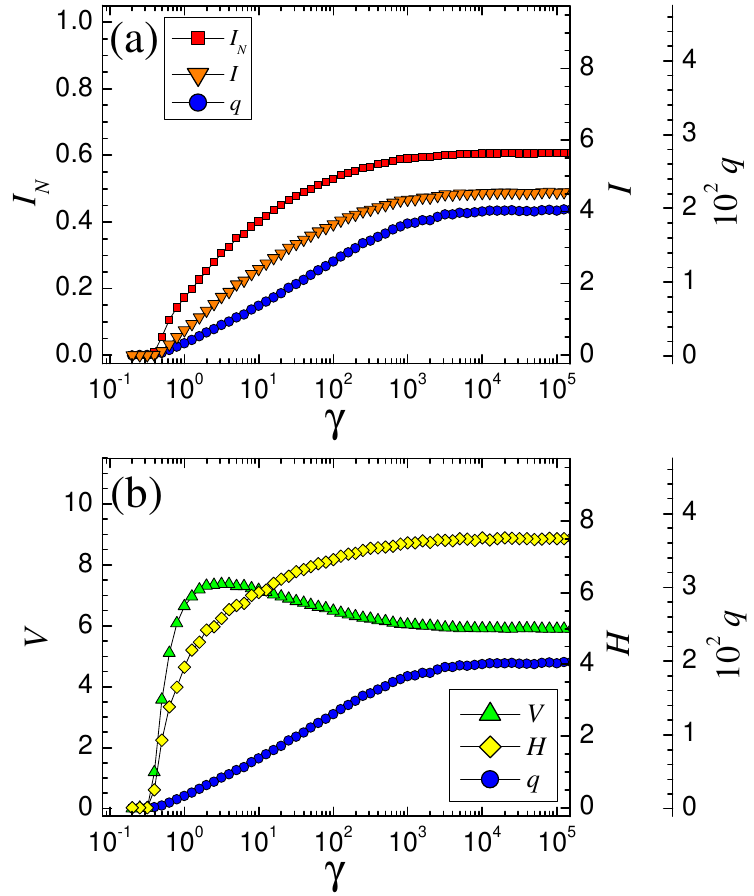}{Panels (a) and (b) show the plots of information measures 
$I_N$, $V$, $H$, and $I$ and the number of clusters $q$ 
(right-offset axes) versus the Potts model weight $\gamma$ 
in \eqnref{eq:ourPottsmodel}.
The LJ system contains $2000$ atoms in a mixture of $80\%$ type A and
$20\%$ type B (Kob-Andersen binary LJ system \cite{ref:kobandersenOne}) 
with a simulation temperature of $T=5$ (energy units) which is well 
\emph{above} the glass transition of $T_c\simeq 0.5$ for this system.
At this temperature, the replicas are significantly less correlated 
than the corresponding low temperature case in \figref{fig:LJTLow}.}
{fig:LJTHigh}{0.84\linewidth}{b}
% end MRA algorithm example figures for metallic glass *******

\begin{figure}[t]
\begin{center}
\subfigure[]{\includegraphics[width=\subfigwidth]{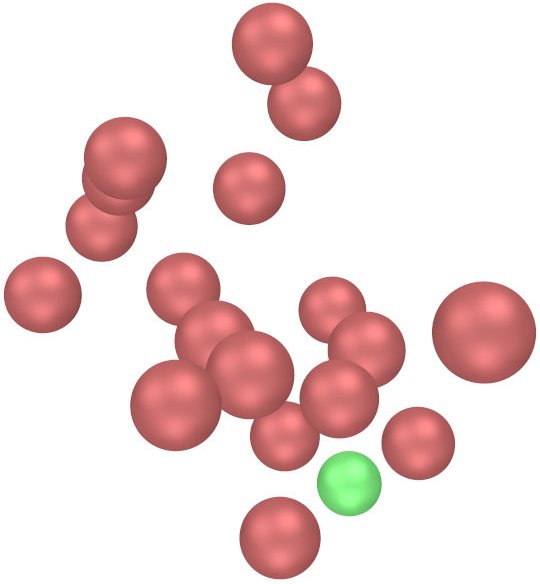}}
\subfigure[]{\includegraphics[width=\subfigwidth]{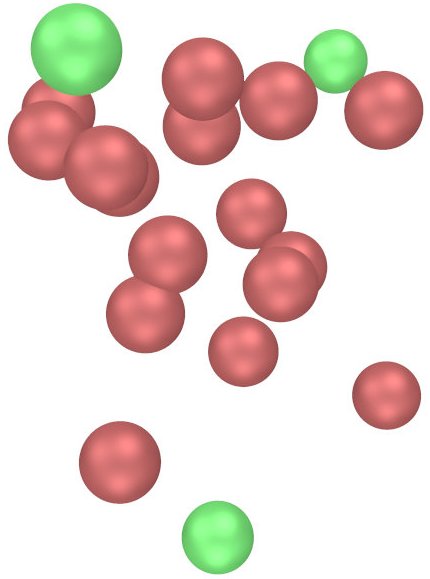}}
\subfigure[]{\includegraphics[width=\subfigwidth]{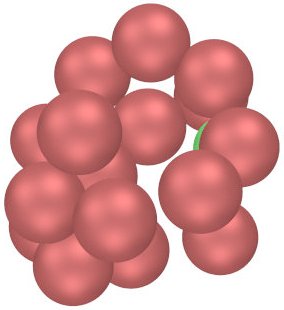}}
\subfigure[]{\includegraphics[width=\subfigwidth]{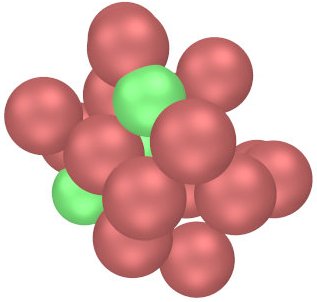}}
\caption{A sample depiction of dispersed clusters for the LJ system  Eq. (\ref{eq:LJVr}) at a temperature of $T = 5$ (in units where $k_{B}=1$).
The shown clusters correspond to the multiresolution plot in \figref{fig:LJTHigh} at  value of the resolution
parameter of $\gamma=10^4$.
These clusters are a sample of high temperature counterparts to the low 
temperature clusters in \figsref{fig:LJTLow}{fig:LJbestclusters}.
Panels (a) and (b) show a more typical example of dispersed clusters at a number of replicas $s=10$.
In some cases, the identified high temperature clusters can be more compact,
but not densely packed. Panels (c) and (d) provide sample solutions for $s=20$ replicas.
An increasing replica ($s$) number (and generally also trial ($t$) number, see text)
required to achieve better solutions is indicative of a greater
computational complexity of the system. Physically,
a larger time is required for the system to realize better
clusters under ideal conditions (in the absence of
quenching and any relaxation time constraints).
{\em The sparsity of the identified clusters in this high temperature system is generally consistent across all clusters 
in the network solution}. (This lies in contrast to the more compact and more strongly correlated structures
found at low temperatures.)
The atomic identities are B (silver) and A (red) in order of increasing 
diameters.}
\label{fig:LJTHighbestclusters}
\end{center}
\end{figure}

In \figsref{fig:LJTLow}{fig:LJTHigh}, panels (a) and (b) show 
the data for the replica information correlations over a range 
of network resolutions.
The lower temperature system at a temperature of $T=5$ (in units of $k_{B} =1$) 
in \subfigref{fig:AlYFeTLow}{a} shows a plateau in NMI at ($i$a) 
with a corresponding VI plateau at ($i$b) which are the local extrema
($V=0$ is a trivial solution with only one cluster in this problem).
\figref{fig:LJbestclusters} depicts a sample of the best clusters, 
including overlapping node memberships, at resolution ($i$) 
for $\gamma_{best}\simeq 10^4$.
The corresponding higher temperature solutions at $\gamma_{best}\simeq 10^4$
(see Figs. \ref{fig:LJTHigh} and\ref{fig:LJTHighbestclusters})
have a lower NMI (indicating a weak agreement among replicas).
The dependence number of replicas (See Fig. \ref{fig:LJTHighbestclusters}) required to achieve high accuracy
underscores the faint agreement between contending solutions and the high temperature complexity
of the problem.  Our identified structures for this LJ model system are consistent 
in terms of the cluster sizes and are almost exclusively SRO 
configurations with simple adjunct-type atoms extending into the 
low end of MRO size structures.

\section{Conclusions} \label{sec:amorphousconclusion}

Our method is a new and very general approach to determine the natural
multi-scale structures of complex physical 
systems.
We do not bias the expected configurations in any way.
The required input is that of inter-particle interactions (or measured
correlations as further detailed in Appendix \ref{sec:gijrapplication}). Information theory extrema (including plateaus) 
between contending solutions give the different 
pertinent structures on all important length scales (lattice scales and correlation lengths) of the system 
in an unbiased unified way.

Apart from benchmarks on general networks and on lattices and spin systems with and without defects
(see Appendices \ref{sec:lattices},\ref{sec:Ising},\ref{sec:LJlattices}), to illustrate the feasibility of this approach, we focused
in this work on structural glasses. The detection of structure in structural glasses is a heavily investigated hard problem.
By the use of our method, we identified consistent SRO or MRO structures 
at temperatures below the glass transition in two different model 
glass formers.
Our analysis evaluates structures in terms the potential energies 
(i.e., the internal binding energies of the clusters).
This approach differs from some other methods of structural analysis
that look strictly at the relative atomic positions.
We briefly comment on the relation between our work and that of simpler
mode analysis of the interactions -- the latter have been indeed been used
to detect a link between spatial structure and local dynamics \cite{ref:widmercooperHF,irr}.
Local heterogeneous dynamics was shown to be correlated with topological defects \cite{reich}
and thus (as a consequence of   \cite{ref:widmercooperHF,irr}) 
with the system modes. (Indeed, a recent work \cite{manning} directly reaffirms such
a connection.)
Similarly, some old variants of graph partitioning methods such as direct spectral clustering \cite{spectral_cluster1,spectral_cluster2}
as well as community detection
employ a normal mode analysis \cite{newman-vector}.  Indeed, a method based on particle dynamics 
in high dimensions enabled multi-scale community detection \cite{dynamicshighd}. Other different yet conceptually
related approaches also include oscillator synchronization analogies \cite{kur,osc}. 
In the above approaches, physical analogies were made.
It is thus natural to suspect that the approach may inverted and that community 
detection will link spatial structure with dynamics in a broad class of physical systems
where the physics based analysis of the interacting multi-particle system is hampered by the shear complexity of the problem. 
The current work indeed fleshes out this link in detail
and introduces a method for the direct detection of general spatio-temporal structures
in rather general physical systems.
Notably, via the use of information theory correlations and extrema as a function
of the Hamiltonian parameters therein ($\gamma$ of our defining Hamiltonian of
Eq. \ref{eq:ourPottsmodel} and $\phi_{0}$ of \secref{sec:application}),
we are able to identify in an unbiased way {\em all of the natural scales} of the system.
In more rudimentary approaches this needs to introduced by hand (e.g., a cutoff on mode 
occupancy in spectral based approaches) and it is not obvious 
how to determine all pertinent structures of a complex physical system.

Our approach identifies MRO as the dominant feature of our ternary model 
glass former with no strongly defined SRO.
In contrast, the LJ system shows a largely SRO structure with adjunct 
atoms that create near-MRO structures.
Compounding the changes in structure that we find by analyzing the 
atomic system at different temperatures and minimizing the energy 
function to determine the optimal division into clusters, there are 
also {\em entropic effects}.
The distribution of optimal partitions becomes wider and less pronounced 
due to these effects as the temperature increases.
This is a validation of our arguments in \secref{phdiag}.

On a lattice, plateaus in information theory correlation steps correspond 
to a cascade of structures starting from the smallest dyads of nodes, 
to basic plaquette structures (square, triangle, etc.), and growing ever 
larger (two joined plaquettes etc.).
In Ising spin systems at different temperatures on a square lattice,
the domains of ``$+$'' and ``$-$'' spins are separated from one another 
by domain walls.
The information theory plateaus correspond similarly to the cascade of small 
plaquette structures found on the lattice itself (\ie{}, the single plaquette, 
two joined plaquettes etc.) up to a cutoff scale set by the domain wall.
This is sensible since no clear structure is found beyond the domain length scale. 
The largest fluctuations occur at the boundaries between different domains.
These domain walls are directly attained by the extrema (those corresponding 
to the \emph{maximum} in VI). % in the information theory correlations.
Physically, they correspond to the scales at which the largest fluctuations occur 
where the large fluctuations lead to poor information theory correlations between 
the different replicas. 
\figref{fig:IsingPictureCL} corresponds to a sample depiction of the system 
at the \emph{maximum} VI.
The correlation lengths -- set by the scale of the largest domain walls -- are thus related to the scale which the fluctuations
in the information theory overlaps are maximal (as manifest via \emph{maxima} in VI).

As we detail in Appendix \ref{sec:action}, a general rigid amorphous solid supports shear and a divergent 
``shear penetration length''. This length scale monotonically increases as a liquid is cooled
to form a rigid glass. Within our graph theoretic method, we may employ
the shear stress correlations as to represent
graph weights to analyze such behavior.

\section*{ACKNOWLEDGMENTS}

We are indebted to M. Widom and M. Mihalkovi\v{c} for help with effective atomic potentials and to 
ongoing work \cite{effpotentials},
to Wolfgang Weiser  for the 2D Ising lattice simulation code made available on his website.
This work was supported in part by the LDRD DR on the physics of algorithms 
at LANL and the Center for Materials Innovation at Washington University 
in St. Louis. ZN thanks N. Goldenfeld, Z. Rotman, G. Tarjus, and P. G. Wolynes for helpful comments. He also wishes 
to thank the KITP and the Lorentz Center
for hosting stimulating workshops.

\appendix

%----------------------------------------------------------------
% moved to here to slightly improve figure placement
\myfig{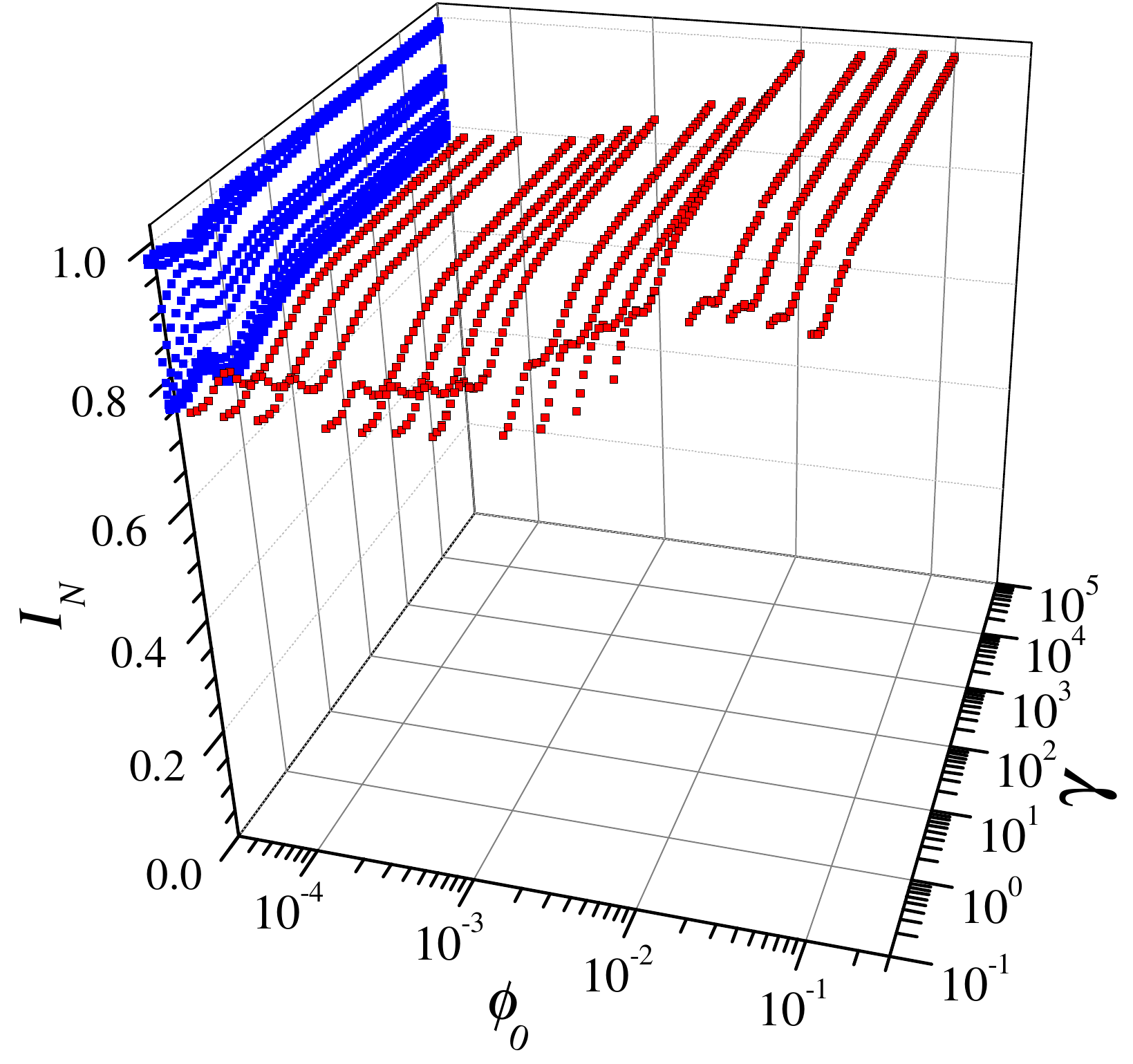}{Plot of NMI $I_N$ as a function of the 
potential shift $\phi_0$ and the Potts model weight $\gamma$ for the 
ternary model system in \secref{sec:modelresults}.
The temperature is $T=300$ K which is \emph{below} the glass 
transition temperature for this system.
This plot shows that the peak is roughly constant across a range 
of potential shifts.
In general systems, we can obtain all natural scales by shifting 
both $\phi_0$ and $\gamma$ and looking for extrema (and plateaus 
in some cases, see \secref{sec:lattices}).
See \figref{fig:AlYFeTLow} for the corresponding 2D plot using 
$\phi_0=\phi_{avg}$.}
{fig:AlYFeTLowshiftgamma}{\figsize}{t}
%----------------------------------------------------------------
%----------------------------------------------------------------
% moved to here to slightly improve figure placement
\myfig{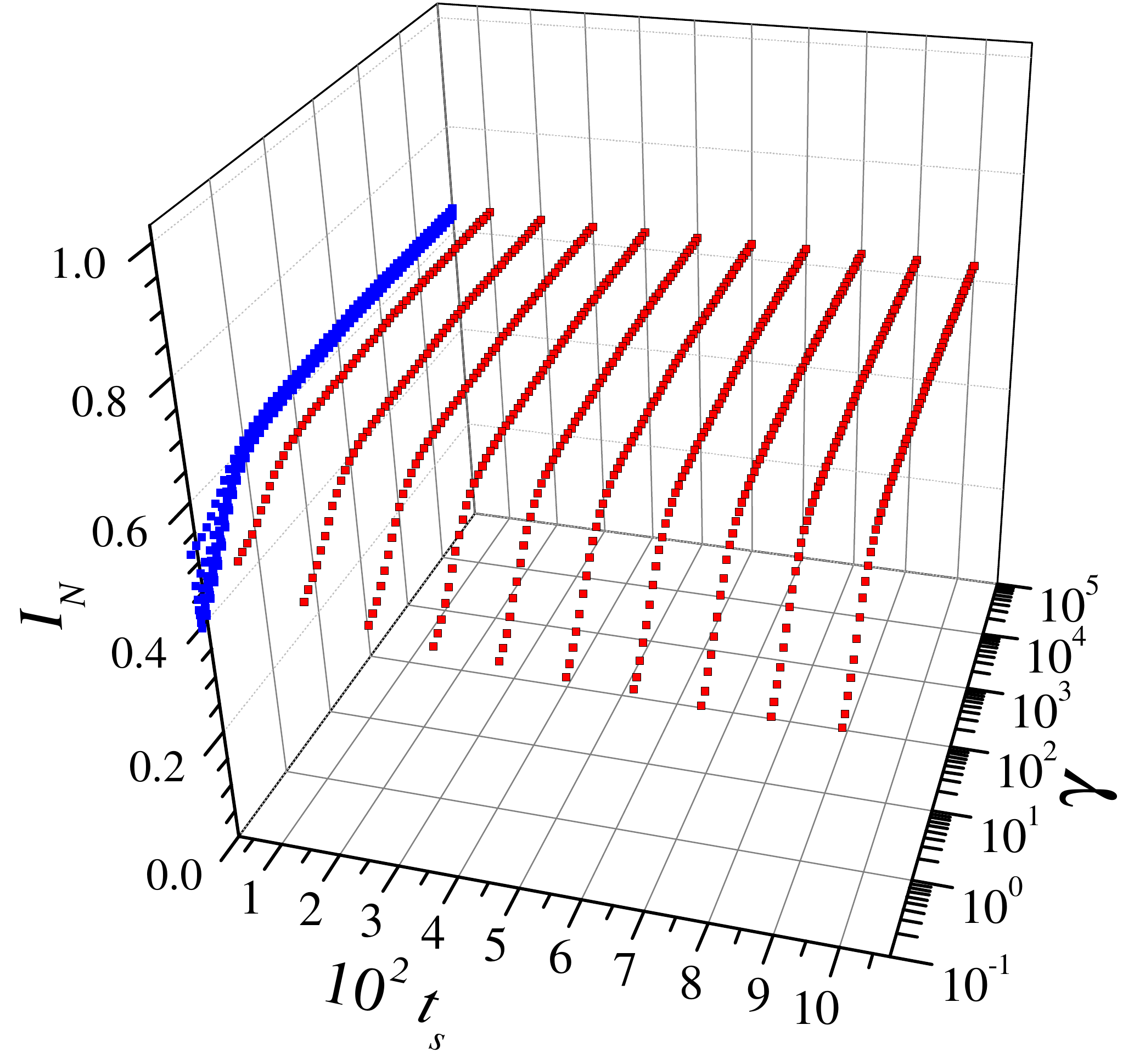}{Plot of NMI $I_N$ as a function of the time 
(MD time steps) between configurations $t_s$ and the Potts model weight $\gamma$ 
for the ternary model system in \secref{sec:modelresults}.
The temperature is $T=1500$ K which is \emph{above} the glass transition 
temperature for this system.
Intuitively, this plot shows that the correlations become weaker as the time 
between configurations is increased.
This process of examining the correlations as a function of $t_s$ and $\gamma$
allows us to examine the relevant time scales in addition to the natural spatial 
scales that we identify in \secref{sec:application}.}
{fig:AlYFeTHightimegamma}{\figsize}{b}
%----------------------------------------------------------------
%----------------------------------------------------------------
% moved to here to slightly improve figure placement
\myfig{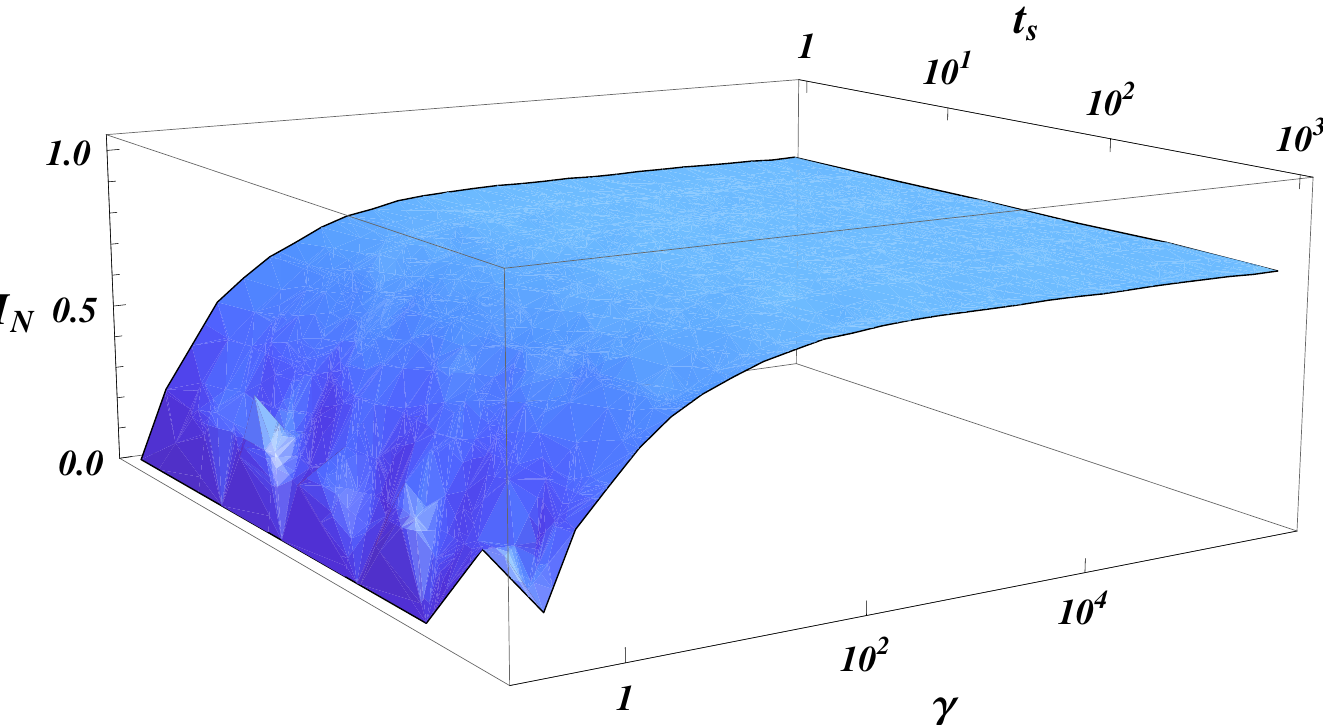}{Plot of NMI $I_N$ as a function of the time 
(units of $10$ MD time steps) between configurations $t_s$ and the Potts 
model weight $\gamma$ for the ternary model system in \secref{sec:modelresults}.
The temperature is $T=0.01$ which is \emph{above} the glass transition 
temperature for this system.
Intuitively, this plot shows that the correlations become weaker as the time 
between configurations is increased.
Intuitively, this plot shows that the correlations become weaker as the time 
between configurations is increased.}
{fig:LJTLowtimegamma}{\figsize}{t!}
%----------------------------------------------------------------
%----------------------------------------------------------------
\begin{figure}[b]
\begin{center}
\subfigure[]{\includegraphics[width=\figsize]{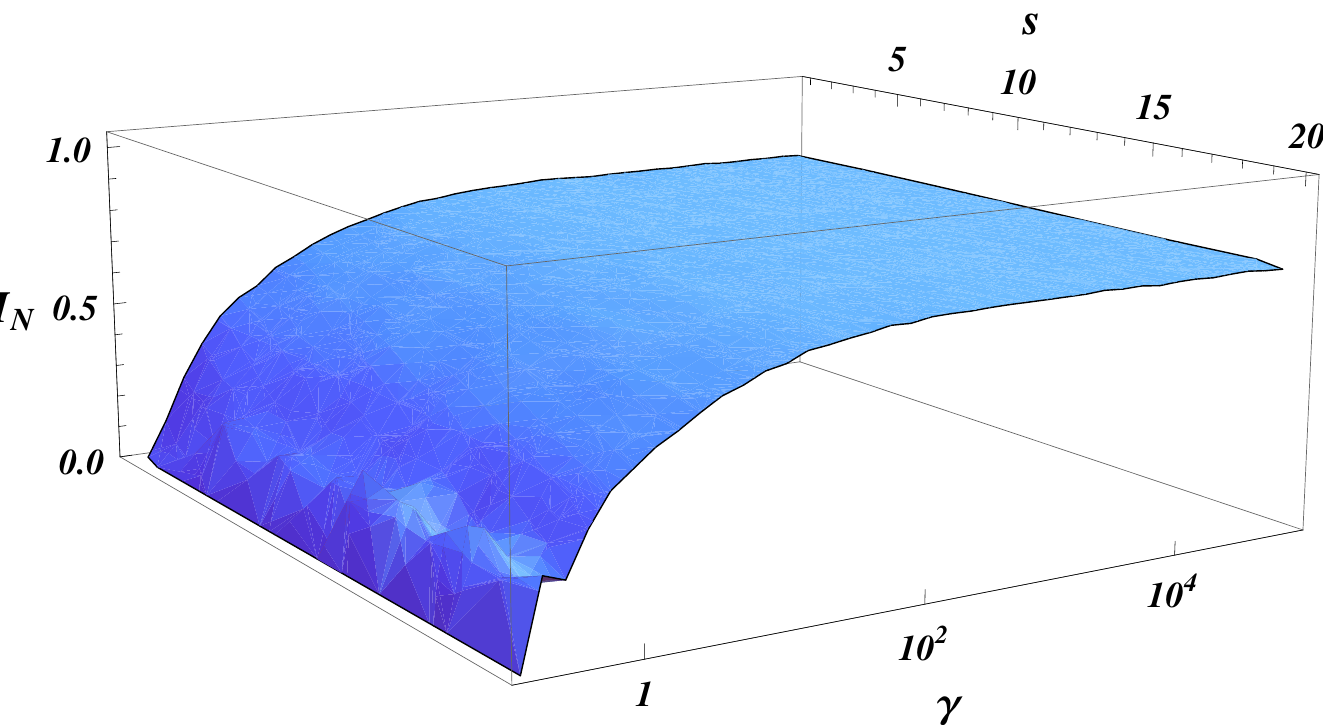}}
\subfigure[]{\includegraphics[width=\figsize]{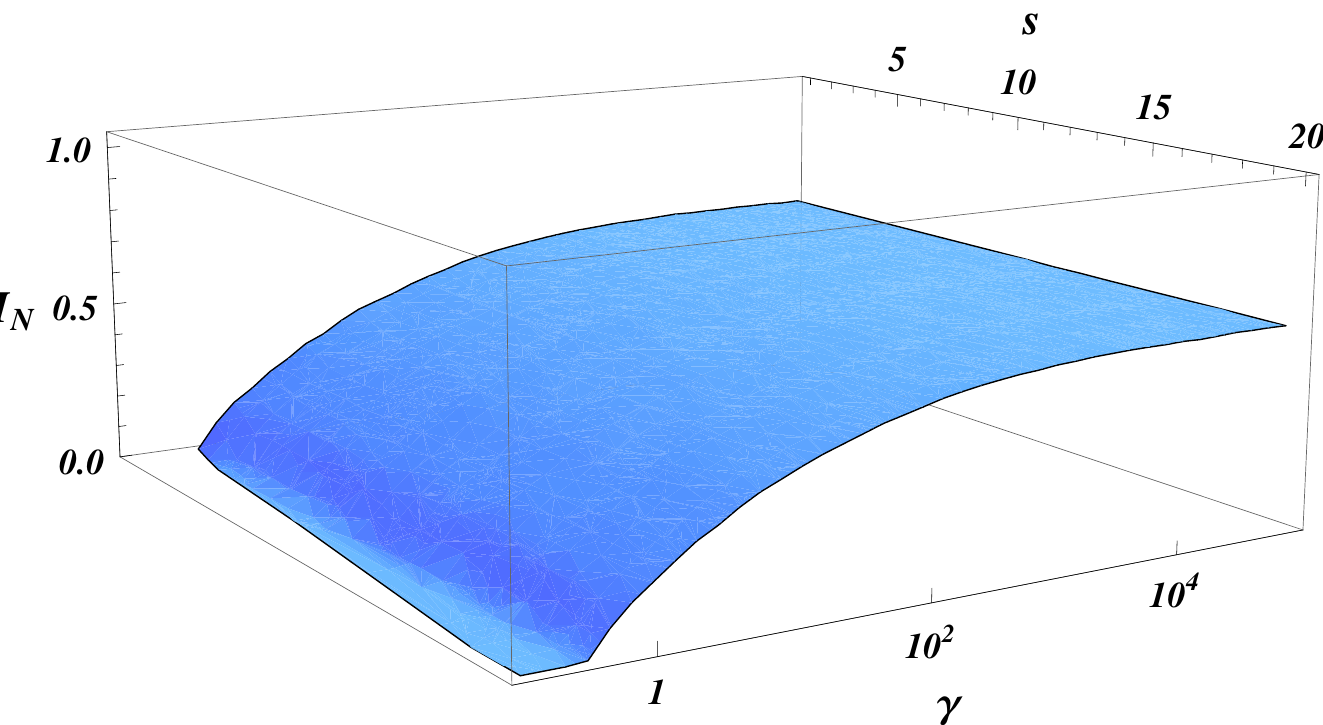}}
\caption{Plot of NMI $I_N$ as a function of the number of trials $s$ 
used to optimize the solution and the Potts model weight $\gamma$ 
for the binary LJ system in \secref{sec:modelresults}.
The temperatures are $T=0.01$ in panel (a) and $T=5$ in panel (b).
The number of trials has a higher effect on the accuracy in the $T=5$ system,
but the number of trials does not result in a drastic improvement to the accuracy 
of our algorithm for either LJ system.}
\label{fig:LJTLowHightrialsgamma}
\end{center}
\end{figure}
%----------------------------------------------------------------

\section{Information theory measures} \label{sec:vi}

We utilize the variation of information \cite{ref:vi} 
and normalized mutual information to measure the strength 
of the correlations among the independent replicas 
in our multiresolution algorithm.
The mutual information $I(A,B)$ between community divisions 
$A$ and $B$ is 
\begin{equation} \label{eq:mi}
  I(A,B) = 
  \sum_{i=1}^{q_A}\sum_{j=1}^{q_B} \frac{n_{ij}}{N} 
  \log\left(\frac{n_{ij} N}{n_i n_j}\right) .
\end{equation}
$q_A$ and $q_B$ are the number of communities in divisions $A$ and $B$,
$n_i$ and $n_j$ are the number of nodes in communities $i$ and $j$,
and $n_{ij}$ is a ``confusion matrix'' that identifies 
the number of nodes in community $i$ of partition
$A$ that are found in community $j$ of partition $B$.
For a single community division $A$, we can determine the 
self mutual information by $H(A) = I(A,A)$ which is identically 
equal to the Shannon entropy.
We use base $2$ logarithms.

We calculate the variation of information $V(A,B)$ by
\begin{equation} \label{eq:vi} 
  V(A,B) = H(A) + H(B) - 2I(A,B).
\end{equation}
The range for VI is $0\le V(A,B)\le\log N$.
The normalized mutual information $I_N(A,B)$ is %defined as
\begin{equation} \label{eq:nmi} 
  I_N(A,B) = \frac{2I(A,B)}{H(A) + H(B)}.
\end{equation}
The range for NMI is $0\le I_N(A,B)\le 1$.
The NMI and VI afford slightly different perspectives on the replica 
correlations in the multiresolution analysis above.

These information measures are based on the cluster definitions %, 
and are not directly related to a thermodynamic entropy.
The replica correlation measures could be improved by incorporating 
information about the assigned ``overlapping'' configurations 
(nodes may belong to more than one cluster) such as in Ref.\ \cite{ref:lanc},
by modifying the measures to account for the fact that same-type atoms 
are indistinguishable in this type of a model, 
or by utilizing a thermodynamic entropy since we are dealing 
with a physical system in the current application.

\section{Physical community detection models} \label{sec:edgediscussion}

Minimizing the Hamiltonian of \eqnref{eq:ourPottsmodel} using the algorithm
briefly explained in \secref{sec:hamiltonian}, models a dynamic community 
detection process where we search for a local energy minimum that indicates 
a ``good'' community partition (in general).
In particular, we have applied a potential energy (PE) model of network edges, 
and we could, in principle, apply other edge definitions to obtain other relevant
configurations with the following caveat.
When we minimize the Hamiltonian using the edge weights, a low energy state 
corresponds to a good partition, so for consistency with the community detection 
problem, any edge definitions should ideally be extremized at the most favorable 
configuration(s).
The subtle distinctions between a few different natural edge weight models can 
result in different calculated clusters (beyond natural fluctuation caused by 
a high configurational entropy).

An inverted PE model fits the above criterion, and it would at first appear 
to be conceptually consistent with a model based on the relative pairwise 
squared node displacement (SND) model.
For both cases, small deflections about the minimum still indicate a good, 
even if not optimal, configuration corresponding to the intuitive notion 
of a bound cluster.
However, the SND model would cause a perfect crystal to be identified as a 
single contiguous cluster.
The PE would not have this effect since more distant nodes have a much lower
potential energy.
In effect, the PE model could identify the smaller scale units in a perfect
crystal that the SND model could not isolate.

Another intuitive edge model for bound clusters is related to the attractive 
force that one may expect to exist between constituent elements of a physical 
cluster.   
Of course, the system forces still cause the formation of a crystalline ground 
under the appropriate conditions, but the forces dictate the system's physical 
\emph{dynamics} over time where the PE model (more) directly indicates the 
relevant ground state(s).
Forces with a minimum such as the LJ model in \eqnref{eq:LJVr} would be zero 
at the optimal ground state configuration (ideal crystal state), and the 
maximum force would be at a larger radius $r$ than the PE minimum.
While a force model of edge weights is also intuitive, it would correspond 
to (perhaps slightly) different clusters in a physical system compared to clusters 
derived from a PE model because the community detection Hamiltonian is extremized 
at different radii.

This discussion of the subtle differences between the edge models leads 
to the natural question of which is the ``best'' physical cluster model.
Another perspective is that the different edge weight models simply answer
different questions.
We have selected the PE definition since it best corresponds to the ideal
ground state of the system in our pairwise interaction model.
We then infer the best local clusters within configurations in local 
(frustrated) equilibrium (in the solid systems).

\section{MRO and structure factor pre-peaks} \label{sec:Sofqprepeaks}

% new text here ----------------------------------------------------------
One experimental approach to ascertain MRO is to look for ``prepeaks'' 
in the scattering data.
That is, one can look for lower amplitude peaks in the structure factor 
$S(q)$ which precede the dominant $S(q)$ peak for wavenumbers $q$.
While the approach may capture general MRO structures, it is possible 
to have MRO structures {\em without significant pre-peak(s)} 
in the structure factor plot. 

We illustrate the basic premise of this
statement with an elementary example -- that of a random arrangement of crystallites.
Even though the Fourier transform of the mass density in each individual 
crystallite has sharp peaks at the reciprocal lattice vectors corresponding 
to these small crystallites, the structure factor obtained from the entire 
system may have those peaks vanish.
We may, for instance, denote the locations of the centers of mass of individual 
grains $i$ by $\vec{R}_{i}$ and denote the location of individual 
atoms $j$ in each grain with respect to its center of mass by $\vec{r}_{j}$.
In that case, the structure factor is
%$S(\vec{k}) = \sum_{\vec{R}_{i}} \sum_{\vec{r_j} \in \mbox{$i$} }
%\exp(i \vec{k} \cdot (\vec{R}_{i} + \vec{r}_{j})$.
\begin{equation}
  S\big(\vec{k}\big) = \sum_{\vec{R}_{i}} \sum_{\vec{r_j} \in \mbox{$i$} }
               \exp\left[i \vec{k} \cdot \left(\vec{R}_{i} + \vec{r}_{j}\right) \right].
\end{equation}
Within each individual grain on its own, 
%$\sum_{\vec{r}_{j} \in \mbox{crystallite $i$}  }
%\exp(i\vec{k} \cdot (\vec{R}_{i} + \vec{r}_{i}) = S_{i}(\vec{k})$ will be
%%$\sum_{\vec{r}_{j} \in \mbox{$i$} }
%%\exp(i\vec{k} \cdot (\vec{R}_{i} + \vec{r}_{j}) = S_{i}(\vec{k})$ 
\begin{equation}
  S_{i}\big(\vec{k}\big) = \sum_{\vec{r}_{j} \in \mbox{$i$} }
               \exp\left[i\vec{k} \cdot \left(\vec{R}_{i} + \vec{r}_{j}\right)  \right]
\end{equation}
will be sharply peaked about the corresponding reciprocal lattice vectors 
of grain $i$. 
However, the complete sum, 
%$\sum_{\vec{R}_{i}} \exp(i \vec{k} \cdot \vec{R}_{i}) S_{i}(\vec{k})$
\begin{equation}
  S\big(\vec{k}\big) = \sum_{\vec{R}_{i}} \exp\left(i \vec{k} \cdot \vec{R}_{i}\right) 
               S_{i}\big(\vec{k}\big),
\end{equation}
may vanish if the relative distances between the different crystallites
are random and lead to phases $\exp(i \vec{k} \cdot \vec{R}_{i})$ which 
obliterate the signatures of order in the individual $S_{i}\big(\vec{k}\big)$.
% new text here ----------------------------------------------------------

\section{Overlapping nodes between different communities} \label{sec:overlapmethod}

We wish to account for the possibility of a given atom being connected 
to more than one physical cluster.
For example, in a cubic lattice, each atom participates in the local 
structure of multiple unit cells.
In community detection, this corresponds to allowing ``overlapping'' 
community memberships where a node can be a member of more than one 
community.
To accomplish this task, we select the lowest energy partition 
at the best resolution(s) of the model network 
[\ie{}, value(s) of $\gamma$ in \eqnref{eq:ourPottsmodel}
corresponding to extrema in $I_N$ or $V$].

First, we fix the initial node memberships including the number 
of communities $q$.
We then sequentially iterate through the community memberships of 
each node and make changes according to the following:
($1$) place the node in any additional (non-member) clusters to which 
it is bound (a negative energy contribution), or 
($2$) remove the node from any member clusters (except for the original 
membership) in which the current net energy contribution is positive.
This process iterates through all nodes as many times as necessary until 
no node additions or removals are found.
The total computational cost is slightly higher than the initial partitioning
cost in \secref{sec:hamiltonian} due to the multiple allowed memberships.
See also \cite{ref:lanc} for another method that allows overlapping 
multiscale network analysis in a general fashion.

\begin{figure}[t]
\begin{center}
\subfigure{\includegraphics[width=\subfigwidth]{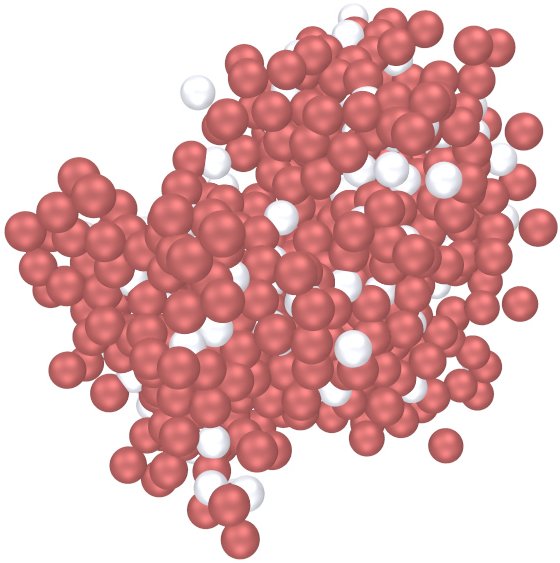}}
\subfigure{\includegraphics[width=\subfigwidth]{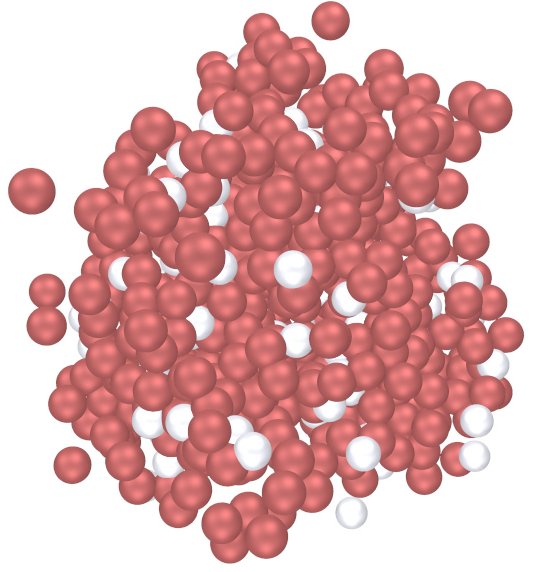}}
\subfigure{\includegraphics[width=\subfigwidth]{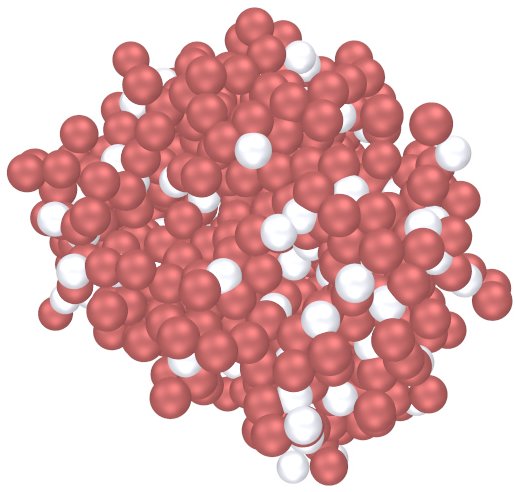}}
\subfigure{\includegraphics[width=\subfigwidth]{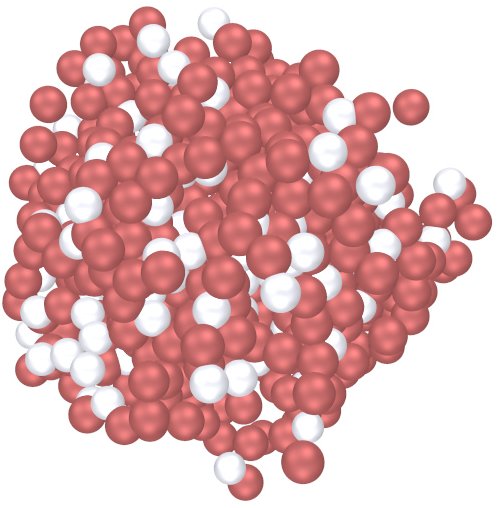}}
\caption{The best clusters found for the peak $V$ for the replica correlation within the high temperature LJ system (see text).
These clusters include overlapping node membership assignments
where each node is required to have a overall negative binding 
energy to the other nodes in the cluster. The peak in the variation of information $V$ generally correlates
with the scale on which fluctuations are the largest. 
The atomic identities are B (silver) and A (red) in order 
of increasing diameters. The diffuse objects found at high
temperatures are no longer as compact as at lower temperatures.
The normalized mutual information
at the peak V is also correspondingly lower than that
for the low temperature system.
%Some corresponding high temperature clusters are shown in 
%\figref{fig:LJVIMaxbestclustersTHigh}.
}
\label{fig:LJVIMaxbestclustersTHigh}
\end{center}
\end{figure}

\myfig{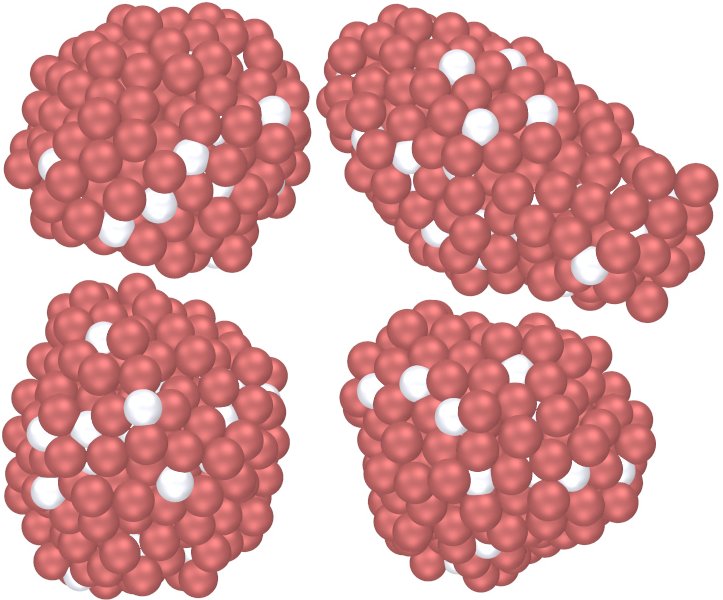}{Some of the best clusters 
for the peak $V$ for the low temperature system shown 
in \figref{fig:LJTLow} in panel (b). The peak in the variation of information $V$ generally correlates
with the scale on which the fluctuations in the system division are most prominent. 
These clusters include overlapping node membership assignments
where each node is required to have a overall negative binding 
energy to the other nodes in the cluster.
The atomic identities are B (silver) and A (red) in order 
of increasing diameters. The resulting configurations constitute
tightly bound objects. 
Representative corresponding high temperature clusters are shown in 
\figref{fig:LJVIMaxbestclustersTHigh}.}
{fig:LJVIMaxbestclustersTLow}{0.85\linewidth}{b!}

% correlation length figure ------------------------------------------
\begin{figure}[b]
\begin{center}
\subfigure[]{\includegraphics[width=\subfigwidth]{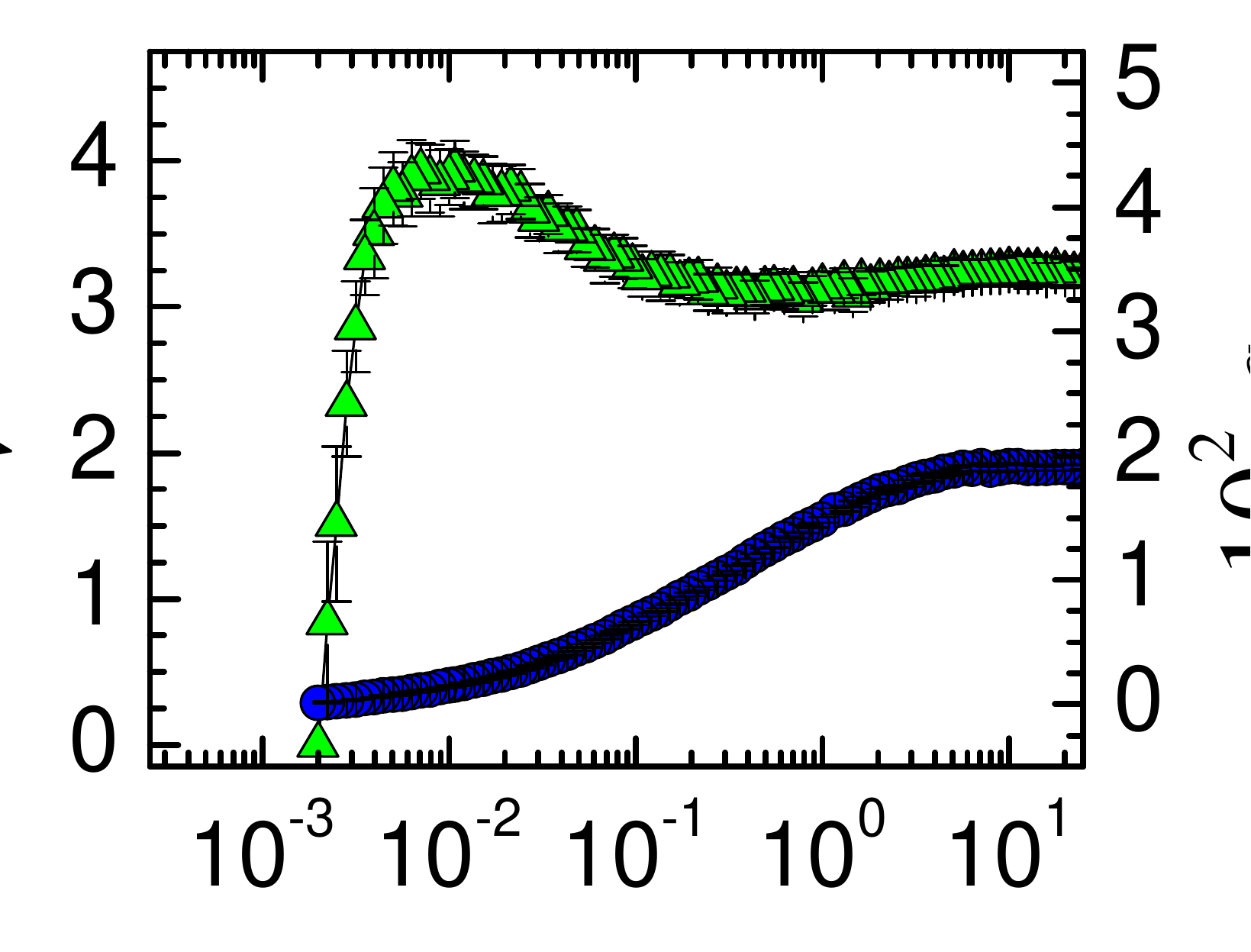}}
\subfigure[]{\includegraphics[width=\subfigwidth]{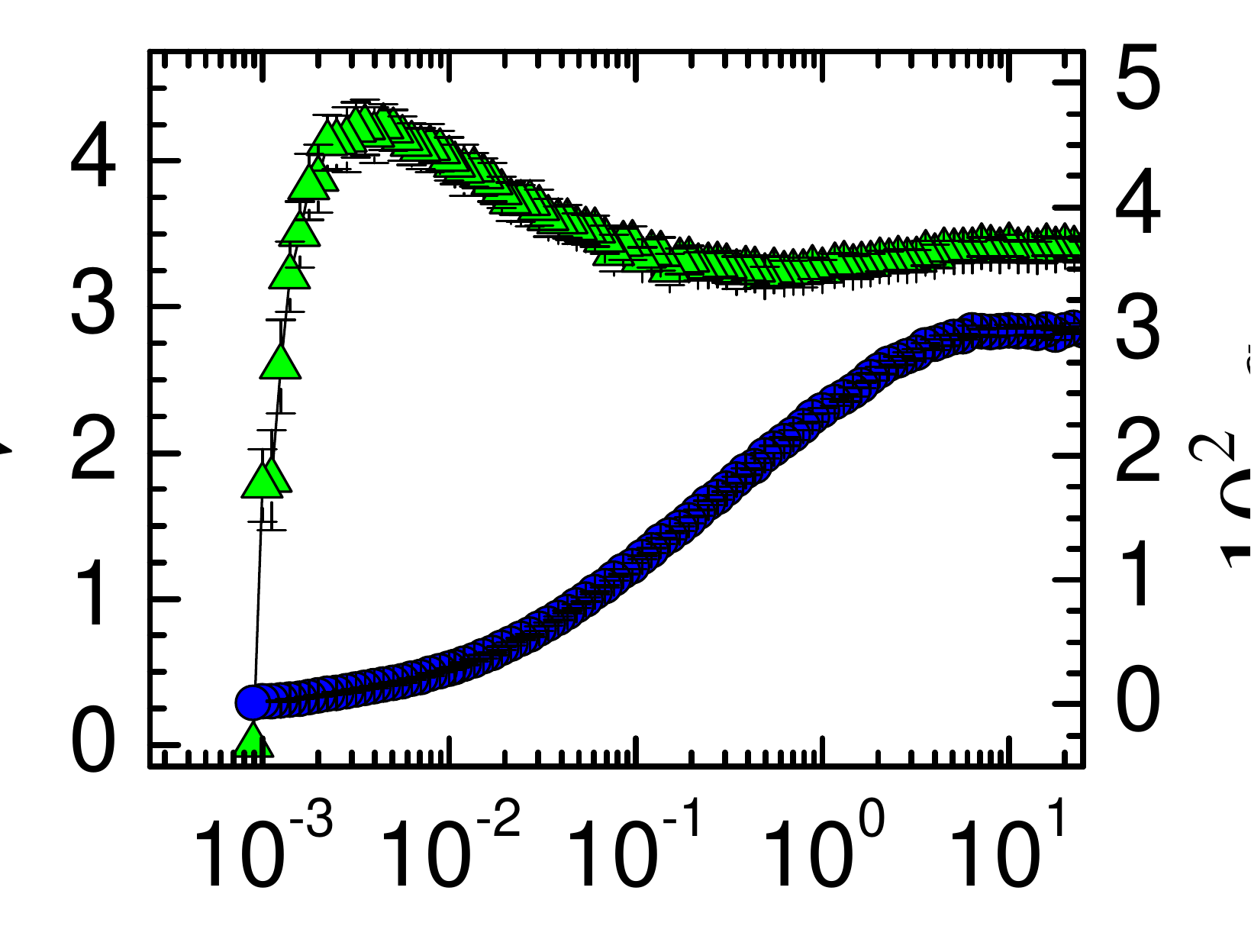}}
\subfigure[]{\includegraphics[width=\subfigwidth]{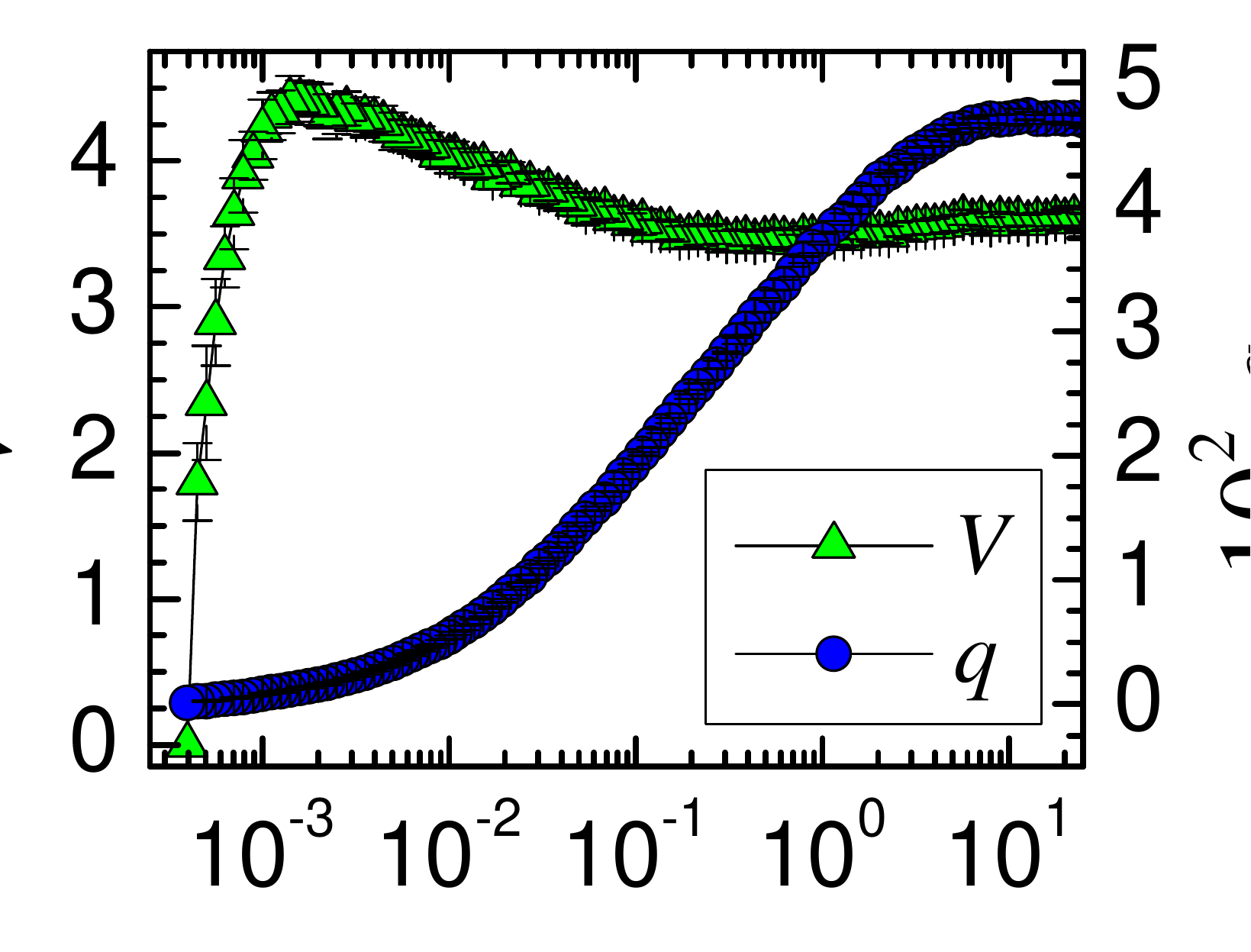}}
\caption{Plots of the variation of information $V$ between replica pairs as a function of the resolution parameter $\gamma$ for several system sizes in a LJ 
simulation (see \secref{sec:LJsimulation}).
Panels (a), (b), and (c) use $N=2000$, $N=4000$, and $N=8000$, respectively.
Note that the value of $\gamma$ corresponding to the peak $V$ scales downward
with the system size (larger structures) which may indicate that it is correlated 
to a diverging length scale.}\label{fig:LJfinitesize}
\end{center}
\end{figure}
% end of correlation length figure -----------------------------------

%--------------------------------------------------------------------
% moved to here to try to improve figure placement
% begin MRA algorithm example figures for static metallic glass *****
\myfig{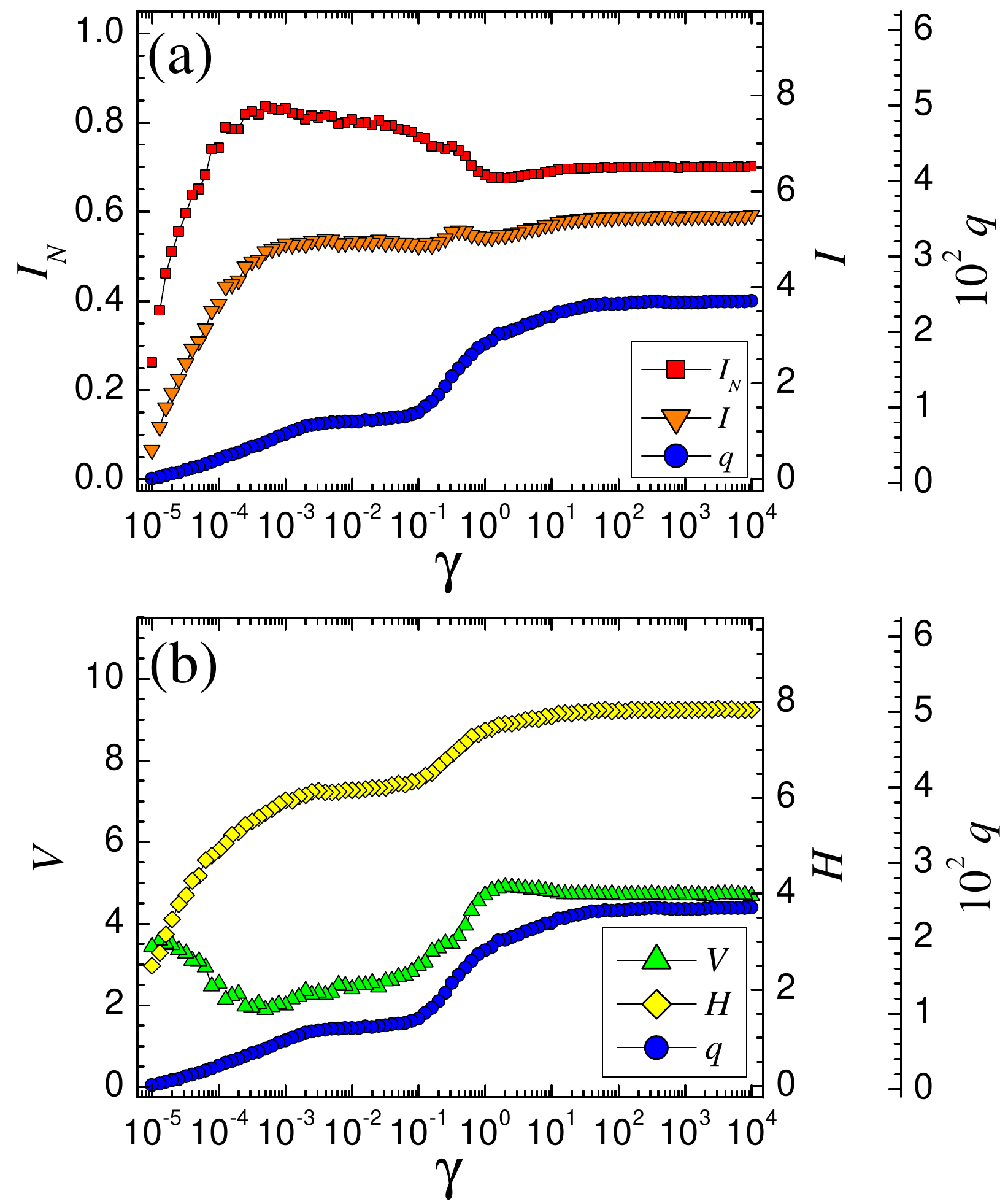}{Panels (a) and (b) show the plots 
of information measures $I_N$, $V$, $H$, and $I$ and the number of clusters $q$ 
(right-offset axes) versus the Potts model weight $\gamma$ 
in \eqnref{eq:ourPottsmodel}.
The ternary model system contains $1600$ atoms in a mixture of $88\%$ type A, 
$7\%$ of type B, and $5\%$ of type C with a simulation temperature of $T=300$ K 
which is well \emph{below} the glass transition for this system.
In this system, we use a single \emph{static} ``snapshot'' of the system 
to analyze what our multiresolution algorithm finds at $T=300$ K.
This low temperature case shows a preferred resolution at low $\gamma$
as evidenced by the information extrema at ($i$) in both panels.
See \figref{fig:AlYFeTHighStatic} for the corresponding high temperature case.
}{fig:AlYFeTLowStatic}{\figsize}{t}

%\myfig{SNAlYFeT1500Vavg.eps}{
\myfig{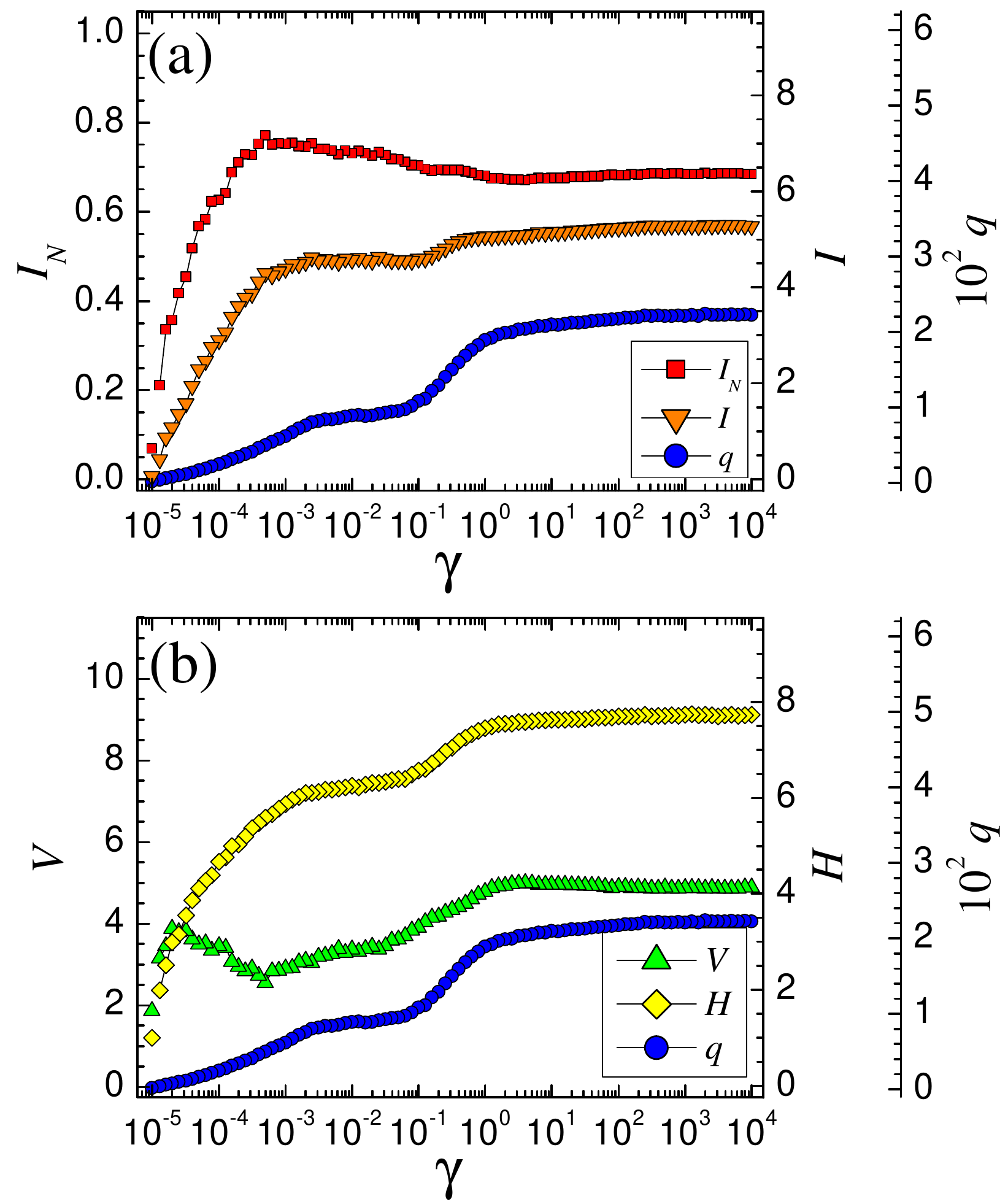}{Panels (a) and (b) show the plots 
of information measures $I_N$, $V$, $H$, and $I$ and the number of clusters $q$ 
(right-offset axes) versus the Potts model weight $\gamma$ 
in \eqnref{eq:ourPottsmodel}.
The ternary model system contains $1600$ atoms in a mixture of $88\%$ type A, 
$7\%$ of type B, and $5\%$ of type C with a simulation temperature of $T=1500$ K 
which is well \emph{above} the glass transition for this system.
In this system, we use a single \emph{static} ``snapshot'' of the system 
to analyze what our multiresolution algorithm finds at $T=1500$ K.
This high temperature case does show a preferred resolution at low $\gamma$
as evidenced by the information extrema at ($i$) in both panels,
but the NMI maxima are somewhat lower.  
The solutions also take longer to solve accurately meaning that 
the community detection energy landscape is more complicated.
See \figref{fig:AlYFeTLowStatic} for the corresponding low temperature 
case.}
{fig:AlYFeTHighStatic}{\figsize}{b!}
% end MRA algorithm example figures for static metallic glass *******
%--------------------------------------------------------------------

%--------------------------------------------------------------------
% moved to here to try to improve figure placement
% begin MRA algorithm example figures for static metallic glass *****
\myfig{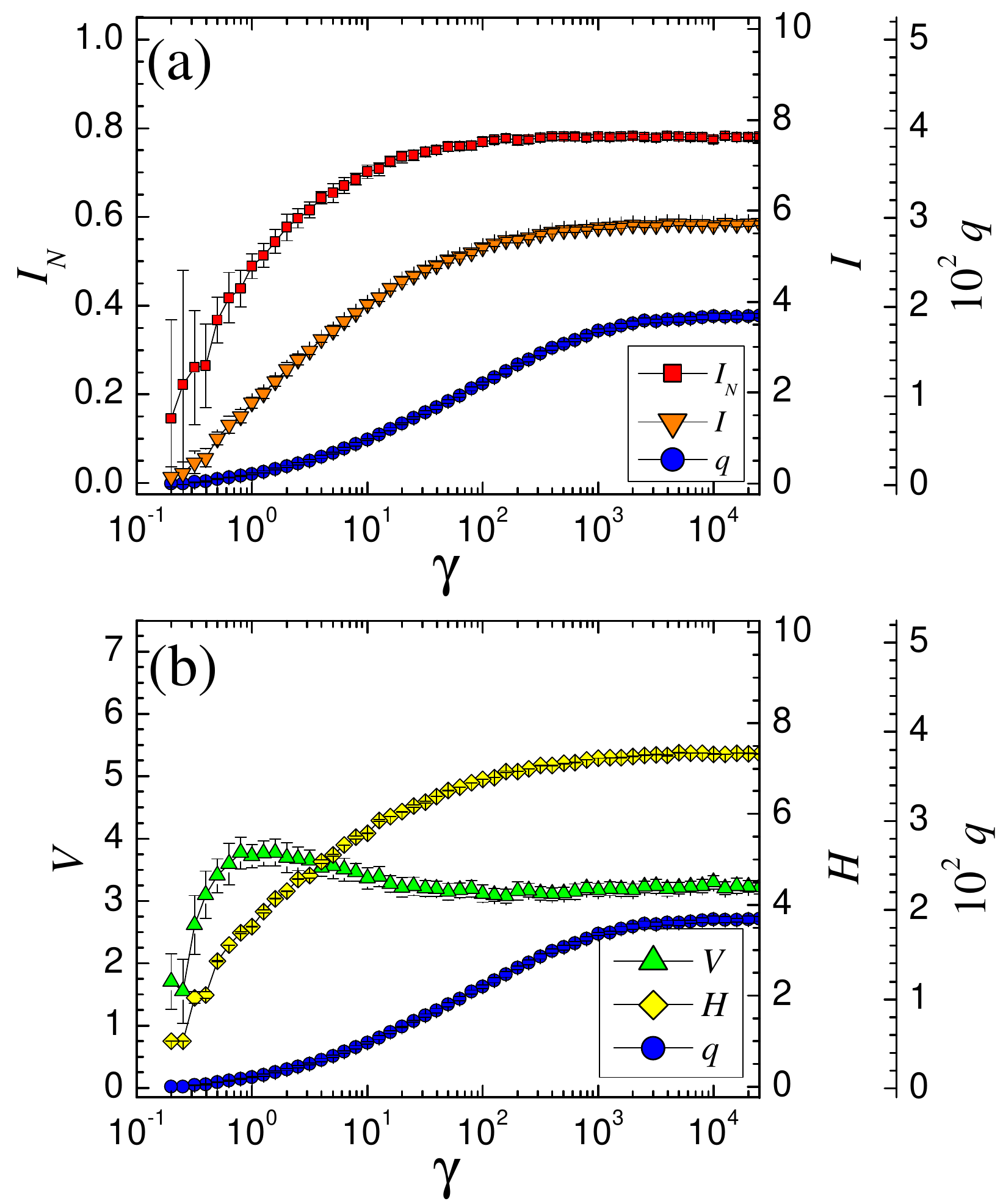}{Panels (a) and (b) show the plots 
of information measures $I_N$, $V$, $H$, and $I$ and the number 
of clusters $q$ (right-offset axes) versus the Potts model weight 
$\gamma$ in \eqnref{eq:ourPottsmodel}.
As in \secref{sec:LJresults}, this LJ system contains $2000$ atoms 
in a mixture of $80\%$ type A and $20\%$ type B with a simulation 
temperature of $T=0.01$ (energy units) which is well \emph{below} the 
glass transition for this system.
In this test, we use a single \emph{static} ``snapshot'' of the system.
This low temperature system has a higher overall level of correlation among 
the replicas than the high temperature case in \figref{fig:LJTLowStatic},
and the overall results are very similar to what was observed for the time
separated replicas in \figref{fig:LJTLow} in \secref{sec:LJresults}.
See \figref{fig:LJTLowStaticbestclusters} for a sample of some of the best
observed clusters.}
{fig:LJTLowStatic}{\figsize}{h} %{t}

\myfig{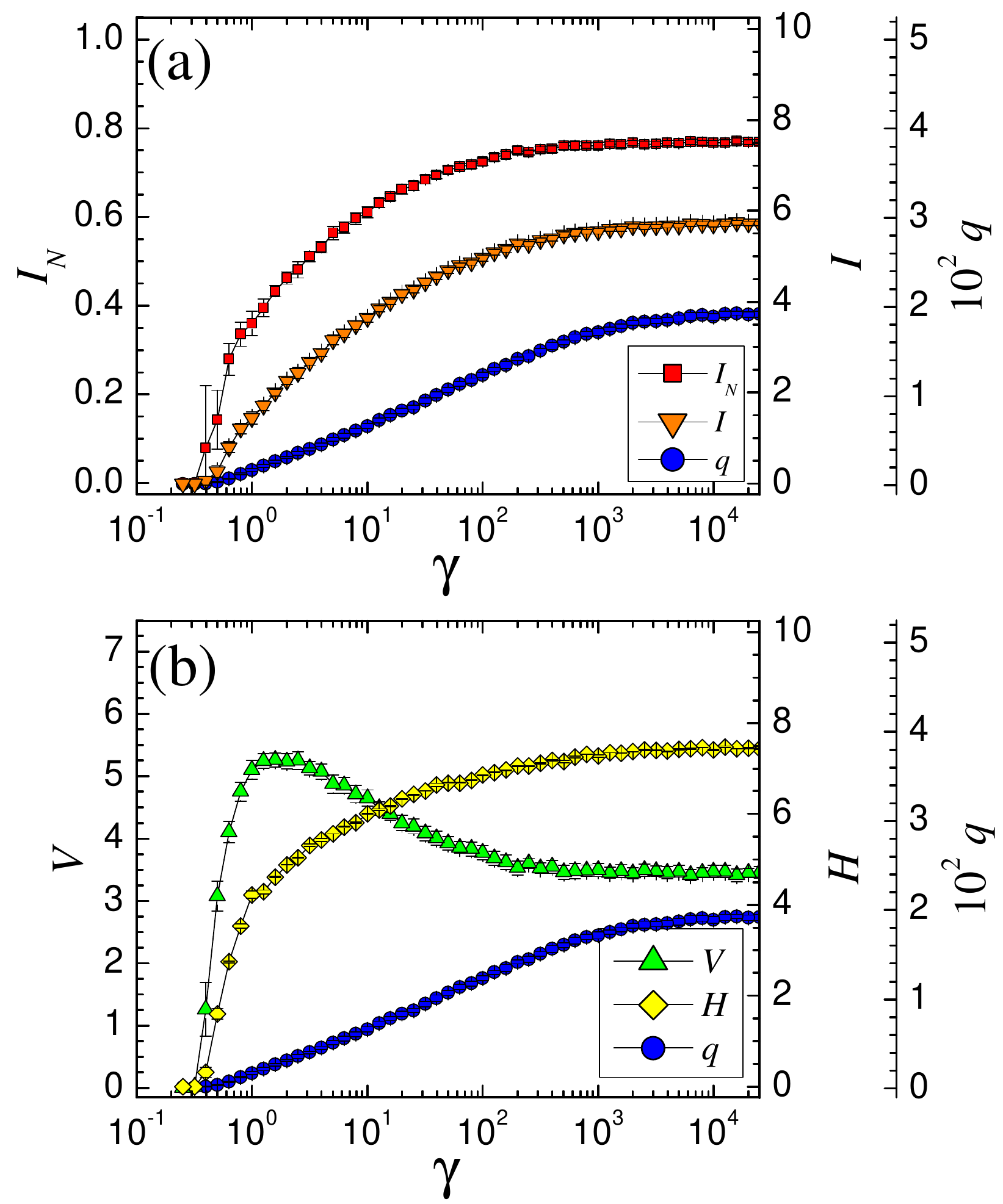}{Panels (a) and (b) show the plots 
of information measures $I_N$, $V$, $H$, and $I$ and the number of clusters $q$ 
(right-offset axes) versus the Potts model weight $\gamma$ 
in \eqnref{eq:ourPottsmodel}.
The LJ system contains $2000$ atoms in a mixture of $80\%$ type A and 
$20\%$ type B with a simulation temperature of $T=5$ (energy units) 
which is well \emph{above} the glass transition for this system.
In this analysis, we use a single \emph{static} ``snapshot'' of the system.
This high temperature system has a lower overall correlation than
the low temperature case in \figref{fig:LJTLowStatic}, and the overall
results are very similar to what was observed for the time separated
replicas in \figref{fig:LJTHigh} of \secref{sec:LJresults}.}
{fig:LJTHighStatic}{\figsize}{h} %{b!}

\myfig{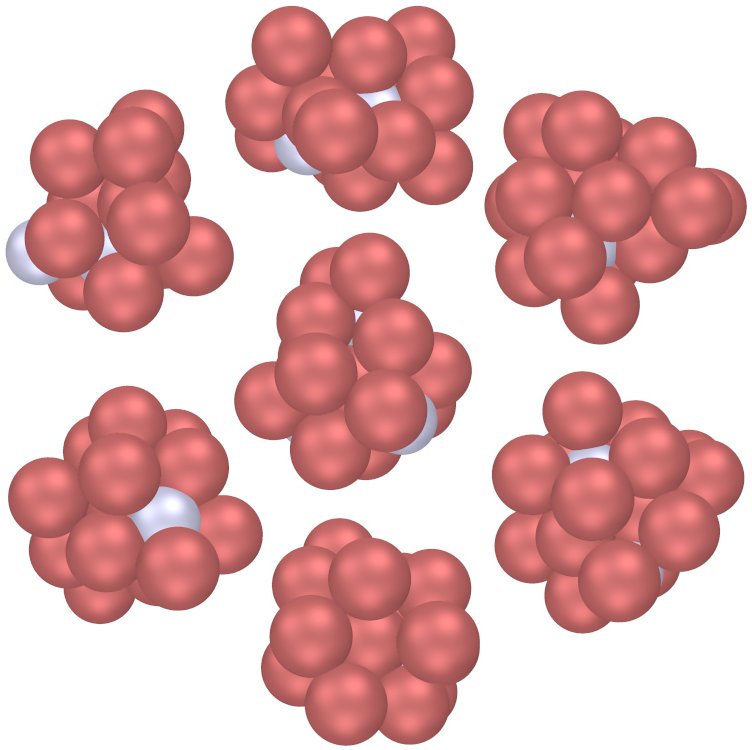}{A depiction of some of the best 
clusters for the peak replica correlation in the static system of \figref{fig:LJTLowStatic}.
These clusters include overlapping node membership assignments
where each node is required to have an overall negative binding 
energy to the other nodes in the cluster.
The atomic identities are A (red) and B (silver).}
{fig:LJTLowStaticbestclusters}{0.85\linewidth}{h} %{t}
% end MRA algorithm example figures for static metallic glass *******
%--------------------------------------------------------------------

\section{Potential shift in the ternary metallic glass} 
\label{sec:ShiftsAlYFemodel}

In \figref{fig:AlYFeTLowshiftgamma}, we show the NMI correlations for a range
of potential shifts $\phi_0$ and the Potts model weights $\gamma$
using $s=12$ replicas and $t=10$ trials for all data.
The NMI peaks are roughly constant over a range of $\phi_0$ for the ternary 
model glass up to $\phi_0=0.1$ (the A-A interaction minimum 
$\phi_\mathrm{min}\simeq 0.24$ eV).

\section{Time correlations in the ternary model glass} 
\label{sec:TimesAlYFemodel}

In \figref{fig:AlYFeTHightimegamma}, we show the NMI correlations for a range
of separation times $t_s$ between replicas (units are in MD time steps) and the 
Potts model weights $\gamma$ using $s=12$ replicas and $t=10$ trials for all data.
This plot intuitively shows that the correlations weaken as we increase the 
separation time between replicas.
Through varying the time step between replicas, we examine the correlations 
as a function of $t_s$ and $\gamma$ which allows us to determine the most 
important time scale(s) (or relative equivalence).
See \figref{fig:AlYFeTLow} for the corresponding 2D plot using $t_s=1000$, and 
the static limiting case is shown explicitly in \secref{sec:StaticAlYFemodel}.

\section{Time correlations in the binary LJ glass} 
\label{sec:TimesLJemodel}

In \figref{fig:LJTLowtimegamma}, we show the NMI correlations for a range
of separation times $t_s$ between replicas (units are in MD time steps) and the 
Potts model weights $\gamma$ using $s=12$ replicas and $t=10$ trials for all data.
This plot shows that the correlations weaken only a little as we increase the 
separation time between replicas up to $10~\!000$ MD time steps.
Through varying the time step between replicas, we examine the correlations 
as a function of $t_s$ and $\gamma$ which allows us to determine the most 
important time scale(s) (or relative equivalence).
See \figref{fig:AlYFeTLow} for the corresponding 2D plot using $t=1000$, and 
the static limiting case is shown explicitly in \secref{sec:StaticAlYFemodel}.

\section{An optimization effect in the binary LJ glass} 
\label{sec:TrialsLJmodel}

In \figref{fig:LJTLowHightrialsgamma}, we show the NMI correlations for a range
of optimization trials $t$ for $s=12$ replicas and the Potts model weights $\gamma$.
The plots show that the number of trials $t$ has a small effect on the overall 
accuracy of the solution for either temperature ($T=0.01$ or $5$), but the effect
is slightly more pronounced in the higher temperature $T=5$ case.
See \figsref{fig:LJTLow}{fig:LJbestclusters} for the corresponding 2D plots using
$s=10$.

The high temperature clusters are generally much more dispersed.
See \figref{fig:LJTHighbestclusters} for sample clusters using $t=10$ and $20$ 
at $T = 5$ where the clusters correspond to the multiresolution plot 
in \figref{fig:LJTHigh} at $\gamma=10^4$.
The corresponding low temperature clusters are analyzed \figref{fig:LJTLow} and 
presented in \figref{fig:LJbestclusters}.
Panels (a) and (b) in \figref{fig:LJTHighbestclusters} display the typical case 
of dispersed clusters at $t=10$.
In some instances, the high temperature clusters can be more compact,
albeit not densely packed, where panels (c) and (d) show two examples at $s=20$.
The clusters that are identified are generally consistent in terms of sparseness 
across all clusters in the solution.
The frequency of occurrence of the more compact high temperature clusters, and 
its dependence on the level of optimization, is a subject for further study.

%----------------------------------------------------------------
% begin MRA algorithm example figures for metallic glass *****
\begin{table}[t]
\begin{tabular}{| c | c | c | c | c | c | c | c |}
\hline
& $a_0$ & $a_1$ & $a_2$ & $a_3$ & $a_4$ & $a_5$  \\
\hline
AA & 2.11* & 9.49* & -32.3* & 3.66* & -10.6* & 6.20* \\
AB & 1.92 & 17.4 & 6.09 & 3.05 & -4.68 & 3.48 \\
AC & 2.38 & 8.96 &-14.9 & 3.11 & -3.88 & 4.38 \\
BB & 2.01* & 4.95* & 5.01* & 2.74* & -2.26* & 3.00* \\
BC & 1.88 & 8.00 &-3.42 & 2.53 & -1.25 & 3.00 \\
CC & 2.75* & 15.3* &-6400* & 2.38* & -4.69* & 8.71* \\
\hline
\end{tabular}
\caption{Fit parameters for \eqnref{eq:VrAlYFe} obtained from fitting 
configuration forces and energies to ab-initio data \cite{ref:mihalkovicEOPP,ref:vaspwww}. 
The units of the parameters are such that given $r$ in $\AA$, $\phi(r)$ is in $eV$.
The same-species (*) data is different from Table \ref{tab:ppfit}.}\label{tab:ppfitNew}
\end{table} 

%\myfig{SNAlYFeT300Vavg.eps}{
\myfig{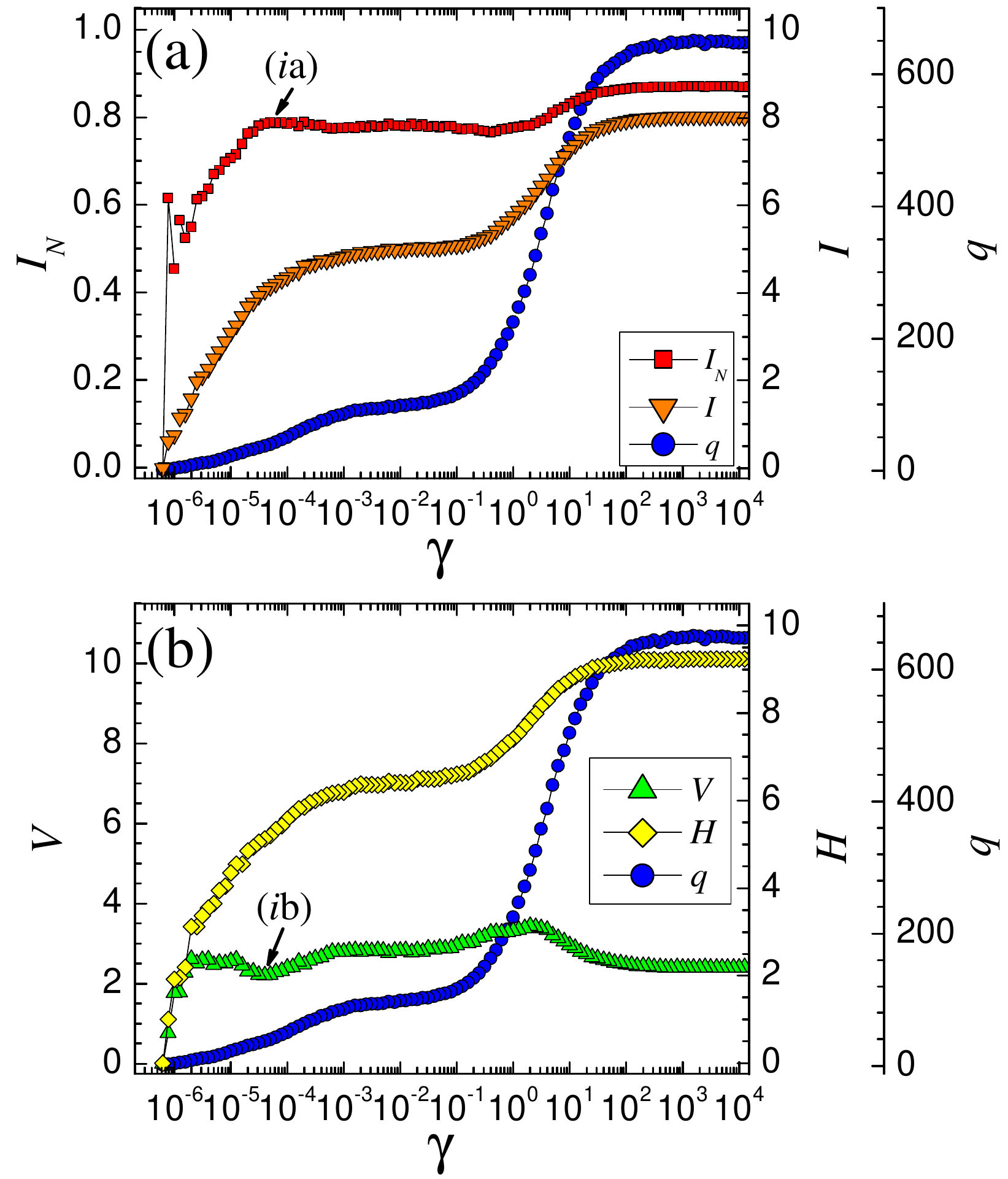}{Panels (a) and (b) show the plots 
of information measures $I_N$, $V$, $H$, and $I$ and the number of clusters $q$ 
(right-offset axes) versus the Potts model weight $\gamma$ 
in \eqnref{eq:ourPottsmodel}.
The ternary model system contains $1600$ atoms in a mixture of $88\%$ type A, 
$7\%$ of type B, and $5\%$ of type C with a simulation temperature of $T=300$ K 
which is well \emph{below} the glass transition for this system.
This alternate system uses the best fit parameter data for the same-species 
interactions as opposed to the pseudo-potential interaction used 
in \figdref{fig:AlYFeTLow}{fig:AlYFebestclusters}.
This system shows a locally preferred resolution as evidenced by the information 
extrema at ($i$) in both panels.
A set of sample clusters for the best resolution at $\gamma\simeq 0.0001$ 
is depicted in \figref{fig:AlYFebestclustersalt}.
The region for $\gamma\gtrsim 100$ actually has a higher correlation,
but the clusters are very small ($n\simeq 5$ nodes) and somewhat dispersed.}
{fig:AlYFeTLowalt}{\figsize}{b}

%\myfig{SNAlYFeT1500Vavg.eps}{
\myfig{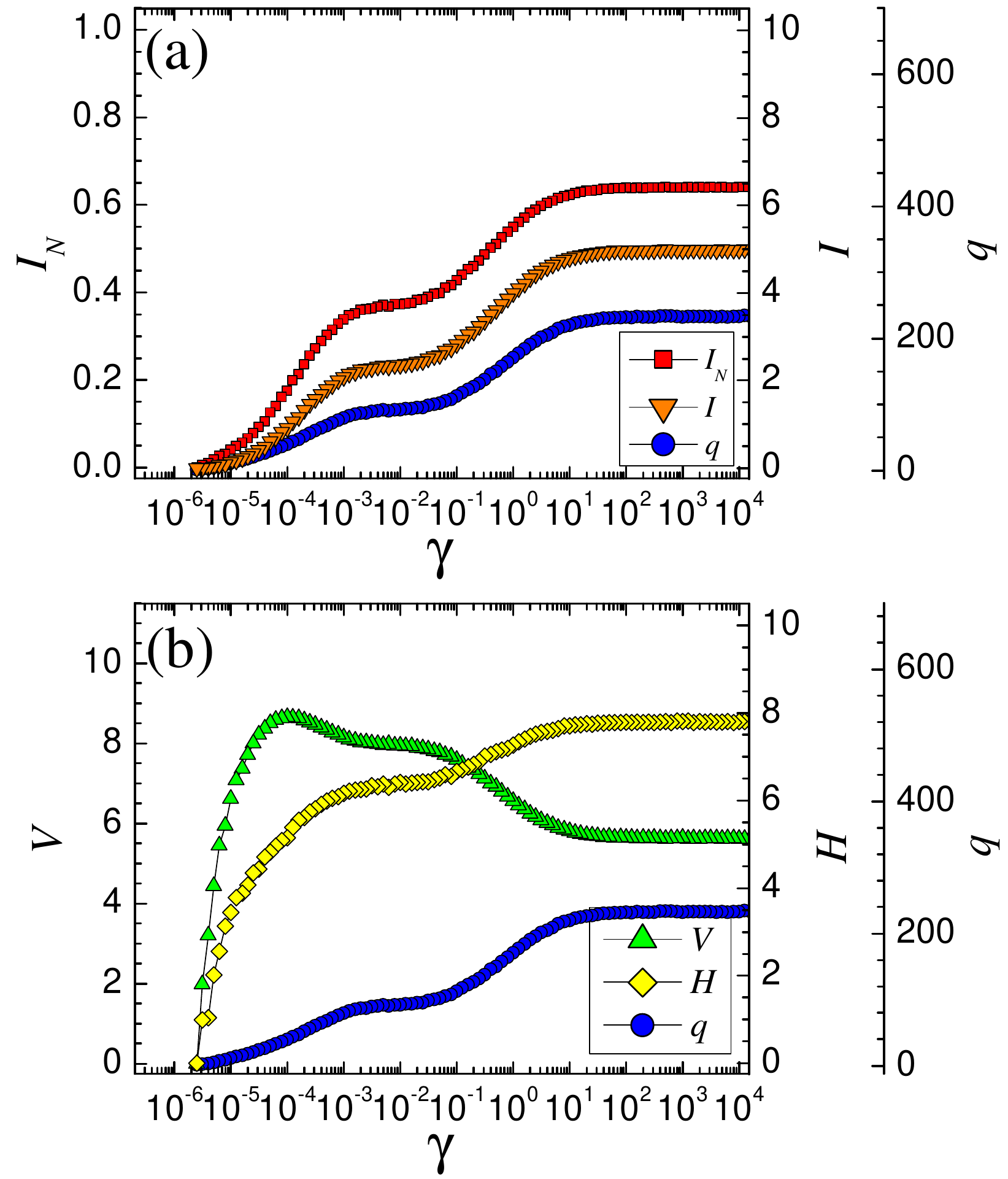}{Panels (a) and (b) show the plots 
of information measures $I_N$, $V$, $H$, and $I$ and the number of clusters $q$ 
(right-offset axes) versus the Potts model weight $\gamma$ 
in \eqnref{eq:ourPottsmodel}.
The ternary model system contains $1600$ atoms in a mixture of $88\%$ type A, 
$7\%$ of type B, and $5\%$ of type C with a simulation temperature of $T=1500$ K 
which is well \emph{above} the glass transition for this system.
This alternate system uses the best parameter fit for the same-species 
interactions as opposed to the pseudo-potential interaction used 
in \figdref{fig:AlYFeTLow}{fig:AlYFebestclusters}.
At this temperature, there is no resolution where the replicas are 
strongly correlated.
See \figref{fig:AlYFeTLowalt} for the corresponding low temperature case
where the replicas are much more highly correlated at $\gamma\simeq 0.001$.}
{fig:AlYFeTHighalt}{\figsize}{h} %{t}
% end MRA algorithm example figures for metallic glass *******

\myfig{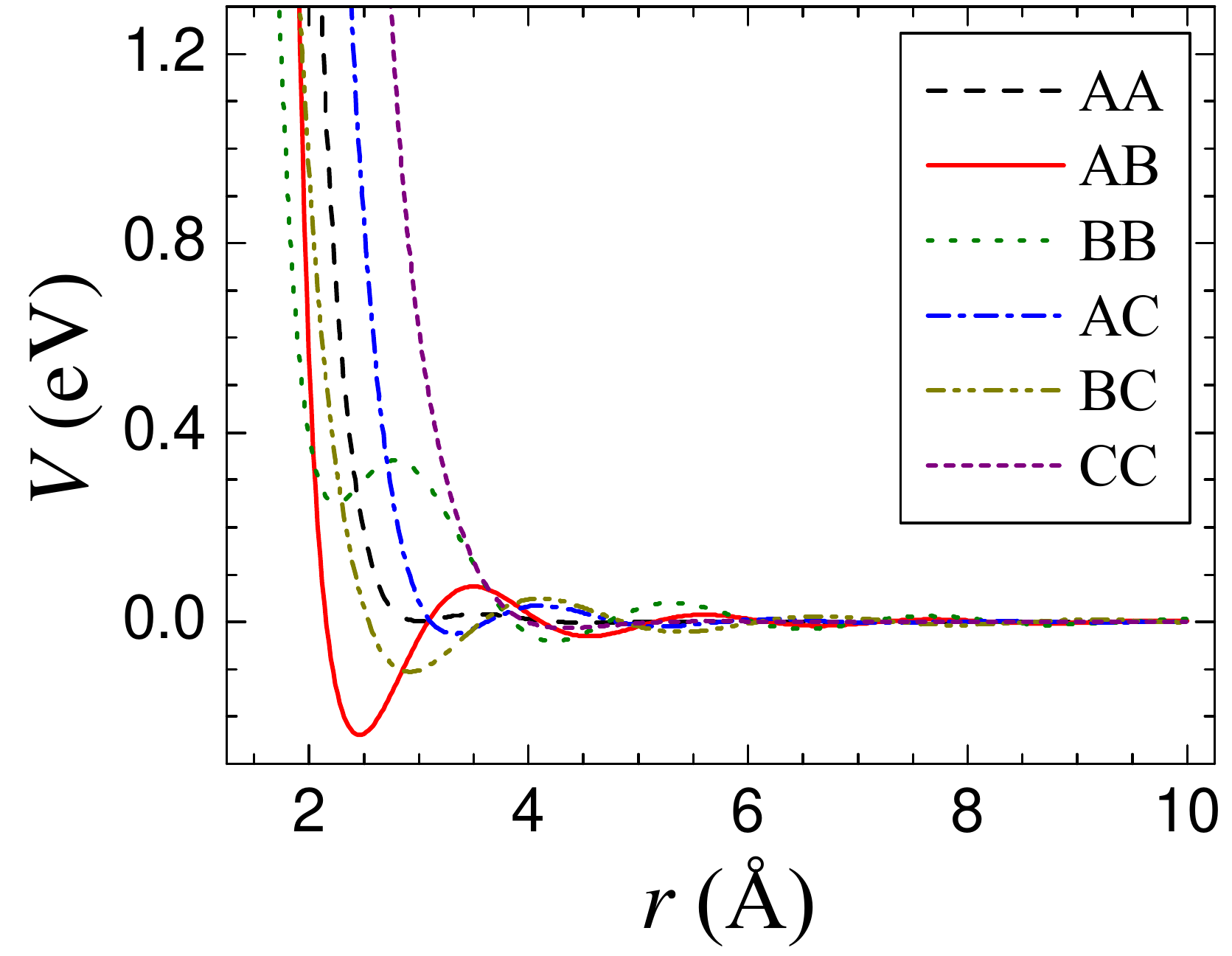}{A plot of the model potentials for our 
three-component model glass former (see \figref{fig:fullAlYFesystem}) 
using the fit data in Table \ref{tab:ppfitNew}.
We indicate the atomic types by ``A'', ``B'', and ``C'' which are included 
with mixture ratios of $88\%$, $7\%$, and $5\%$, respectively.
The units are given for a specific candidate atomic realization (AlYFe)
discussed in the text.  
Here, we apply the ab initio fit data for the same-species interactions, 
as opposed to the GPT \cite{ref:moriartyGPT} model from Table \ref{tab:ppfit} 
in \figref{fig:ppfit}.}
{fig:ppfitNew}{0.85\linewidth}{b}

%\myfig{AlYFeClusters.eps}{
\myfig{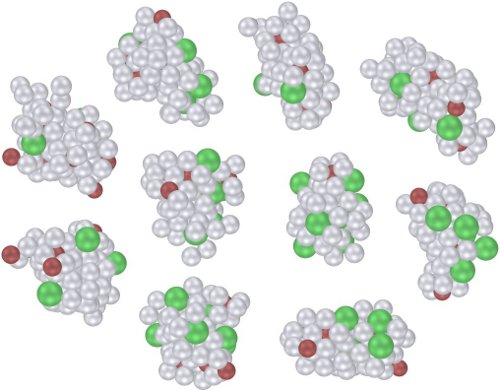}{A depiction of some of the best clusters 
for the peak replica correlation at feature ($i$) in \figref{fig:AlYFeTLow}.
These clusters include overlapping node membership assignments
where each node is required to have an overall negative binding 
energy to the other nodes in the cluster.
The atomic identities are C (red), A (silver), B (green) in order 
of increasing diameters.
These clusters are generally larger than $1/2$ the simulated system width;
therefore, it would beneficial to test their consistency in a larger 
simulation (requiring substantially longer computational time).}
{fig:AlYFebestclustersalt}{\figsize}{h} %{t}
% end MRA algorithm example figures for metallic glass *******
% ------------------------------------------------------------

\section{Multiresolution VI maxima in the LJ system} 
\label{sec:LJVImaxima}

In \figref{fig:LJTLow}, we show the multiresolution correlations as a function 
of the Potts model weight $\gamma$ for a LJ system of $N=2000$ nodes
using $s=12$ replicas and $t=10$ trials for all data.
The value of $\gamma$ corresponding to the peak in $V$ corresponds to the 
``maximum complexity'' of the energy landscape which is often correlated 
to the system size (see Appendix \ref{sec:LJfinitesizeeffect}).
\figref{fig:LJVIMaxbestclustersTLow} shows a sample of the best clusters 
corresponding to the VI peak which are roughly $300$ atoms in size 
($7-8$ atoms in diameter) which is approximately $1/7$ of the size 
of the system.
\figref{fig:LJVIMaxbestclustersTHigh} shows some corresponding high temperature 
$T=5$ clusters which we note are approximately twice the size ($n\simeq 600$ nodes)
and more dispersed.

As an additional note, we remark that, intuitively, one would expect that the VI maxima are 
directly related to the NMI minimum in this region of $\gamma$.
This is not the case in this region because the VI metric does to zero 
as the network collapses to one cluster.  
It is rapidly shifting from a region of maximum complexity at a peak VI value,
where most of the information comes from the sum of the Shannon entropies $H(A)$
and $H(B)$, for two partitions $A$ and $B$, to minimum complexity at a value of $V=0$.
NMI has a different behavior here, as the ratio $I_N =2I(A,B)/\left[H(A) + H(B)\right]$,
because the mutual information between different replicas becomes smaller and approaches 
zero as the system collapses into larger communities.
The low temperature LJ system in \figref{fig:LJTLow} shows a transitional very low 
peak in NMI which corresponds to a near bisection of the system.
However, in this case the overlapping cluster configurations collapse to the entire 
system.

\section{Finite size effects in the LJ system} 
\label{sec:LJfinitesizeeffect}

In \figref{fig:LJfinitesize}, we show the VI correlations as a function 
of the Potts model weight $\gamma$ for several system sizes of $N=2000$, 
$N=4000$, and $N=8000$ in panels (a), (b), and (c), respectively.
We also used $s=12$ replicas and $t=10$ trials for all data.
The value of $\gamma$ corresponding to the peak in $V$ scales downward
with the system size (larger structures) indicating that the maximum 
fluctuations are scaling with the size of the simulation box.  
In the limit of an infinite size simulation box, the VI peak would tend
to $\gamma=0$.

\section{Static Multi-resolution analysis on ternary model glass} \label{sec:StaticAlYFemodel}

The inter-replica NMI/VI correlations change with the time separation interval between the 
replicas where the longer time separations intuitively result in poorer
correlations.
The other limiting case is for a \emph{static} analysis of the system. 
Thus, we are interested in determining what our multiresolution analysis
finds when we examine a single-time snapshot of the system.
A single snapshot may, potentially, capture a transient feature of the system. 
This is illustrated in panel (a) of Fig. \ref{fig:MRAreplicaspic} in which 
the time separation between different replicas is set to zero. 
We applied the same algorithm as in \secref{sec:MRAandCD} except that
all replicas correspond to the same system time, and we solved the systems 
with $s=8$ replicas and $t=4$ trials per replica.
In \figsref{fig:AlYFeTLowStatic}{fig:AlYFeTHighStatic}, our analysis 
identifies a \emph{static length} associated with the instantaneous 
configuration.
At both high and low temperatures, the static multi-resolution analysis display information 
extrema at low $\gamma$ (large spatial scales). The higher temperature system displays 
weaker correlations with sparser structures whereas the structures of the low temperature
system display more significant correlations (and more compact structures).
As seen in  \figsref{fig:AlYFeTLowStatic}{fig:AlYFeTHighStatic} the high temperature system displays weaker
correlations in the vicinity of the peak NMI than those of the low temperature system. That is, the high temperature solutions require more 
effort to solve accurately (there are far more metastable minima and more trials are required in order
to find better contending solutions). (A similar occurrence was found for the high temperature LJ system in Fig. \ref{fig:LJTHighbestclusters}.)
The solved clusters (not depicted) are comparable to those identified
for the time-separated replicas in \secref{sec:application} for both system 
temperatures.

\section{Static Multi-resolution analysis on LJ glass} \label{sec:StaticLJmodel}

As in Appendix \ref{sec:StaticAlYFemodel} for the ternary system, we further 
tested a static version of the binary LJ model where the results are similar 
to the dynamic case in \secref{sec:LJresults} both in the MRA plot and the 
resulting clusters.
A notable exception is that the distinction between the high and low temperature
cases becomes more subtle in the multiresolution plots.

We applied the same algorithm as in \secref{sec:MRAandCD} except that
all replicas correspond to the same system time, and we solved the systems 
with $s=12$ replicas and $t=10$ trials per replica.
In \figsref{fig:LJTLowStatic}{fig:LJTHighStatic}, we show our multiresolution 
analysis associated with the instantaneous configurations.
These two static plots are similar, but the high temperature case has slightly 
weaker correlations (lower NMI peak and higher VI minimum).
In \figsref{fig:LJTLowStatic}{fig:LJTHighStatic} where we use time separated 
replicas, the contrast in the MRA plots at different temperatures is stronger.
\figref{fig:LJTLowStaticbestclusters} shows a sample of the best clusters
in the low temperature case which are comparable to those identified
for the time-separated replicas in \figref{fig:LJbestclusters} 
in \secref{sec:LJresults}.

\section{An analysis of yet another ternary system} \label{sec:altAlYFemodel}

In \figsref{fig:AlYFeTLowalt}{fig:AlYFeTHighalt}, 
we repeat the analysis of \secsref{sec:ternarysimulation}{sec:modelresults}, 
for an alternate ternary model glass system (AlYFe).
In particular, we use the best parameter fit for the same-species interactions 
that are tabulated in Table \ref{tab:ppfitNew} as opposed to implementing the 
GPT \cite{ref:moriartyGPT} as used 
in \figdref{fig:AlYFeTLow}{fig:AlYFebestclusters} and tabulated in Table \ref{tab:ppfit}.
A plot of the alternate potentials is shown in \figref{fig:ppfitNew}.
These interactions conform to those 

The lower temperature system at $T=300$ K in \subfigref{fig:AlYFeTLowalt}{a} 
shows a peak NMI at ($i$a) with a corresponding VI minimum at ($i$b).
Following \secref{sec:modelresults}, \figref{fig:AlYFebestclustersalt} depicts 
a sample of the best clusters at $\gamma_{best}\simeq 0.001$ where we include 
overlapping node memberships (the replicas correlations are calculated on partitions).
We used $s=12$ replicas and $t=20$ trials per replica.
The corresponding $T=1500$ K high temperature solutions have a much lower 
NMI at $\gamma_{best}\simeq 0.001$ indicating very poor agreement among 
the replicas.
At $T=300$ K, the best structures have consistent cluster sizes that 
are MRO or a little larger which are generally larger than $1/2$ the 
simulated system width.
Therefore, it would beneficial to test the consistency of the clusters
in a larger system requiring a substantially longer computational time.

In this system the NMI plateau at $\gamma\gtrsim 100$ is actually higher 
than the configuration at ($i$), but the clusters are almost exclusively 
small ($n\simeq 5$ nodes) and are not completely contiguous.
The distinction in the results between the different potential models
is likely due to the longer range Al-Al minimum in the ab initio potential 
parameter fit data as compared to the GPT \cite{ref:moriartyGPT} minimum.

\myfig{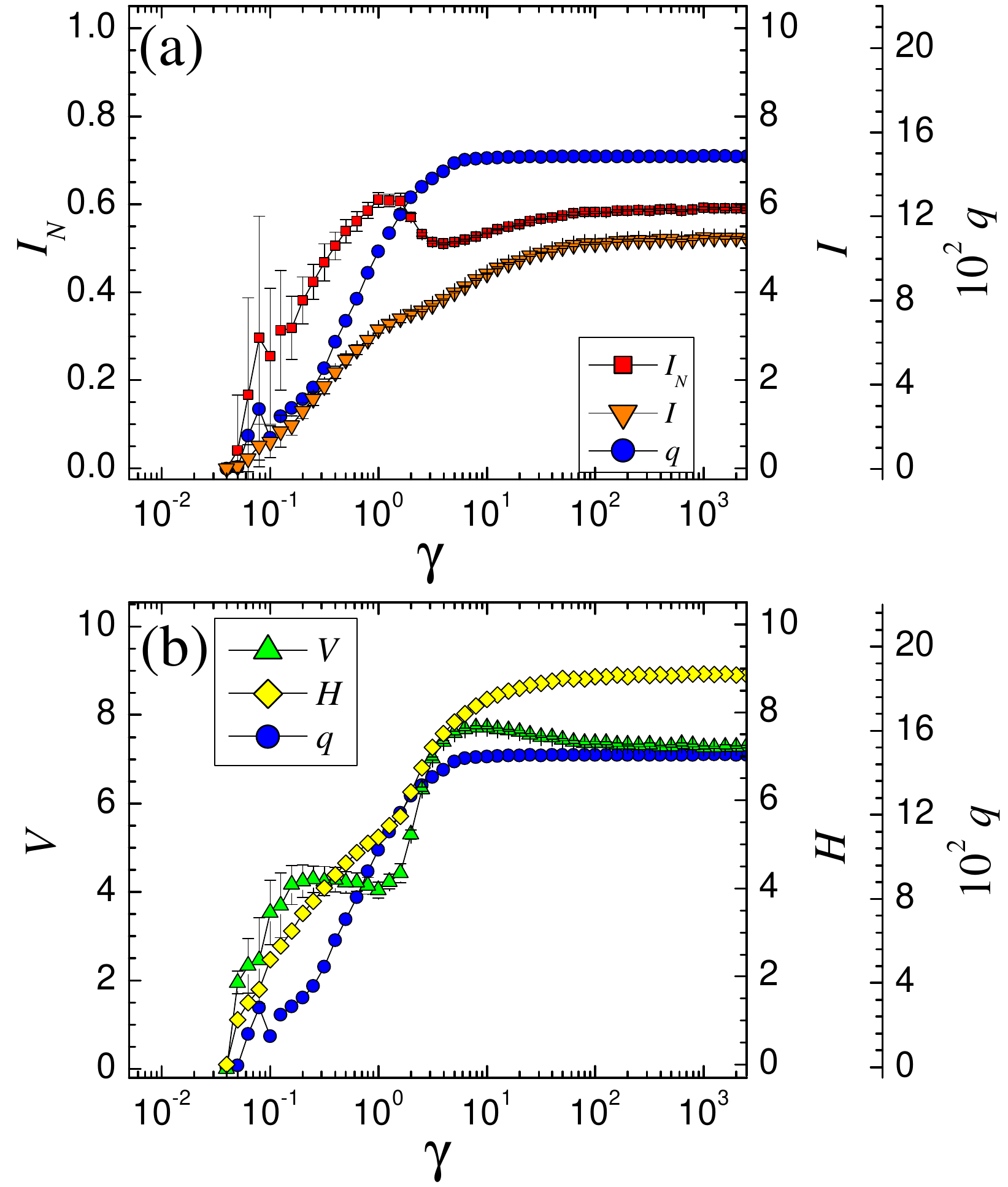}{A plot for the multiresolution analysis 
of a binary Zr$_{80}$Pt$_{20}$ system at $250$ K below the liquidus.
Atomic configurations were generated using conventional RMC methods that 
are consistent with the experimentally determined scattering data for liquid 
Zr$_{80}$Pt$_{20}$ at $250$ K below the liquidus.
The plot shows a poorly correlated, but nevertheless well-defined, peak in NMI 
near $\gamma\simeq 1$.}
{fig:ZrPtgijr}{\figsize}{h} %{b!}

%\myfig{ZrPt_gijr_N7500_Clusters.jpg}{A sample of clusters determined by a multiresolution
\myfig{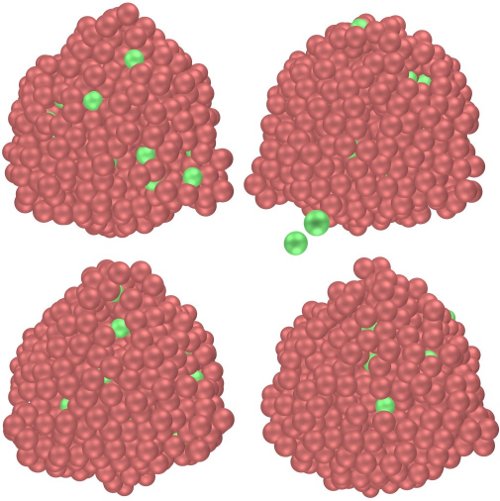}{A sample of clusters determined by a multiresolution
clustering analysis based on the pair correlation function in \figref{fig:ZrPtgijr}.
The system is a Zr$_{80}$Pt$_{20}$ system at $250$ K below the liquidus (see text)
where Zr atoms are depicted as red, and Pt atoms are depicted as green.
Each cluster has approximately $700-800$ atoms.}
{fig:ZrPtgijrclusters}{\figsize}{h} %{t!}

\section{Multi-resolution analysis via measured pair correlation function} \label{sec:gijrapplication}

As stated in \secref{sec:background}, we may apply the same multiresolution network 
clustering ideas for other structure models.
In particular, in \figref{fig:ZrPtgijr}, we apply the method to experimentally adduced pair correlation functions
$g_{ab}(r)$ in an amorphous ZrPt system between the different of the different components $a$ and $b$
(i.e., Zr-Zr, Zr-Pt, and Pt-Pt). That is, we set in Eq. (\ref{eq:ourPottsmodel}),  $A_{ij} = -(g_{ij} +\overline{g}_{ij})$ if
$(g_{ij} +  \overline{g}_{ij})<0$
and $B_{ij} = (g_{ij} - \overline{g}_{ij})$ if $(g_{ij} - \overline{g}_{ij})>0$. Here, 
$g_{ij}$ denotes the pair correlation between the atoms corresponding to nodes $i$ and $j$,
$\overline{g}_{ij}$ is a background average (which we set to zero). 
Some sample clusters then found by our method are seen in \figref{fig:ZrPtgijrclusters}.
Due to the large system size, we used $s=8$ replicas and $t=4$ trials per replica.

Amorphous systems with extended atomic order beyond the nearest neighbor shell 
provide an excellent framework to test the identification of natural structural 
elements.
The ZrPt is a system that has been shown to have MRO in both the glassy and liquid 
state 
\cite{ref:nakamuraZrPt,ref:saidaZrPt,ref:sordeletZrPt,ref:wangwangZrPt,ref:maurokeltonZrPt}.
MRO, or correlations beyond direct chemical bonding manifest as pre-peaks in the 
static structure factor extracted from scattering studies of liquids and glasses.  
The dominant structural elements at the heart of such an ordering are of extreme 
interest.
Atomic configurations that are consistent with the experimentally determined 
scattering data for liquid Zr$_{80}$Pt$_{20}$ at $250$ K below the liquidus 
were generated using conventional Reverse Monte Carlo (RMC) methods 
\cite{ref:mcgreevyULS,ref:keenRMC,ref:kimkelton}.
The result of analysis uses unconstrained RMC which may or may not emulate the 
precise microscopic structure.
We then analyzed a representative system with $N=7500$ nodes using the algorithm 
outlined in \secref{sec:MRA}.

Specifically, we analyze the system as a \emph{static} model.
That is, each replica is based on the one particular system representation since 
the data is obtained by RMC methods rather than by a dynamical simulation.
The multi-resolution analysis (MRA) is seen in \figref{fig:ZrPtgijr}.  
A few of the relatively uniform large clusters are shown in \figref{fig:ZrPtgijrclusters}
where each cluster is approximately $700-800$ atoms in size.
Interestingly, the ZrPt system shows a well-defined secondary MRA NMI peak near 
$\gamma\simeq 1$.  
Although the peak NMI value has a poor overall correlation among the replica
solutions (low $I_N$), it is notable that the system displays this secondary 
peak since it is entirely absent in the LJ binary liquid results 
in \secref{sec:LJresults}.

\section{Multiresolution application to lattice systems} 
\label{sec:lattices}

We define several uniform lattices systems for the purpose of comparing 
the results to the model glasses where %. %, initially unweighted, 
we use relatively small systems for presentation purposes.
We would normally just treat respective model networks as unweighted networks, 
but here we wish to maintain a consistent analysis across all systems 
in the paper, so we further apply a ``potential'' shift $\phi_0$ which 
corresponds to the negative of the average weight over all pairs of nodes 
(any non-neighbor has a weight of $B_{ij}=1$).

We allow zero energy moves for the lattice solutions and perform a more
strenuous optimization using $s=20$ replicas and $t=40$ trials. 
These representative networks result in ``imperfect'' tilings of the favored 
local structures due to the constraints imposed by the perfect symmetry 
in the Hamiltonian and by the local solution dynamics.
In the depictions, different colors represent distinct clusters (best 
viewed in color) and edges \emph{between} clusters are made partially 
transparent as a visualization aid.
No overlapping nodes are assigned in the lattice depictions.

\subsection{Square lattice} \label{sec:squarelattice}

We define a uniform, initially unweighted, square lattice with $N=400$ 
nodes. 
Edges are assigned to each neighbor in the $x$ and $y$ directions 
with periodic boundary conditions.
The ``potential'' shift is $\phi_0=0.97995$.
We then perform the same multiresolution analysis to the graph as 
in the previous systems.
In \figref{fig:squareLatticePlot}, we see that where are three dominant 
plateaus in the information measures.

Except for the plateaus, the overall multiresolution pattern resembles 
the LJ system in \figref{sec:LJresults}, except for the presence of the 
plateaus in the lattice plot, which suggests that the LJ system is 
``more ordered'' than the metallic glass model.
However, a purely random graph \cite{ref:rzmultires} also shows a similar 
pattern, including a plateau, which indicates that there may exist an analogy 
between purely ``random'' and perfectly ``ordered'' systems in our analysis.
However, these data are not alone sufficient to be conclusive.

We select a configuration at $\gamma=60$ corresponding to the center plateau. 
The lattice is depicted in \figref{fig:squareLatticePicture}
with $q=120$ clusters of $78$ squares, $4$ triads, and $38$ dyads.
At this resolution, the square dominates the configuration which shows
that our algorithm is able to isolate the basic unit cells of the lattice
in a natural fashion.

The plateau for $\gamma\gtrsim 100$ corresponds to essentially all 
dyads, and the plateau for $\gamma\lesssim 30$ corresponds to a mixture 
of dyads, squares, and tight six-node configurations (a square plus two 
adjacent nodes).
The lower $\gamma$ plateau favors the six-node configuration in terms
of the cluster energies, but the larger features are even more difficult 
with which to tile the lattice than for squares.

\myfig{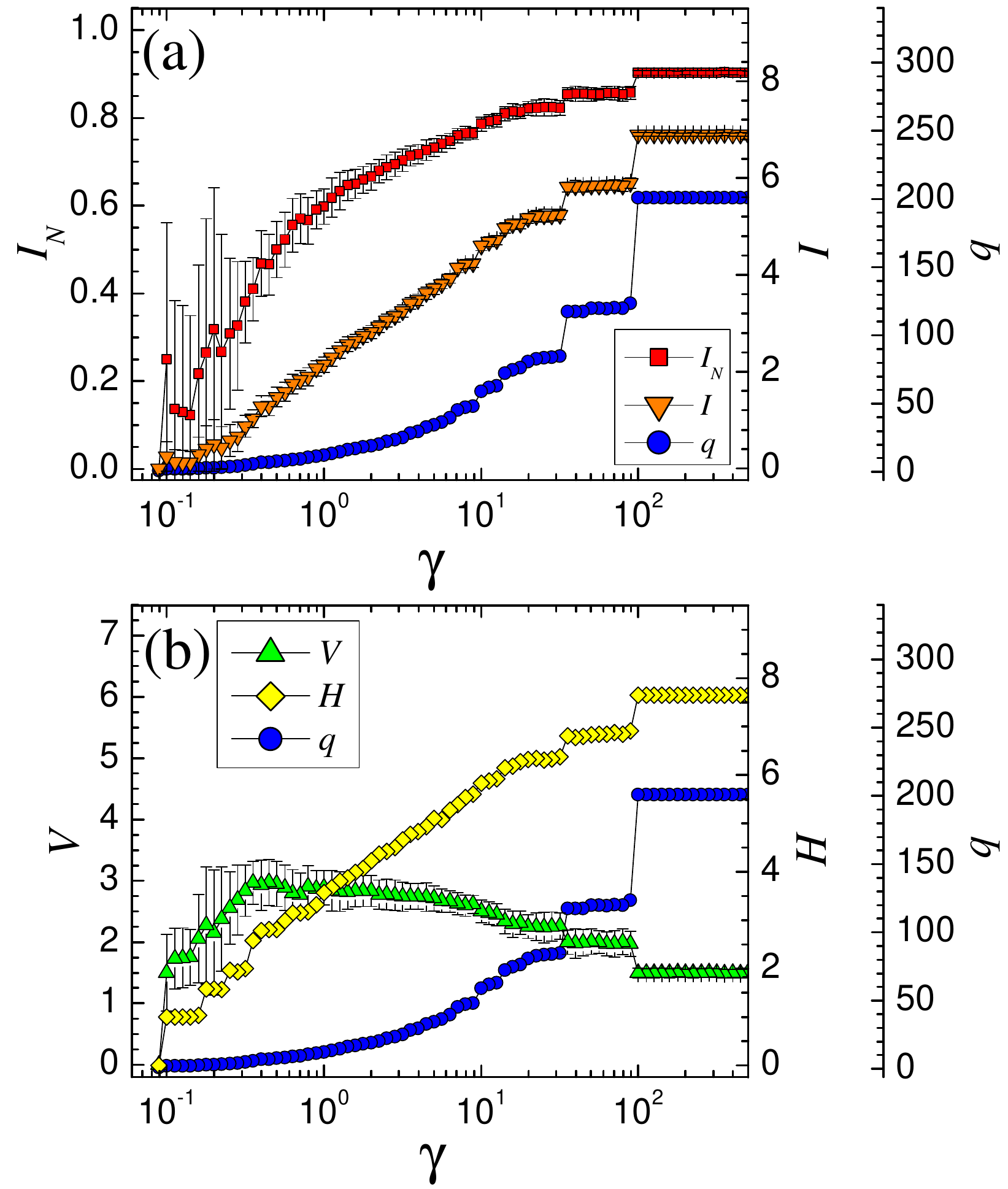}{The multiresolution analysis 
of a square lattice with periodic boundary conditions.
Neighbors have an initial weight of $A_{ij}=1$ and non-neighbors 
have an initial weight of $B_{ij}=1$.
We further apply a ``potential'' shift of $\phi_0=0.97995$ in order 
to be consistent with our previous analysis on glasses.
There are three distinct plateaus in the information measures.
A depiction of the system at $\gamma = 60$ for the center plateau 
is shown in \figref{fig:squareLatticePicture}.}
{fig:squareLatticePlot}{\figsize}{h} %{b!}

\myfig{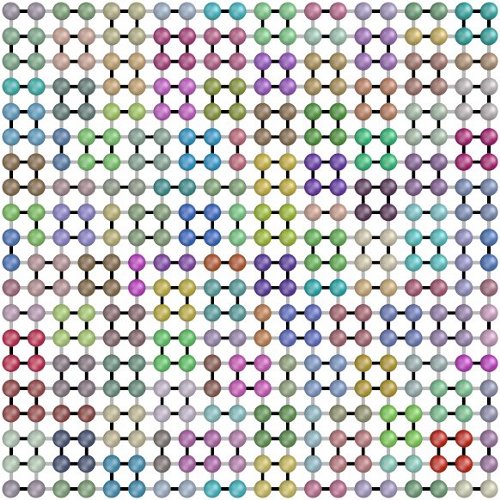}{A partition of a square 
lattice with periodic boundary conditions.
The corresponding multiresolution plot is seen in \figref{fig:squareLatticePlot}.
We use the algorithm described in the paper at $\gamma = 60$ 
to solve the system.
To aid in visualization of the clusters, neighbor links 
\emph{not in the same cluster} are made partially transparent.
In this configuration, there were $q=120$ clusters with $78$ 
squares, $4$ triads, and $38$ dyads which indicates that square 
configuration dominates the partition, and it shows how our algorithm
can naturally identify the basic unit cells of the square lattice.}
{fig:squareLatticePicture}{\figsize}{h} %{t!}
%\clearpage

\subsection{Triangular lattice} \label{sec:trilattice}

Similar to the square lattice we define a uniform, initially unweighted, 
triangular lattice with $N=400$ nodes.
Edges are assigned to each triangular neighbor using periodic boundary conditions.
The ``potential'' shift is $\phi_0=0.969925$.
We again perform the same multiresolution analysis to the graph with the
results shown in \figref{fig:triLatticePlot}.
There are three dominant plateaus in the information measures; and the 
overall multiresolution pattern resembles both the square lattice and 
the LJ systems except for the presence of the plateaus here.

We select a configuration at $\gamma=50$ corresponding to the ``left'' plateau.
The lattice is depicted in \figref{fig:triLatticePicture}
with $q=104$ clusters of $17$ triads, and $86$ ``diamonds'', and $1$ five-node
cluster.
The five node cluster is a result of a preference being given for an isolated 
node to join a diamond as opposed to forming its own single-node cluster.
At this resolution, the diamond configuration dominates.

Two plateaus to the right of $\gamma\gtrsim 65$ are both strongly dominated
by triads of nodes.
The distinction between the two is that the central plateau favors an
isolated node joining a triangle, to form a rare diamond, rather than 
forming its own single-node cluster.
Together, the different plateaus show that our algorithm is able to isolate 
the basic unit cells of the lattice in a natural fashion.

\myfig{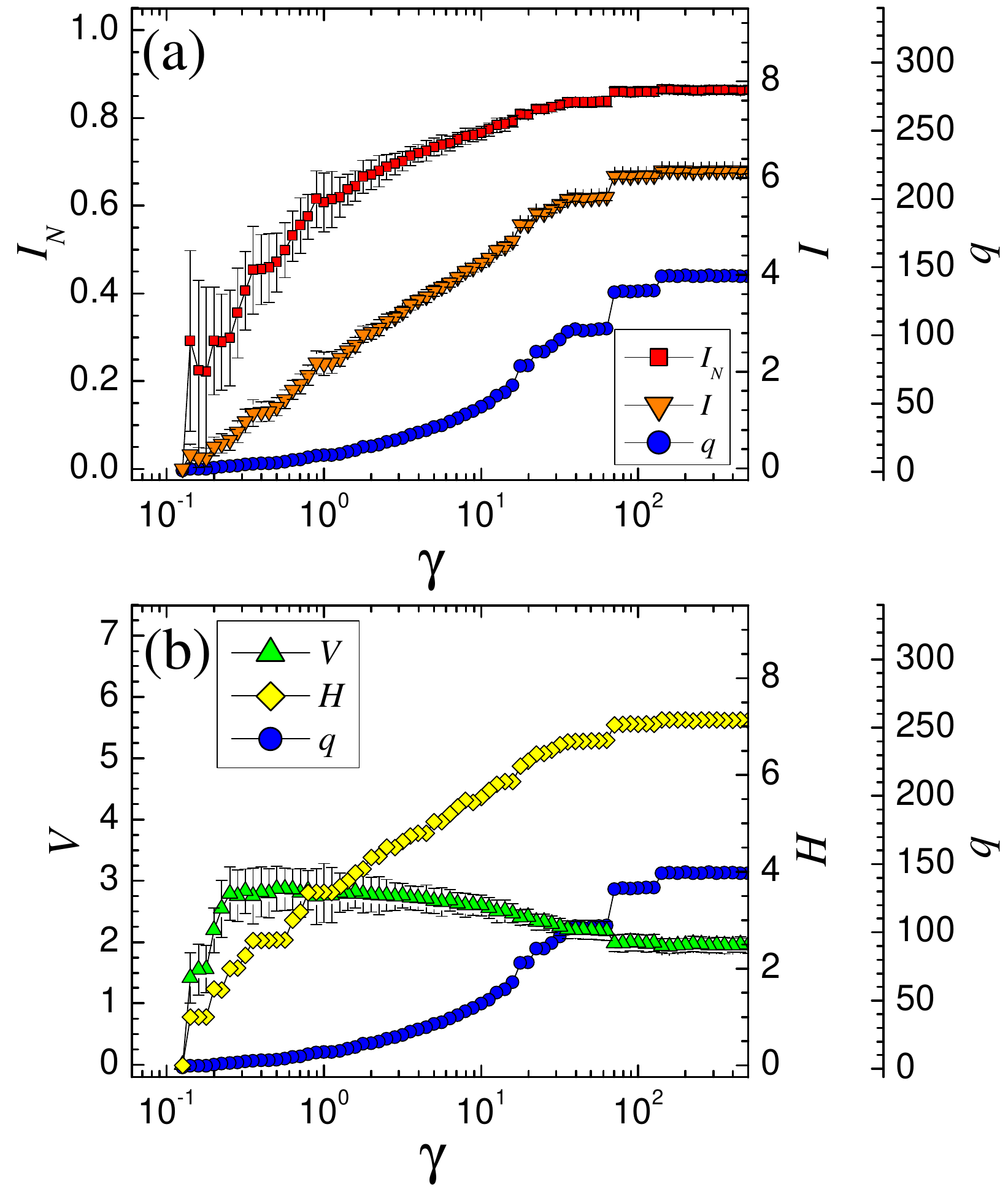}{The multiresolution analysis 
of a triangular lattice with periodic boundary conditions.
Neighbors have an initial weight of $A_{ij}=1$ and non-neighbors 
have an initial weight of $B_{ij}=1$.
We further apply a ``potential'' shift of $\phi_0=0.969925$ in order 
to be consistent with our previous analysis on glasses.
There are three distinct plateaus in the information measures, 
but the latter two are closely related (see text).
A depiction of the system at $\gamma = 50$ for the left plateau 
is shown in \figref{fig:triLatticePicture}.}
{fig:triLatticePlot}{\figsize}{h} %{b}

\myfig{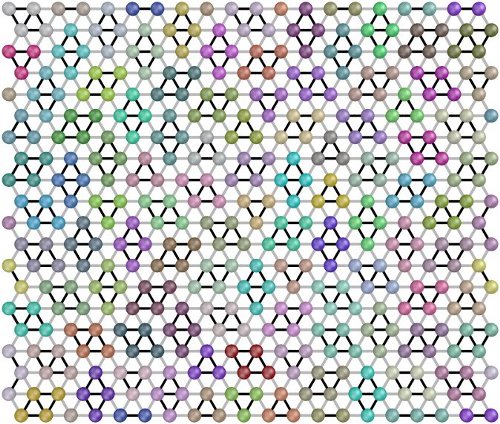}{A partition of a triangular 
lattice with periodic boundary conditions.
The corresponding multiresolution plot is seen in \figref{fig:triLatticePlot}.
We use the algorithm described in the paper at $\gamma = 50$
(the ``left'' plateau) to solve the system.
To aid in visualization of the clusters, neighbor links 
\emph{not in the same cluster} are made partially transparent.
In this configuration, there were $q=104$ clusters with $17$ triads, 
$86$ ``diamonds,'' and $1$ five-node configuration (see text) which 
indicates that the diamond configuration dominates the partition.
Our algorithm can naturally identify the different basic unit 
cells of the lattice. By varying $\phi_{0}$ (and the corresponding $\gamma$)
we detect natural structures on varying scales.}
{fig:triLatticePicture}{\figsize}{h} %{t}
%\clearpage

\subsection{Cubic lattice} \label{sec:cubiclattice}

We further define a uniform, initially unweighted, cubic lattice
with $N=1000$ nodes.
Edges are assigned to each neighbor in the $x$, $y$, and $z$ directions 
using periodic boundary conditions.
The ``potential'' shift is $\phi_0 = 0.987~988$.
We perform the same multiresolution analysis to the graph as in the 
previous systems where the results are summarized in 
\figsref{fig:cubicLatticePlot}{fig:cubicLatticePicture}.

Out of $q=177$ clusters, we identified $66$ squares, $51$ six-node 
configurations (square plus two adjacent nodes), $45$ cubes, 
and $15$ other assorted configurations smaller than cubes.
At $\gamma\simeq 50$, the cube is the preferred cluster in terms 
of the cluster energy, but they consist of only slightly more than 
$25\%$ of the clusters because the large cube configuration is more 
difficult to identify due to the perfect network symmetry and local
constraints imposed by the evolution of the community structure during 
the algorithm dynamics.
The ``middle'' plateau represents a square-dominated region ($201$ out 
of $294$ clusters are squares), and the ``right'' plateau consists 
of dyads of neighbor nodes almost exclusively.

% cubic figures ----------------------------------------------------
\myfig{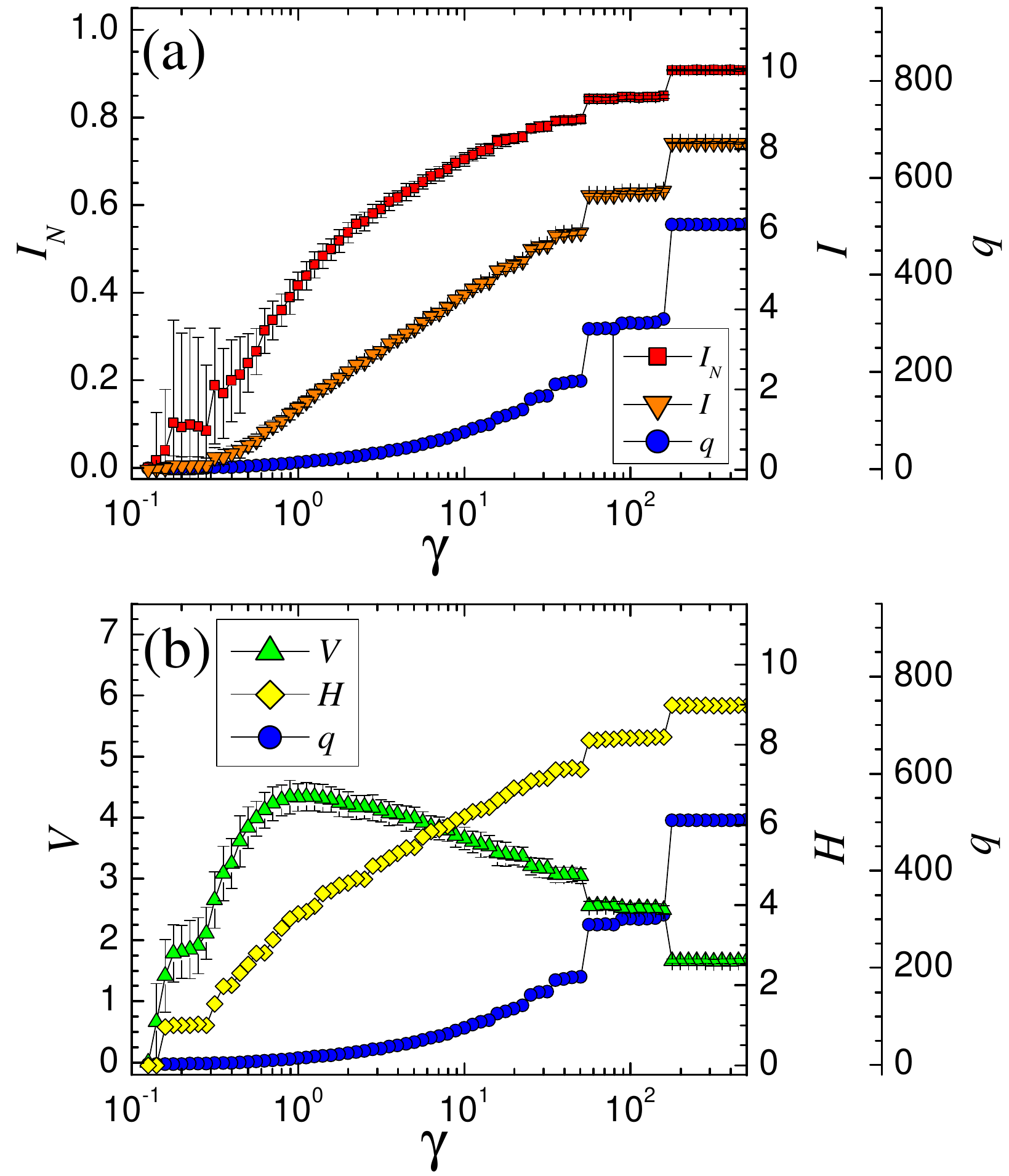}{The multiresolution analysis 
of a cubic lattice with periodic boundary conditions.
Neighbors have an initial weight of $A_{ij}=1$ and non-neighbors 
have an initial weight of $B_{ij}=1$.
We further apply a ``potential'' shift of $\phi_0 = 0.987~\!988$ in order 
to be consistent with our previous analysis on glasses.
There are three distinct plateaus in the information measures.
The leftmost short plateau is the cube preferred resolution (in terms 
of cluster energy) with $45$ out of $177$ clusters are cubic 
clusters (no configurations larger than cubes are found).
A depiction of the system at $\gamma = 50$ for the left plateau is 
shown in \figref{fig:cubicLatticePicture}.}
{fig:cubicLatticePlot}{\figsize}{h} %{b}

\myfig{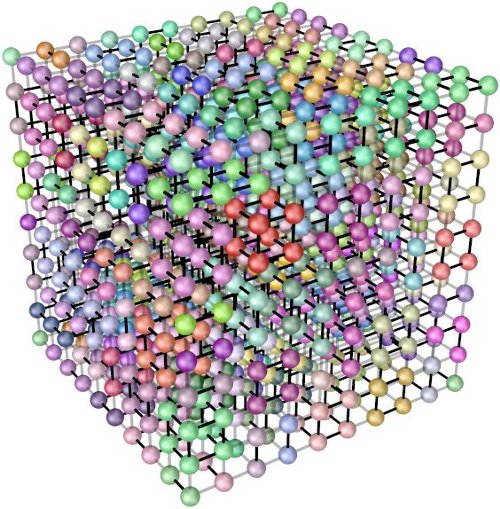}{A partition of a cubic 
lattice with periodic boundary conditions.
The corresponding multiresolution plot is seen in \figref{fig:cubicLatticePlot}.
We use the algorithm described in the paper at $\gamma = 50$
(the short ``left'' plateau) to solve the system.
To aid in visualization of the clusters, neighbor links 
\emph{not in the same cluster} are made partially transparent.
In this configuration, there were $q=177$ clusters with $66$ squares, 
$51$ six-node configurations, $45$ cubes, and $15$ other assorted 
configurations.
The cubic cluster is the preferred partition (in terms of cluster energy), 
but it is difficult to fill the system with cubes in practice due to the 
perfect symmetry of the network.}
{fig:cubicLatticePicture}{\figsize}{h} %{t}
% end cubic figures ------------------------------------------------

\section{Multiresolution application to a 2D Ising lattice} 
\label{sec:Ising}

We define two-dimensional square lattice of Ising spins using the Hamiltonian
\begin{equation}
  H=-\sum_{ij}\sigma_i\sigma_j
\end{equation}
for all pairs of neighbor spins $i$ and $j$.
Neighbors with the same spin have an initial weight of $A_{ij}=1$, 
neighbors with opposite spins have an initial weight of $B_{ij}=1$, 
and non-neighbors have an initial weight of $B_{ij}=0$.
There is an unavoidable inherent discontinuity in defining the 
interactions for the Ising system (only \emph{neighbors} can interact) 
and the continuous systems elsewhere in this paper (\emph{all} pairs 
of nodes interact).
With this caveat, we analyze this Ising system according to the same 
multiresolution analysis, and we then compare the results to the other 
tested systems.
We apply $s=8$ replicas and $t=8$ trials.

We could effectively treat the above Ising interactions using an 
\emph{unweighted} network model since we can factor our the weight, 
but we again wish to maintain a consistent analysis across all systems 
in this paper.
With this in mind, we further apply the average ``potential'' shift $\phi_0$
which varies on each defined network and which corresponds to the negative 
of the average weight over \emph{all} pairs of nodes (including non-neighbors).
We allow zero energy moves for the solution dynamics.
No overlapping nodes are assigned in the lattice.

In \figref{fig:IsingPlot}, there are two main plateaus for $\gamma\gtrsim 100$
and a noticeable configuration ``shift'' at $\gamma\simeq 100$.
The ``right'' plateau for $\gamma\gtrsim 500$ consists largely of same-spin
dyads except where a given spin has no matching neighbors.
The ``middle'' plateau for $100\lesssim\gamma\lesssim 500$ corresponds 
to a ``natural'' grouping of medium size clusters.
In \figref{fig:IsingPicture}, different colors represent distinct clusters 
(best viewed in color) and edges \emph{between} clusters (neighbor spins 
with the \emph{same sign} but that are in \emph{different} clusters) are 
depicted in gray.  Missing edges are not shown

Here, the plateaus correspond to a cascade of structures starting 
from the smallest dyads of nodes, to basic plaquette structures 
(square, triangle, etc.), and growing ever larger (two joined plaquettes etc.).
In Ising spin systems at different temperatures on a square lattice,
the domains of ``$+$'' and ``$-$'' spins are physically separated from one 
another by domain walls.
The plateaus here similarly correspond to the cascade of small plaquettes 
found on the lattice up to a cutoff scale set by the domain walls.

The domain walls are closely related to the \emph{maximum} in VI
(NMI displays a different, non-extremal, behavior where the standard 
deviation is large).
Physically, they correspond to the scales at which the largest fluctuations 
occur where the large fluctuations result in poor information correlations 
between the replicas.
\figref{fig:IsingPictureCL} shows a sample depiction of the system 
at this VI peak.
Correlation lengths are thus likely related to poorly correlated replicas 
best indicated by a VI \emph{maximum}.

\myfig{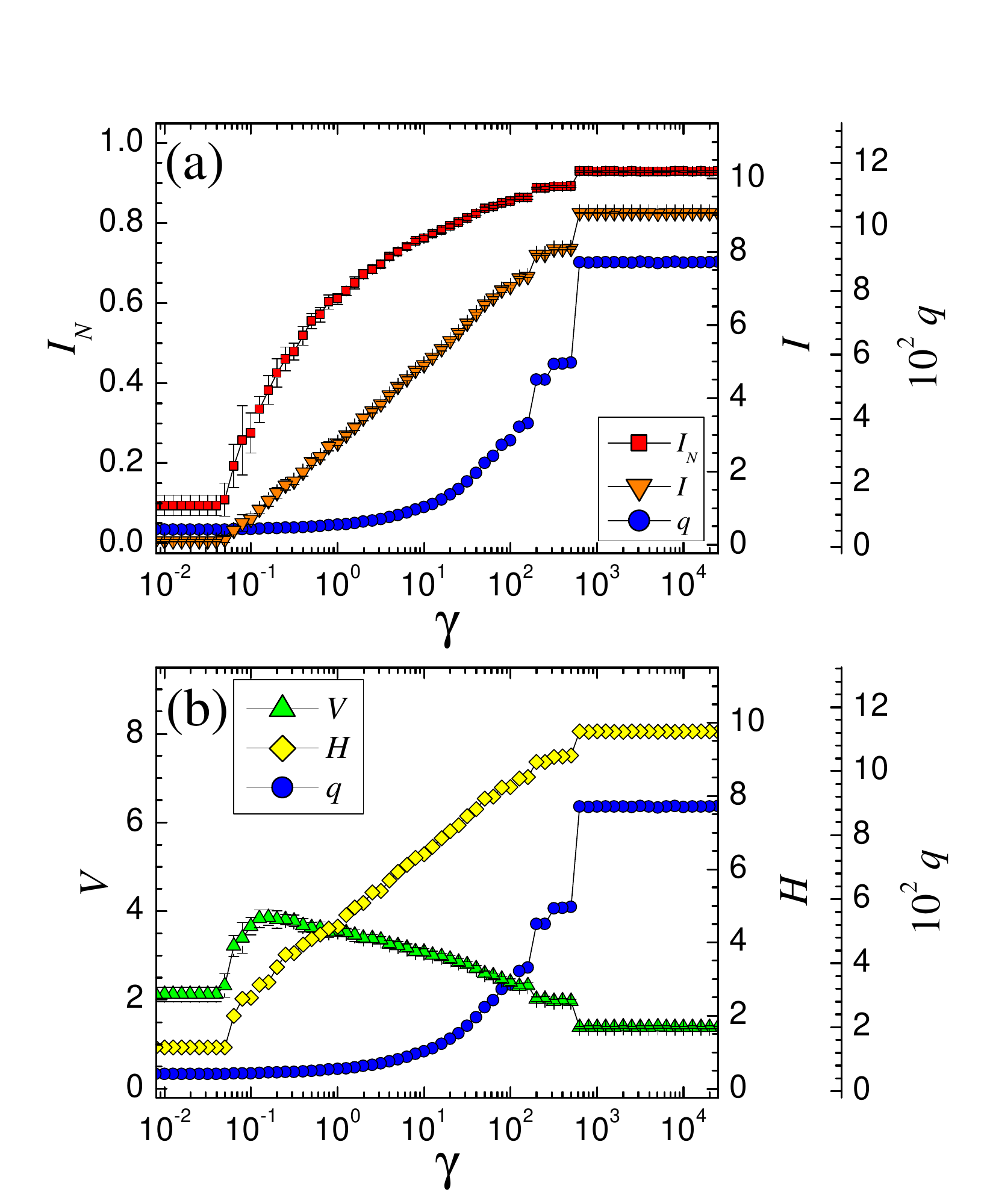}{A plot for the multiresolution analysis 
of a square lattice of Ising spins with periodic boundary conditions at a 
simulation temperature of $T=2.269$ (close to the transition temperature).
Neighbors with the same spin have an initial weight of $A_{ij}=1$, 
neighbors with opposite spins have an initial weight of $B_{ij}=1$, 
and non-neighbors have an initial weight of $B_{ij}=0$.
There are two distinct plateaus in the information measures for
$\gamma\gtrsim 100$ with a significant configuration shift near $\gamma\simeq 100$.
A depiction of the system at $\gamma = 400$ for the center plateau 
is shown in \figref{fig:IsingPicture}.}
{fig:IsingPlot}{\figsize}{h} %{b!}

\myfig{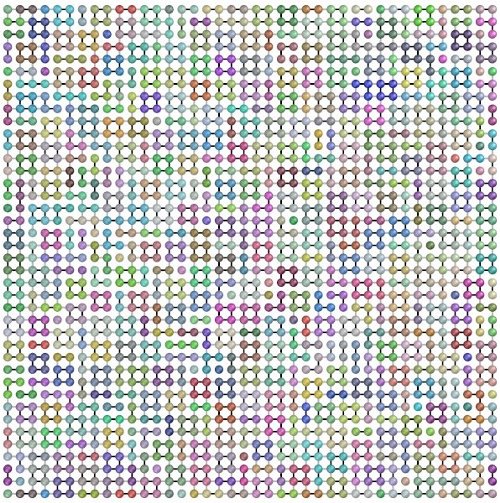}{A depiction of a partition of a square 
lattice of Ising spins with periodic boundary conditions at a simulation 
temperature of $T=2.269$ (proximate to the critical temperature).
The corresponding multiresolution plot is seen in \figref{fig:IsingPlot}.
We use the algorithm described in the paper at $\gamma = 400$ (corresponding
to the transition to the plateau with the highest NMI / lowest VI value)
  to solve the 
system.
For presentation purposes, we depict all spins by spheres and links that 
cross a community boundary as grey edges. As with the multi-resolution 
analysis of the square lattice, this multiresolution analysis 
shows that the dominant communities are square 
plaquettes within same-spin domains.}
{fig:IsingPicture}{\figsizelarge}{h} %{t!}

\myfig{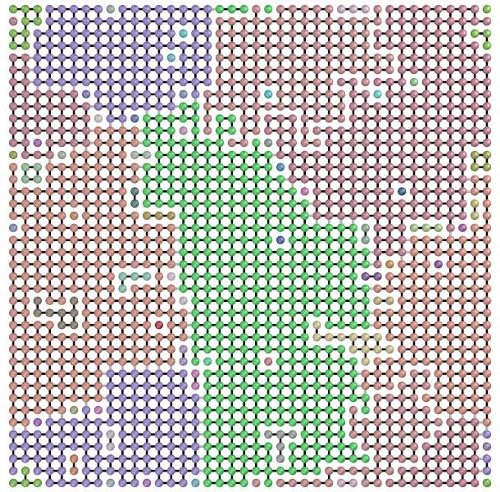}{A partition of a 
square lattice of Ising spins with periodic boundary conditions at a 
simulation temperature of $T=2.269$.
The corresponding multiresolution plot is seen in \figref{fig:IsingPlot}.
We use the algorithm described in the paper at $\gamma = 3.60\times 10^{-4}$,
corresponding to the \emph{peak} VI, to solve the system.
For presentation purposes, we depict all spins by spheres and links that 
cross a community boundary as grey edges.
The identified clusters vary substantially across the replicas (maximum
fluctuations), but they are of the scale of the size of the spin domains.
This qualitatively implies that the poor information correlations
are naturally correlated with the individual domains of ``+'' and ``-'' spins.
At this scale,  the fluctuations are largest (and hence the related 
weakest information theory correlations). This scale on which fluctuations are maximal
(coinciding with the Ising domain walls) should be contrasted with that of Fig. \ref{fig:IsingPicture}
for which the fluctuations were minimal (lowest variation of information (VI)) and the normalized
mutual information (NMI) was the highest.}
{fig:IsingPictureCL}{\figsizelarge}{h} %{b!}

\section{Multiresolution analysis of Lennard-Jones systems with elastic defects}
\label{sec:LJlattices}

We define a uniform lattice based on a two-dimensional Lennard-Jones
system using the LJ potential in \eqnref{eq:LJVr} where the lattice
is constructed as the ideal ground state of the system using periodic
boundary conditions.
The main purpose is to check for consistency of our multiresolution method, 
but the lattice also allows us to perform a preliminary investigation 
of the behavior of our analysis in how it places lattice defects that are
randomly inserted into the system.
We used $s=8$ replicas and $t=8$ trials.

The ground state of the LJ system is given by a triangular lattice with
a lattice constant that is equal to the distance that minimizes
the LJ interaction between two atoms. We define a uniform triangular lattice with $N=2943$ nodes where we 
create random ``defects'' by removing some nodes.
Edges are again assigned and weighted according to the LJ potential 
in \eqnref{eq:LJVr} applying periodic boundary conditions.
(The ``potential'' shift is,as before, set to $\phi_0=0.969925$.)
We perform the multiresolution analysis on the graph where the
results are shown in \figref{fig:LJLatticeVoidsPlot}.

Next, we select a configuration at $\gamma\simeq 31.6$ from the peak 
corresponding to the largest natural communities for this
system at this specific lattice spacing.
An example of the best lattice partition is depicted 
in \figref{fig:LJLatticeVoidsPicture}.
Defects appear to occur most likely \emph{near the boundary of neighboring 
communities}.
%\clearpage

\section{Multiresolution analysis in space-time} \label{sec:actionmodel}

In addition to the potential interaction energies, we may model 
our multiresolution community detection analysis using the mechanical action
\begin{equation}
\label{S_elastic}
S = \frac{1}{2} \int dt~\Big[\sum_{i=1}^{N} M_{i} 
    (\frac{d {\vec{r}}_{i}}{dt})^{2} - \frac{1}{2} \sum_{i \neq j}
     \phi({\vec{r}}_{i}, {\vec{r}}_{j}) \Big].
\end{equation}
Here, $M_{i}$ is the mass of particle $i$ and $\vec{r}_{i}(t)$ denotes its location as a function
of time. As throughout, $\phi$ is the two particle interaction. 
Below, we refer to the case of a system in $D$ spatial dimensions.  
In a discretized form (allowing, in particular, for discrete time steps $\Delta t$ along the time axis),
the action
\begin{equation}
  S= \sum_{\alpha\beta} W_{\alpha\beta}
  \label{eq:Selastic}
\end{equation} 
with the nodes $\alpha$ and $\beta$ representing nodes in space-time and $W_{\alpha\beta}$ 
an effective ``interaction weight'' between them.
Thus, we represent our physical system as a network embedded in $(D+1)$ dimensional
space-time.
We extremize this action ``S'' corresponding to 
the Hamiltonian of \eqnref{eq:ourPottsmodel}.
The spatial links are determined by $\phi({\vec{r}}_{i}, {\vec{r}}_{j})$ similar to before (apart 
from a sign inversion) while the links parallel to the time axis 
are determined by the particle masses.
As before, extrema of the information theory correlation pinpoint the 
natural structures and scales in the system.
In this case, there is only one tuning parameter $\gamma$ as applied 
to different replicas of complete space-time networks.
Thus, the space and time scales cannot be independently adjusted and are 
intertwined and are related by a ratio which is the typical speed of sound 
$c$ (or an appropriate average of such speeds) in equilibrated solids.

\section{Continuum elasticity about local minima and the shear penetration depth} \label{sec:action}

The focus of our article is on a detailed bona fide description of amorphous materials
that invokes graph theory methods to ascertain general structure. The various examples
shown have hopefully clearly outlined the strengths of our approach. In this section, we depart
from this approach and outline a continuum formalism.

Continuum elasticity in regular solids does not assume detailed knowledge of the underlying 
atomic constituents. In this section, we follow suit and outline general considerations
for amorphous solids. In an ideal crystalline system, any small deformation 
about the crystalline ground state will raise the energy
in a harmonic manner. The increase of energy (for non-plastic deformations)
is captured by the elastic constants of the material. Nothing prevents, in principle, 
the application of these considerations for deformation about
local energy minima -- as these pertain to an amorphous solid 
in a local energy minimum (an {\em inherent} structure
as it would have pertained to supercooled liquids). In what briefly follows, we consider the harmonic expansion of
Eq. (\ref{S_elastic}) about such a local energy minimum state. We then review general considerations about
``shear penetration depth''- the length scale on which the medium reacts to shear.
 Earlier treatments considered the shear penetration depth in homogeneous media \cite{ref:zaanenNM,CNMZ1}.
Here, we consider what may occur when replicating these considerations for deformations about a general non-uniform local
ground state. 

In what follows, we label the displacement of the $i$-th atom, $\vec{u}_i$ about its position
in a given state, $\vec{R}_i$ by
\begin{eqnarray}
\vec{u}_i=\vec{r}_i-\vec{R}_i.
\label{rR}
\end{eqnarray}
In Eq. (\ref{rR}), $\vec{r}_i$ denotes the location 
of the displaced atoms. Henceforth, we consider a coarse grained description in which $\vec{u}$ is a field defined in the continuum. (That is, 
we replace the lattice indices $i$ in Eq.(\ref{rR}) by continuous spatial coordinates $x$.) The displacement vector $\vec{u}$ is a $D$-dimensional
vector in real space. We apply field theoretic ideas introduced by \cite{kleinert}. In particular, we next follow the procedure of \cite{ref:cvetkovicNZ,ref:zaanenNM,CNMZ1}
and write the corresponding action of the system in a manner similar to that of the energy of an anisotropic solid in $(D+1)$ dimensions.  We will  label the time direction
as the $\alpha =0$ direction and denote all spatial Cartesian coordinates by $\alpha = 1,2, ..., D$.
We will use Latin indices $a=1,2, ..., D$ to label the spatial directions alone. 
In the continuum limit, 
expanding the energy to harmonic order about any 
local minima (inherent state), %and denote by $\vec{u}_{i}$ the displacement of atom $i$ about an equilibrium position 
the action reads
 \cite{ref:zaanenNM,ref:cvetkovicNZ},
\begin{equation}
  S = \frac{1}{2} \int d^{D}x \int dt ~\partial^{\alpha} u^a C_{\alpha \beta ab}(x,t)\partial^{\beta} u^b.
      \label{eq:ZNM}
\end{equation}
In Eq. (\ref{eq:ZNM}), repeated indices are to be summed over.
The quantities $C_{\alpha \beta ab}$ derive from effective elastic moduli. In regular 
solids, $C_{\alpha \beta ab}$ are space (and time) independent quantities.
In an amorphous 
medium, however, $C_{\alpha \beta ab}(x,t)$ are generally {\em space (and time) dependent}.
The effective elastic constant $C_{00ab}$ in Eq. (\ref{eq:ZNM}) is set by the mass
density, i.e., $C_{00ab} = \rho(x,t)$. \cite{ref:cvetkovicNZ,ref:zaanenNM,CNMZ1}
When $\alpha$ and $\beta$ both assume values between $1$ and $D$,
the quantities $C_{\alpha \beta ab}$ 
are set by the elastic moduli of the solid (e.g., the shear modulus $\mu$).
The square root of the typical ratio between the usual spatial elastic moduli 
and those along the time axis for the space-time representation of the solid 
in \eqnref{eq:ZNM} is given by a typical speed of sound in the system
$c = {\cal{O}}(\sqrt{\mu/\rho})$. 

One approach to defining defects such as dislocation and disclinations in regular solids
is to consider the net change in the displacement or local orientation associated with
closed trajectories. Defect densities relative to a given inherent state may
be defined in a similar way. More formally, we can dualize (via a Hubbard Stratonovich transformation) 
the quadratic action of Eq.(\ref{eq:ZNM}) and associate the dual fields
with defect current densities in space-time. Replicating the analysis performed for
regular solids in \cite{ref:cvetkovicNZ,ref:zaanenNM,CNMZ1} to amorphous solids
with an invertible space-time dependent elastic moduli $C_{\alpha \beta ab}$, 
we obtain the same result obtained for regular solids.
Namely, dislocation and disclination current densities are given by
\begin{eqnarray}
J_{\alpha \beta}^a=\epsilon_{\alpha \beta \lambda\rho}\partial^{\lambda}\partial^{\rho}u^a\nonumber\\
T^{\alpha\beta}_{\gamma \delta}=\epsilon_{\gamma\delta\lambda\rho}\partial^{\lambda}\partial^{\rho}\omega^{\alpha\beta},
\end{eqnarray}
with the local rotation given by
\begin{eqnarray}
\omega_{\alpha\beta}=\frac{1}{2}\epsilon_{\alpha \beta \lambda a}\partial_{\lambda}u^a.
\end{eqnarray}
The temporal direction ($\alpha =0$) components of these
tensors are the usual (dislocation and disclination) defect densities of elasticity,
\begin{eqnarray}
\alpha_i^a=\epsilon_{ijk}\partial^j\partial^ku^a\nonumber\\
\Theta_i^a=\frac{1}{2}\epsilon_{ijk}\epsilon^{abc}\partial^j\partial^k\partial_bu^c.
\label{def_den}
\end{eqnarray}
The condition of a divergence-less stress tensor ($\sigma$) in the absence
of an applied force, enables us to write it as a curl of
a gauge field,
\begin{eqnarray} 
\sigma_\alpha^a=\epsilon_{\alpha \beta \lambda}\partial^{\beta}B^{\lambda a}.
\end{eqnarray}
Elastic shear is mediated by the gauge field $B$. 
When a plastic component of the displacement field $\vec{u}$ exists, it couples minimally to
the stress-gauge fields via 
\begin{eqnarray}
S_{disl} = \int d^{D}x ~dt ~B^{\alpha a} J_{\alpha a}
\end{eqnarray}
for the coupling between the dislocation currents and gauges.
Mathematically, the structure is very similar to that in electromagnetism
when currents couples minimally to gauge fields. Just as charges screen
applied electromagnetic fields (as, e.g.,  in the Debye screening of plasmas), defect charges can screen the elastic shear.
In the uniform medium, when a defect condensate $(|\Psi_0| \neq 0$) appears, it gives rise a screening of
the elastic shear \cite{ref:cvetkovicNZ,ref:zaanenNM}, with a screening length
set by 
\begin{eqnarray}
\lambda_{shear} = \frac{c_{T}}{|\psi_{0}| \sqrt{\mu}}.
\label{higgs}
\end{eqnarray}
In Eq. (\ref{higgs}), $c_{T}$ is the transverse sound velocity and $\mu$ is the  
shear modulus.
In the case of simple dilation of the elastic moduli when $C_{\alpha \beta a b}(x,t) = w(x,t) K_{\alpha \beta a b}$
where $K_{\alpha \beta}$ is a space-time independent tensor and $w$ is a dilation function,
the calculation of the elastic shear penetration depth can be replicated for the amorphous
system to yield Eq.  (\ref{higgs})  when the quantities refer to the local values of
$\lambda_{shear}$, $c_{T}$, $\psi_{0}$, and $\mu$. 
A similar effect appears for general space-time dependent elastic moduli \cite{CVZ}
when considering elastic deformations about a local energy minimum.  
As is seen from Eq. (\ref{higgs}), {\em when the defect density tends to zero,
the elastic shear penetration depth diverges} 
\begin{eqnarray}
\lambda_{shear} \to \infty
\end{eqnarray}
as the system becomes rigid
throughout. In order to ascertain
detailed shear response in a system,
we may employ the shear stress correlation functions
as link weights within our community detection
algorithm. This is topic of a future study.

% lattice figures manual formatting ------------------------------------------------
%\clearpage %manual formatting
%manual formatting due to lots of figures %%%%%%%%%%%%%%%%%%%%%%%%%%%%%%%%%%%%%%%%%%
%\clearpage
%manual formatting due to lots of figures %%%%%%%%%%%%%%%%%%%%%%%%%%%%%%%%%%%%%%%%%%
%\noindent 

\myfig{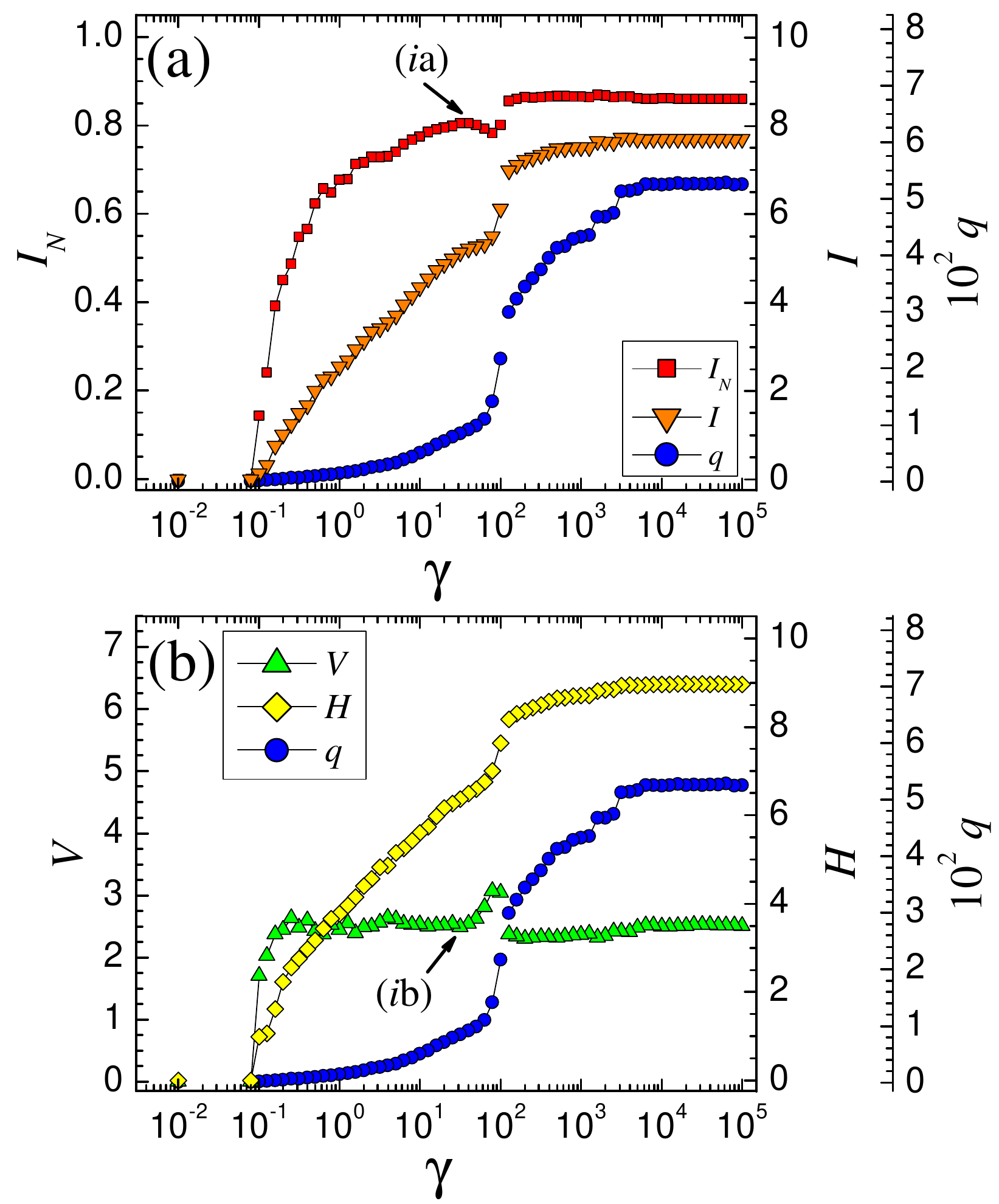}{A multiresolution analysis 
of a 2D triangular LJ lattice with periodic boundary conditions.
Edges are weighted according the LJ potential in \eqnref{eq:LJVr}
(with a potential shift $\phi_0=0.969925$).
There are two preferred regions: a small peak on the left and a large plateau 
on the right, where the peak here corresponds to the largest possible 
``natural'' clusters.
The defects make only a small alteration to the multiresolution plot.
A depiction of the system at $\gamma\simeq 31.6$ for the left peak
is shown in \figref{fig:LJLatticeVoidsPicture}.}
{fig:LJLatticeVoidsPlot}{\figsize}{h} %{t!}

\myfig{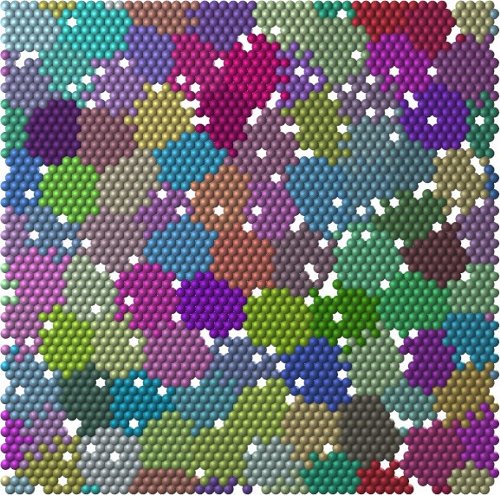}{A depiction of a partition of a 2D LJ 
triangular lattice with periodic boundary conditions.
The corresponding multiresolution plot is seen in \figref{fig:LJLatticeVoidsPlot}.
We use the algorithm described in the paper at $\gamma\simeq 31.6$
(the left peak) to solve the system.
Our algorithm places defects generally near the boundary of the communities
in order to minimize the energy cost of the defects in the community assignments.}
{fig:LJLatticeVoidsPicture}{\figsize}{h} %{b!}
% end LJ triangular lattice figures ----------------------------
\clearpage

% ********************** END DOCUMENT TEXT **************************
%\bibliography{mybiblio}
%\bibliography{/Users/Peter/Desktop/mybiblio}

\begin{thebibliography}{28}

\bibitem{brief} P. Ronhovde, S. Chakrabarty, M. Sahu, K. K. Sahu, K. F. Kelton, N. Mauro, and Z. Nussinov,
http://arxiv.org/pdf/1101.0008 (2010).

\bibitem{ref:physicsofAS}
R. Zallen, The Physics of Amorphous Solids, John Wiley \& Sons, Inc.,  pages 23--32 (1983).

\bibitem{ref:BMGintroMRS} A. L. Greer and E. Ma, MRS Bulletin {\bf 32}, 611 (2007).

\bibitem{ref:pharmasolubility} B. C. Hancock and M. Parks, Pharmaceutical Research {\bf 17}, 397 (2000).

\bibitem{ref:caseforBMG} M. Telford, Materials Today {\bf{7}}, 36 (2004).



\bibitem{wut} M.  Wuttig,  and N. Yamada, Nature Materials {\bf 6}, 824 (2007).



 \bibitem{jan}J. Zaanen, Nature {\bf 404}, 714  (2000).
 
\bibitem{rmp_steve} S.A. Kivelson, I. P. Bindloss, E. Fradkin, V. Oganesyan. J. M. Tranquada, A. Kapitulnik, and  C. Howald,  Rev. Mod. Phys. {\bf 75}, 1201 (2003).










\bibitem{ref:schenkISRO} T. Schenk, D. {Holland-Moritz}, V. Simonet, R. Bellissent,
           and D. M. Herlach, 
           Phys. Rev. Lett. {\bf 89}, 075507 (2002).
           
\bibitem{ref:keltonfirst} K. F. Kelton, G. W. Lee, A. K. Gangopadhyay, R. W. Hyers,
           T. J. Rathz , J. R. Rogers, M. B. Robinson and D. S. Robinson, Phys. Rev. Lett. {\bf 90}, 195504 (2003).

\bibitem{ref:nelsonGF} D. R. Nelson, ``Defects and Geometry in Condensed Matter Physics'', Cambridge University Press, Cambridge (2002).

\bibitem{ref:sadocmosseri} J. F. Sadoc and R. Mosseri,  ``Geometrical Frustration'', Cambridge University Press, Cambridge (1999).

        \bibitem{ref:tarjusAPT} G. Tarjus, S. A. Kivelson, Z. Nussinov, and P. Viot,
           J. Phys.: Condens. Matter {\bf 17}, R1143 (2005).
           
                \bibitem{ref:nussinovAPT} Z. Nussinov,
Phys. Rev. B {\bf 69}, 014208 (2004). 

          
               
           \bibitem{ref:lubchenkowolynes} V. Lubchenko and P. G. Wolynes,
           Annu. Rev. Phys. Chem.
           {\bf 58}, 235 (2007).
           
                 \bibitem{ref:angelaniPEL} L. Angelani, G. Parisi, G. Ruocco, and G. Viliani,
           Phys. Rev. E {\bf 61}, 1681 (2000). 
           

   \bibitem{ref:parisiglasstrans}      G. Parisi, Physica A
           {\bf 280}, 115 (2000).  

           \bibitem{ref:sastry} S. Sastry, P. G. Debenedetti, and F. H. Stillinger,
            Nature {\bf 393}, 554 (1998).
           
     
           \bibitem{ref:debenedetti} P. G. Debenedetti and F. H. Stillinger,
             Nature {\bf 410}, 259 (2001).
            

  \bibitem{ref:lubchenkoaging} V. Lunchenko and P. G. Wolynes,
            J. Chem. Phys. {\bf 121}, 2852 (2004).



            \bibitem{ref:johnsonMRS} W. L. Johnson, M. D. Demetriou, J. S. Harmon,
            M. L. Lind , and K. Samwer
            MRS Bulletin {\bf 32}, 644 (2007). 
           
           \bibitem{ref:doyestructures} Jonathan P. K. Doye, David J. Wales,
            Fredrik H. M. Zetterling, and Mikhail Dzugutov,
    J. Chem. Phys. {\bf{118}},
 2792 (2008).


   
          \bibitem{ref:ktw} T.R. Kirkpatrick, D. Thirumalai, and P. G. Wolynes,  Physical Review A {\bf 40}, 1045 (1989).

  \bibitem{ref:tm} M. Tarzia and M. A. Moore, 
Physical Review E {\bf 75},  031502 (2007). 

           \bibitem{ref:mode_coupling} W. Gotze, J. Phys.: Condes.
Matter {\bf 11}, A1 (1999).

\bibitem{ref:davidr} P. Mayer, K. Miyazaki, and D.R. Reichman, Phys. Rev. Lett. {\bf 97}, 095702 (2006). 
           
                      \bibitem{ref:dyn_con} J. P. Garrahan, J. Phys.: Condens. Matter {\bf 14}, 1571 (2002).



        
           \bibitem{ref:dk}
           D. Kivelson, S. A. Kivelson, X. Zhao, Z. Nussinov, and G. Tarjus,
Physica A {\bf 219}, 27 
(1995).
           
                
          \bibitem{ref:ritortsollich} F. Ritort and P. Sollich, Adv. Phys. {\bf 52}, 219 (2003).
            
            
            \bibitem{ref:cvetkovicNZ}
            V. Cvetkovic, Z. Nussinov, and J. Zaanen, 
Philos. Mag. {\bf 86}, 2995 (2006).
            
            \bibitem{ref:aharonov} 
 E. Aharonov, E. Bouchbinder, H. G. E. Hentschel,
            V. Ilyin, N. Makedonska, I. Procaccia, and N. Schupper,
  Euro. Phys. Lett. {\bf 77}, 56002 (2007). 
  
  
 \bibitem{ref:rev_bert}    L. Berthier and G. Biroli,  arXiv: 1011.2578 (2010). 
 
 \bibitem{ref:chandler} D. Chandler and J. P. Garrahan, Ann. Rev. Phys. Chem.  {\bf 61}, 191 (2010).

  \bibitem{ref:montanariCL}
  A. Montanari and G. Semerjian,  Journal of Statistical Physics {\bf 125}, 23-54 (2006).


    \bibitem{ref:tanakacritical} H. Tanaka, T. Kawasaki, H. Shintani, and K. Watanabe,
        Nature Materials {\bf 9}, 324 (2010).

 \bibitem{ref:mosayebiCLSGT}    M. Mosayebi, E. D. Gado, P. Iig, and H. C. Ottinger, Phys. Rev. Lett. {\bf 104}, 205704 (2010).

 \bibitem{ref:berthierCL} L. Berthier, G. Biroli, J.-P. Bouchaud, L. Cipelletti, D. El
Masri, D. L'Hote, F. Ladieu, and M. Pierno, Science {\bf 310},
1797 (2005).

\bibitem{ref:karmakarsastry} S. Karmakar, C. Dasgupta, and S. Sastry, Proc. Natl.
Acad. Sci. U.S.A. {\bf 106}, 3675 (2010).

\bibitem{ref:biroliverrocchio} G. Biroli, J.-P. Bouchaud, A. Cavagna, T. S. Grigera, and P. Verrocchio, Nature Physics {\bf 4}, 771 (2008).

\bibitem{ref:birolibouchaud} G. Biroli and J. -P. Bouchaud, e-print arXiv:0912.2542 (2009).

\bibitem{ref:kurchanlevineCL}
J. Kurchan and D. Levine, e-print arXiv:0904.4850
(2009).

\bibitem{ref:bernalrandom} J. D. Bernal,  Nature {\bf 185}, 68 (1960).

\bibitem{ref:miracleECP} D. B. Miracle, W. S. Sanders, and O. N. Senkov, Philos. Mag. {\bf 83}, 2409 (2003).


\bibitem{ref:miraclestructure} D. B. Miracle, 
Nature Materials {\bf 3}, 697-702 (2004).

\bibitem{ref:miraclestructuralMRS}
D. B. Miracle, T. Egami, M. K.  Flores, and K. F. Kelton,  MRS Bulletin {\bf 32},
629 (2007).

\bibitem{ref:miracleclusters} D. B. Miracle, E. A. Lord, and S. Ranganathan, Materials Transactions {\bf{47}},
1737 (2006).

\bibitem{ref:luoISROglass}
W. K. Luo, H. W. Sheng, F. M. Alamgir, J. M. Bai, J. H. He, and E. Ma, 
Phys. Rev. Lett. {\bf 92}, 145502 (2004). 

\bibitem{ref:ganeshISRO} P. Ganesh and M. Widom, Phys. Rev. B {\bf 77}, 0145205 (2008).

\bibitem{ref:shenicosahedral} Y. T. Shen, T. H. Kim, A. K. Gangopadhyay, and K. F. Kelton, Phys. Rev. Lett. {\bf 102}, 057801 (2009).

\bibitem{ref:delGadoIS}
E. D. Gado, P. IIg, M. Kroger, and H. C. Ottinger, Phys. Rev. Lett. {\bf 101}, 095501 (2008). 

\bibitem{ref:kobDH} W. Kob, C. Donati, S. J. Plimpton, P. H. Poole, and S. C. Glotzer, Phys. Rev. Lett. 79, 2827 (1997).


\bibitem{ref:weeksDH}  E. R. Weeks, J. C. Crocker, A. C. Levitt, A. Schofield, and D. A. Weitz, Science {\bf  287}, 627 (2000).

\bibitem{ref:widmercooperHF} A. Widom-Cooper, P. Harrowell, and H. Fynewever, Phys. Rev. Lett. {\bf 93}, 135701 (2004). 

\bibitem{ref:stevensonSW} J. D. Stevenson, J. Schmalian, and P. G. Wolynes, Nature Physics {\bf 2}, 268 (2006). 

\bibitem{ref:glassesoverview} L. Berthier, G. Biroli, J.-P. Bouchaud, and R. L. Jack, e-print:1009.4665 (2010).

\bibitem{ref:sheng} H. W. Sheng, W. K. Luo, F. M. Alamgir, J. M. Bai, and E. Ma, 
Nature {\bf 439}, 419 (2006).

\bibitem{ref:finneyRSP} J. L. Finney, Proc. R. Soc. London, Ser. A
 {\b{319}}, 479 (1970).
 
    
           \bibitem{ref:honeycuttHA}  J. Dana Honeycutt and Hans C. Andersen,  J. Phys. Chem. {\bf 91}, 4950 (1987).
           
           \bibitem{ref:steinhardtBOO} P. J. Steinhardt, D. R. Nelson and M. Ronchetti, Phys. Rev. B {\bf 28}, 784  (1983).

\bibitem{ref:HufnagelSROMRO} T. C. Hufnagel and S. Brennan, Phys. Rev. B {\bf 67}, 014203 (2003). 

\bibitem{ref:treacyMRO} M. M. J. Treacy, J. M. Gibson, L. Fan, D. J. Paterson, and I. McNulty, Rep. Prog. Phys. {\bf 68}, 2899 (2005). 

   \bibitem{ref:rzmultires}
          P. Ronhovde and Z. Nussinov, Phys. Rev. E {\bf 80}, 016109 (2009).
          


          \bibitem{ref:rzlocal}
          Peter Ronhovde and Zohar Nussinov, Physical
Review E {\bf 81}, 046114 (2010).




\bibitem{GN} M. Girvan and M. E. J. Newman, Proc. Natl. Acad. Sci. USA {\bf 99}, 7821 (2002).
\bibitem{fortunato} S Fortunato, Physics Reports {\bf 486}, 75 (2010).
\bibitem{fortunato1} A. Lancichinetti and S. Fortunato, Phys. Rev. E {\bf 80}, 056117 (2009).
\bibitem{newman_girvan}  M. E. J. Newman and M. Girvan, Phys. Rev. E {\bf 69}, 026113 (2004). 
\bibitem{blondel} V. D. Blondel, J.-L. Guillaume, R. Lambiotte, and E. Lefebvre, J. Stat. Mech. {\bf 10}, P10008 (2008). 
\bibitem{newman_fast} M. E. J. Newman, Phys. Rev. E {\bf 69},  066133 (2004). 
\bibitem{RB}   J. Reichardt and S. Bornholdt, Phys. Rev. E {\bf 74}, 016110 (2006).
\bibitem{newman-vector} M. E. J. Newman, Phys. Rev. E {\bf 74}, 036104 (2006).
\bibitem{dynamicshighd}  V. Gudkov, V. Montelaegre, S. Nussinov, and Z. Nussinov,
Phys. Rev. E  {\bf 78}, 016113 (2008).
\bibitem{osc} A. Arenas,  A. D«õaz-Guilera, and C. J. P«erez-Vicente, Phys. Rev. Lett. {\bf 96}, 114102 (2006).
\bibitem{RosB} M. Rosvall and C. T. Bergstrom, Proc. Natl. Aca. Sci. U.S.A. 105, 1118-1123 (2008).
\bibitem{book_comm} Book chapter by U. Brandes, D. Dellng, M. Gaertler, R. Gorke, M. Hoefer, Z. Nikoloski, and D. Wagner, ``On finding graph clusterings with maximum modularity'' in the book ``Graph-thoeretic concepts in computer science'', Lecture notes in Computer Science, Springer Berlin/Heidelberg, DOI: 10.1007/978-3-540-74839-7 (2007).







\bibitem{ref:vi} M. Meila, J. Multivariate Anal. {\bf 98}, 873 (2007).



\bibitem{explain_scales} This includes, e.g., (i) relaxation processes of structural glass formers relating to
slow global ($\alpha$) process and local rapid rearrangments ($\beta$ processes),
and (ii) systems with multiple degrees of freedom (structural, magnetic, and other)  that may exhibit transitions
at different temperatures. 



\bibitem{ref:mezardRTGT} M. Mezard and A. Montanari, 
J. Stat. Phys. {\bf 124}, 1317 (2006). 

\bibitem{ref:franzKGM} S. Franz and A. Montanari, 
J. Phys. A. {\bf 40}, F251 (2007). 

\bibitem{ref:sahuAlYFe}
K. K. Sahu, N. A. Mauro, L. Longstreth-Spoor, D. Saha, Z. Nussinov, M. K. Miller, and K. F. Kelton, Acta Materialia {\bf 58}, 4199 (2010).

\bibitem{ref:egamiatomistic} T. Egami, J. Non-Cryst. Solids {\bf 317}, 30 (2003). 

\bibitem{ref:imd} J. Stadler, R. Mikula, and H. R. Trebin, Int. J. of Mod. Phys. C {\bf 8}, 1131 (1997). 

\bibitem{ref:mihalkovicEOPP} M. Mihalkovi\v{c}, C. L. Henley, M. Widom, and P. Ganesh, 
e-print arXiv:cond-mat.mtrl-sci/0802.2926 (2008). 

\bibitem{ref:vaspwww} VASP website:  http://cms.mpi.univie.ac.at/vasp/

\bibitem{ref:moriartyGPT} J. A. Moriarty and M. Widom, Phys. Rev. B {\bf 56}, 7905 (1997). 

\bibitem{ref:kobandersenOne} W. Kob and H. C. Andersen, Phys. Rev. E {\bf 51}, 4626 (1995).


\bibitem{ref:valdesMixing} L. C. Valdes, F. Affouard, M. Descamps, and J. Habasaki,
J. Chem. Phys. {\bf 130}, 154505 (2009).



\bibitem{commentv}
For overdamped systems (such as particles in viscous liquids) obeying ``Aristotelian dynamics'' 
($\vec{F} \propto \vec{v}$), clusters in which the force
on all their individual particles is nearly the same, will move
with uniform velocity. This leads to a high inter-particle correlations between
particles belonging to the same cluster. 

\bibitem{ref:huCDPT} 
D. Hu, P. Ronhovde, and Z. Nussinov, e-print arXiv:1008.2699 (2010).

\bibitem{ref:multires}
 A. Arenas, A. Fernandez, and S. Gomez, New. J. Phys. 10, 053039 (2008).


\bibitem{ref:kumpulamultires} J. M. Kumpula, J. Saramaki, K. Kaski, and J. Kertesz, 
Fluct. Noise Lett. {\bf 7}, L209 (2007).


\bibitem{ref:lanc} A. Lancichinetti, S. Fortunato, and J. Kertesz, New J. Phys. {\bf 11}, 033015 (2009).

\bibitem{ref:muchaSCI} P. J. Mucha, T. Richardson, K. Macon, M. A. Porter, and J. -P. Onnella, Science {\bf 328}, 876 (2010).

\bibitem{ref:fenndynamic} D. J. Fenn, M. A. Porter, M. McDonald, S. Wiliams, and N. F. Johnson,
Chaos {\bf 19}, 033119 (2009). 


\bibitem{irr} Asaph Widmer-Cooper, Heidi Perry, Peter Harrowell, and David R. Reichman, 
Nature Physics {\bf 4}, 711  (2008).


\bibitem{reich} J. X. Lin, C. Reichhardt, Z. Nussinov, L. P. Pryadko, and
          C. J. Olson Reichhardt, Physical Review E {\bf 73}, 061401 (2006).
          
          \bibitem{manning}
          M. Lisa Manning and Andrea J. Liu,   arXiv:1012.0064 (2010).



\bibitem{spectral_cluster1} W. Donath and A. Hoffman, IBM Journal of Research and Development {\bf 17(5)}, 420 (1973).

\bibitem{spectral_cluster2} M. Fiedler,  Czech. Math. J. {\bf 23(98)}, 298 (1973).



\bibitem{kur}Y. Kuramoto, Chemical Oscillations, Waves and Turbulence (Springer-Verlag, Berlin, Germany) (1984).



\bibitem{griffiths} R. B. Griffiths, Phys. Rev. Lett. {\bf 23}, 17 (1969).


\bibitem{effpotentials} S. Chakrabarty, M. Widom, M. Mihalkovi\v{c}, K. F. Kelton, and Z. Nussinov, in preparation


\bibitem{ref:nakamuraZrPt}
T. Nakamura, E. Matsubara, M. Sakurai, M. Kasai, A. Inoue, and Y. Waseda,
J. Non-Cryst. Solids {\bf 312-314}, 517 (2002).

\bibitem{ref:saidaZrPt}
J. Saida, K. Itoh, K. Sato, M. Imafuku, T. Sanada, and A. Inoue, J. Phys.:Cond. Matt. {\bf 21}, 375104 (2009).

\bibitem{ref:sordeletZrPt}
D. J. Sordelet, R. T. Ott, M. Z. Li, S. Y. Wang, M. F. Besser, A. C. Y. Liu, and M. J. Kramer, Metallurgical and Materials Transactions A {\bf 39}, 1908 (2007).

\bibitem{ref:wangwangZrPt}  S. Y. Wang, C. Z. Wang, M. Z. Li, L. Huang, R. T. Ott, M. J. Kramer, D. J. Sordelet, and K. M. Ho,
Phys. Rev. B {\bf 78}, 184204 (2008). 

\bibitem{ref:maurokeltonZrPt} N. A. Mauro and K. F. Kelton, submitted to Phys. Rev. B.


\bibitem{ref:mcgreevyULS} R. L. McGreevy, J. Phys.: Condens. Matter {\bf 3} F9 (1991).

\bibitem{ref:keenRMC} D. A. Keen and R. L. McGreevy, Nature {\bf 344}, 423-5 (1990).

\bibitem{ref:kimkelton} T. H. Kim and K. F. Kelton, 
J. Chem. Phys. {\bf 126}, 054513 (2007).

\bibitem{ref:zaanenNM}
J. Zaanen, Z. Nussinov, and S. I. Mukhin, 
Annals of Physics {\bf 310}, 181 (2004).

\bibitem{CNMZ1}
V. Cvetkovic, Z. Nussinov, S. Mukhin, and J. Zaanen, 
Europhys. Lett. {\bf 81}, 27001 (2008)


\bibitem{kleinert} H. Kleinert, ``Gauge fields in condensed matter, vol. II: Stresses and defects, Differential Geometry, Crystal Defects'',
World Scientific, Singapore (1989) 




\bibitem{CVZ}
V. Cvetkovic,  M. Vasin, and Z. Nussinov, in preparation.


\end{thebibliography}

% ********************* END COMPLETE DOCUMENT *********************
\end{document}